%% file: VPipe.tex
\pdfoutput=1
\documentclass[namedreferences]{SolarPhysics}
\usepackage[optionalrh,solaenum]{spr-sola-addons} 
\usepackage{color}                       
\usepackage{url}                     

\usepackage[margin=1.25in]{geometry}
\usepackage[pdfborder={0 0 0 },urlcolor=blue,breaklinks]{hyperref}
\ifx \doiurl    \undefined \def \doiurl#1{\href{http://dx.doi.org/#1}{\textsf{\textsf{DOI}}}}\fi
\ifx \adsurl    \undefined \def \adsurl#1{\href{http://adsabs.harvard.edu/abs/#1}{\textsf{\textsf{ADS}}}}\fi
\ifx \arxivurl  \undefined \def \arxivurl#1{\href{http://arxiv.org/abs/#1}{\textsf{\textsf{arXiv}}}}\fi
\ifx \urlurl    \undefined \def \urlurl#1{\href{http://#1}{\textsf{#1}}}\fi

\usepackage{graphicx}                    
\usepackage{caption}
\usepackage{listings}
\usepackage{lineno}
\usepackage{hyperref}
\hypersetup{colorlinks=true, citecolor=black, filecolor=black, linkcolor=black}

\newcommand{\arcsec}{{\prime\prime}}

\begin{document}
\begin{article}

\setcounter{section}{0}


\input{VP1-Intro.txt}

\input{VP2-Pipeline.txt}
\input{VP2A-HARP.txt}

\input{VP3-Inversion.txt}

\input{VP4-Disambiguation.txt}
\input{VP5-HARP.txt}
\input{VP6-Uncertainties.txt}
\input{VP7-Summary.txt}

\section*{Acknowledgement}
We thank the numerous team members who have contributed to the success of the SDO mission and
particularly to the HMI instrument. This work was supported by NASA Contract NAS5-02139 (HMI) to
Stanford University. Some of the research described here was carried out by staff of the Jet 
Propulsion Laboratory, California Institute of Technology. HAO/NCAR is supported by the 
National Science Foundation. Efforts at NWRA were also supported through NASA Contract NNH09CF22C
and by PO\# NNG12PP28D/C\# GS-23F-0197P from NASA/Goddard Space Flight Center. J. Schou acknowledges
support from EU FP7 Collaborative Project {\it Exploitation of Space Data for Innovative Helio- 
and Asteroseismology} (SPACEINN).

\appendix
\input{VA1-Segments.txt}
\input{VA2-1332e15.txt}

\input{VA3-SHARPDetails.txt}

\bibliographystyle{spr-mp-sola}
\bibliography{VPipe}

\end{article}
\end{document}

%% file: VP1-Intro.txt
\begin{opening}

\title{The Helioseismic and Magnetic Imager (HMI) Vector Magnetic Field Pipeline: Overview and Performance}


\author{J.~Todd~\surname{Hoeksema}$^{1}$}
\author{Yang~\surname{Liu}$^{1}$}
\author{Keiji~\surname{Hayashi}$^{1}$}
\author{Xudong~\surname{Sun}$^{1}$}
\author{Jesper~\surname{Schou}$^{1,5}$}
\author{Sebastien~\surname{Couvidat}$^{1}$}
\author{Aimee~\surname{Norton}$^{1}$}
\author{Monica~\surname{Bobra}$^{1}$}
\author{Rebecca~\surname{Centeno}$^{2}$}
\author{K.D.~\surname{Leka}$^{3}$}
\author{Graham~\surname{Barnes}$^{3}$}
\author{Michael~\surname{Turmon}$^{4}$}


\runningauthor{J.T. Hoeksema \textit{et al.}}
\runningtitle{HMI Vector Magnetic Field Observations}


\institute{$^{1}$ W.W. Hansen Experimental Physics Laboratory, Stanford University, Stanford, CA 
email: \href{mailto:jthoeksema@sun.stanford.edu}{JTHoeksema@sun.stanford.edu}}

\institute{$^{2}$ High Altitude Observatory (NCAR), Boulder, CO }

\institute{$^{3}$ Northwest Research Associates, Inc., Boulder, CO }
\institute{$^{4}$ Jet Propulsion Laboratory, Pasadena, CA }
\institute{$^{5}$ Max-Planck-Institut f\"ur Sonnensystemforschung,
Justus-von-Liebig-Weg 3, 37077 G\"ottingen, Germany}


\begin{abstract}

The \textit{Helioseismic and Magnetic Imager}
(HMI) began near-continuous full-disk
solar measurements on 1 May 2010 from the \textit{Solar Dynamics Observatory}
(SDO). An automated processing pipeline keeps pace with observations to
produce observable quantities, including the photospheric vector magnetic
field, from sequences of filtergrams. The basic vector-field frame list 
cadence is 135 seconds, but to reduce noise the filtergrams are combined
to derive data products every 720 seconds. The primary 720s observables were
released in mid 2010, including Stokes polarization parameters measured at 
six wavelengths as well as intensity, Doppler velocity, and the line-of-sight 
magnetic field. More advanced products, including the full vector
magnetic field, are now available. Automatically identified HMI
Active Region Patches (HARPs) track the location and shape of magnetic
regions throughout their lifetime.

The vector field is computed using the Very Fast Inversion of the Stokes Vector
(VFISV) code optimized for the HMI pipeline; the remaining $180^\circ$ azimuth
ambiguity is resolved with the Minimum Energy (ME0) code. The Milne-Eddington
inversion is performed on all full-disk HMI observations. The disambiguation,
until recently run only on HARP regions, is now implemented for the full
disk. Vector and scalar quantities in the patches are used to derive active
region indices potentially useful for forecasting; the data maps and indices
are collected in the SHARP data series, {\sf hmi.sharp\_720s}. Definitive SHARP
processing is completed only after the region rotates off the visible disk;
quick-look products are produced in near real time. Patches are provided in
both CCD and heliographic coordinates.

HMI provides continuous coverage of the vector field, but has modest spatial,
spectral, and temporal resolution. Coupled with limitations of the analysis
and interpretation techniques, effects of the orbital velocity, and instrument
performance, the resulting measurements have a certain dynamic range and sensitivity
and are subject to systematic errors and uncertainties that are characterized
in this report.

\end{abstract}

%
\keywords{Magnetic fields, Photosphere; HMI: Vector Field; Solar Active Regions}

\end{opening}


\section{Introduction}\label{sec:Introduction}

The \textit{Helioseismic and Magnetic Imager}
(HMI, \opencite{Schou-calib2012}) provides the
first uninterrupted time series of space-based, full-disk, 
vector magnetic field observations of the Sun with a 12-minute cadence.
This report summarizes the HMI vector magnetic field analysis pipeline, provides an
assessment of the quality of the vector data, and characterizes some of
the known uncertainties and systematic errors. 

HMI was launched 11 February 2010 as part of the \textit{Solar Dynamics Observatory} (SDO,
\opencite{Pesnell2012}) and began taking continuous solar measurements
on 1 May. The HMI investigation \cite{Scherrer2012} studies the origin and
evolution of the solar magnetic field and links to the dynamics of the corona
and heliosphere by producing long uninterrupted time series of the global
photospheric velocity, intensity, and magnetic fields.

\begin{figure}
\begin{center}
\includegraphics[angle=0, width=0.95\textwidth]{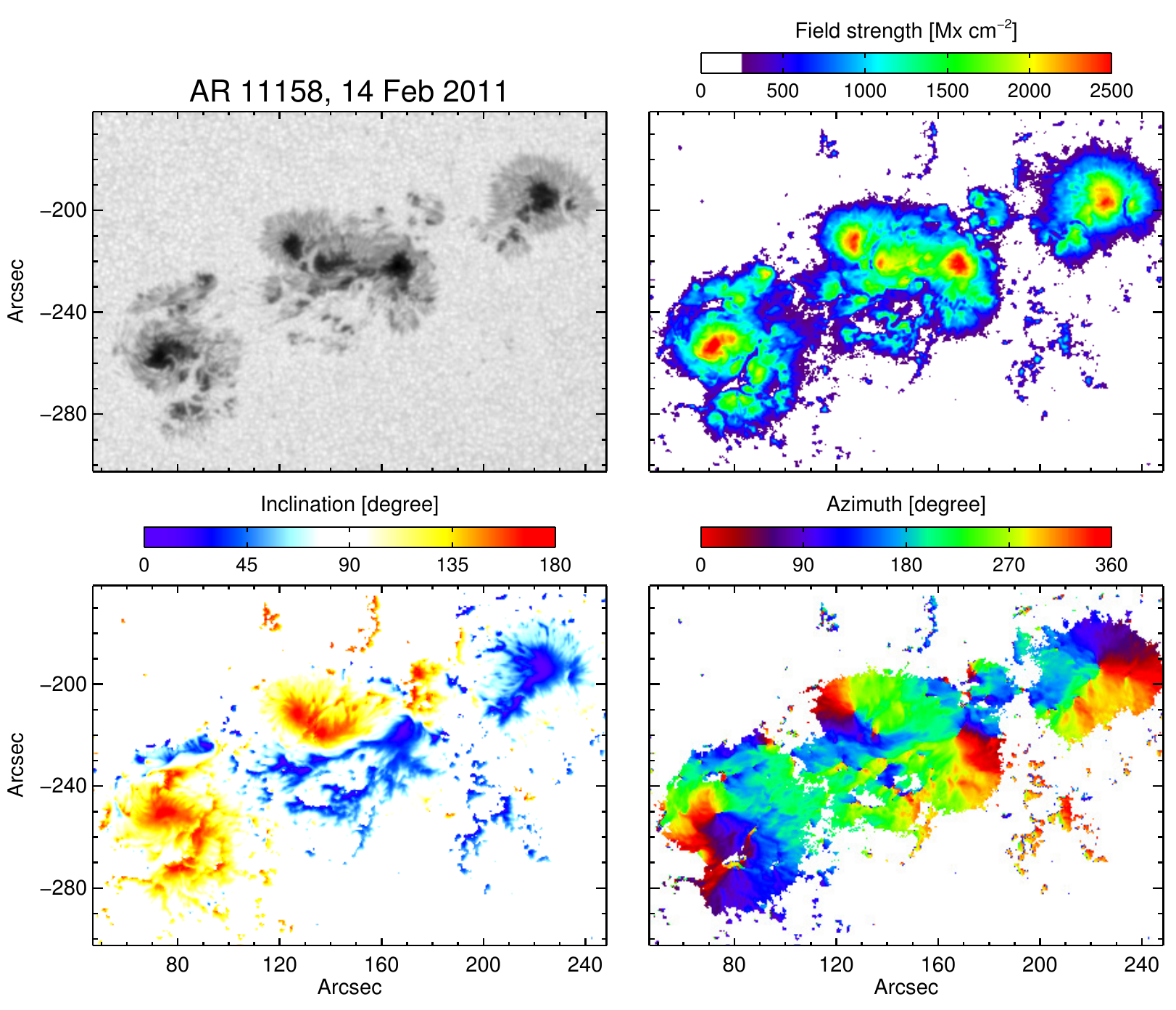}
\caption{
HMI Active Region Patch 377 (NOAA AR\,\#11158) produced the first X-class
flare of Solar Cycle 24 early on 15 February 2011. This figure shows the
intensity and magnetic field components observed near disk center a few hours 
earlier, at 19:00 TAI on 14 February. The panel in the upper left shows
the continuum intensity in CCD coordinates labeled in arc seconds from disk
center. The image has been rotated $180^\circ$ to put solar north near the
top. The upper right panel shows the total field strength,
for pixels with total field strength above 250~Mx~cm$^{-2}$,
saturating red at 2500~Mx~cm$^{-2}$.  The angle of inclination relative to
the line of sight appears in the lower left panel, with field directed toward
the observer shown in blue; transverse field in white, and field directed
away from the observer in red.
The lower right panel shows the azimuth angle, adjusted to give the angle relative to
the direction of rotation (\textit{i.e.} west).  Only pixels above a 250\,G threshold
are plotted.  Data are from the series {\sf hmi.sharp\_720s}.
}
\label{fig:AR11158}
\end{center}
\end{figure}

HMI measures full-disk scalar quantities -- Doppler shift, longitudinal
magnetic field, continuum intensity, line depth, and line width -- using a
repeating sequence of narrow-band images recorded every 45 seconds with a 
$4096 \times 4096$ camera called the ``Doppler'' camera.
This paper focuses primarily on the analysis of data from a
second, identical ``Vector'' camera that measures linear and circular polarization with a 135\,s cadence
frame list \cite{Schou-polarization2012}. 
From these filtergrams the vector magnetic
field and thermodynamical parameters are determined. See Figure \ref{fig:AR11158}. 

Scientific observables are produced by a series of software modules operating
in an analysis pipeline.  HMI collects filtergrams almost continuously.
The Level-0 reconstruction of the instrument data and computation of most quick-look
observables is completed within a couple of minutes.  Some preliminary analysis
is completed using this near-real-time (NRT) data. Validation and filling
of gaps in the scientific and housekeeping data stream typically takes a
little more than a day, after which the final calibration and processing is
initiated to produce definitive Level-1 filtergrams.

The filtergrams are calibrated further to correct a variety of instrument
distortions and remove trends. 
Temporal and spatial interpolations are applied to remove the
effects of cosmic rays and data gaps, and to correct for temporal evolution
and solar rotation. Combining the calibrated filtergrams produces
the scientific observables. The primary vector magnetic observable is the
Stokes vector data series, {\sf hmi.S\_720s}, in which the I, Q, U,
and V polarimetric quantities are determined every 12 minutes at six
wavelengths (Section \ref{sec:Stokes}). Table \ref{tab:DataSeries}
lists relevant HMI data series.

Active regions emerge quickly and span a large range of size and
complexity. An automated code identifies and tracks HMI Active Region
Patches (HARPs) using the HMI 720-second line-of-sight (LoS) magnetic field and
computed continuum intensity images (\opencite{Turmon2013} and Section \ref{sec:HARP}).
HARPs are analogous to NOAA Active Regions, but there are many more, smaller
patches. HARPs aim to identify complete flux complexes, and so may be larger
and incorporate more than one NOAA AR. HARPs follow a region during
its entire lifetime or disk passage, beginning one day before the first flux
emerges and continuing one day after the region has decayed. The HARP data
series, {\sf hmi.Mharp\_720s}, defines the geometry of a rectangular 
bounding box on the
CCD that encloses the maximum heliographic area achieved by a patch during
its life time. At each time step a rectangular bitmap identifies which CCD 
pixels lie within the patch.

To recover useful information about the Sun's vector magnetic field an
inversion of the Stokes vector must be performed that requires specific
assumptions about the solar atmosphere. Such inversion can be 
computationally intensive. Section \ref{sec:MEInversion} briefly describes how the
{\tt fd10} version of the VFISV code \cite{vfisv2010,Centeno2013} that is 
based on a relatively simple Milne-Eddington atmosphere has been adapted 
and optimized for use in the HMI pipeline. All images since 1 May 2010 have 
been processed using a version of {\tt fd10} ({data series: \sf hmi.ME\_720s\_fd10}).

The inversion applied to each pixel cannot resolve the inherent $180^\circ$ azimuth 
ambiguity in the transverse field direction \cite{Harvey1969}, so a non-local
minimization is applied \cite{Metcalf1994,Barnesetal2012}. The disambiguation
is sensitive to noise in the inverted values, which varies spatially and
temporally in
the HMI data. Section \ref{sec:disambiguation} describes the scheme in
more detail and Section \ref{sec:magnoise} explains the method by which
the noise threshold is determined. The disambiguation requires computing
resources that increase with the size of the region and when
spherical geometry is required; so the near-real-time pipeline routinely 
processes smaller active region patches in rectilinear coordinates. 
Disambiguation in spherical coordinates of larger patches
has been accomplished using the same module as the definitive data.
Pipeline full-disk disambiguation of definitive data has begun for
data collected after 19 December 2013.

The disambiguated field is being computed for each active region patch every 12 minutes.
From the vector field a representative set of indices is computed that may be
of interest for space weather forecasting purposes. The indices include total
unsigned flux, current helicity, and mean horizonal field gradient, to name just three;
see Section \ref{sec:SHARP}. The SHARP (Space-weather HARP) data series
({\sf hmi.sharp\_720s}) includes for each numbered HARP at each time record 
all of the indices as
keywords, as well as data arrays providing maps of nearly all the HMI scalar
and vector observables and uncertainties.  The pipeline produces a version
of SHARPs in the standard CCD coordinates and another that is remapped and
projected onto heliographic cylindrical equal area coordinates centered on
the HARP, ({data series: \sf hmi.sharp\_cea\_720s}).
Figure~\ref{fig:AR11158} shows the vector field components and derived continuum
intensity for HARP\,377 (NOAA AR\,\#11158) observed at 19:00 TAI on 14 February 2011.
HMI times are given in International Atomic Time (TAI), which is 35s ahead of
UTC in 2013.

\begin{table}
\begin{center}
\caption{HMI Magnetic Field and Related Data Series Available in JSOC}
\label{tab:DataSeries}
\begin{tabular}{rp{0.6\textwidth}}
\hline
Data Series\tabnote{A description of JSOC Data Series is provided in Appendix A} ~Name & HMI Data Product\\
\hline
 & {\bf Scalar Observables} \\
{\sf hmi.S\_720s} & Level 1 Stokes-Polarization Images - 720s Averages \\
 & 24 images: I\,Q\,U\,V at six wavelengths \\
{\sf hmi.M\_720s}\,$^\dag$ & Full-disk Line-of-sight Magnetic Field \\
{\sf hmi.M\_45s} & 45-second Full-disk Line-of-sight Magnetic Field \\
{\sf hmi.V\_720s} & Full-disk Line-of-sight Doppler Velocity \\
{\sf hmi.Ic\_720s} & Full-disk Computed Continuum Intensity \\
{\sf hmi.Ic\_noLimbDark\_720s}\,$^\dag$ & Ic with Limb Darkening Removed\\
{\sf hmi.Lw\_720s} & Full-disk Computed Line Width \\
{\sf hmi.Ld\_720s} & Full-disk Computed Line Depth \\
{\sf hmi.Mharp\_720s}\,$^\dag$ & HMI Active Region Patch ({\sc harp}) Geometry Information \\
 & Tracked Bitmap Computed from LoS Magnetic Field \\
\hline
 & {\bf Vector Observables} \\
{\sf hmi.ME\_720s\_fd10}\,$^\dag$ & Full-Disk VFISV {\tt fd10} Milne-Eddington Inversion \\
 & Magnetic and Plasma Parameters with Uncertainties \\
{\sf hmi.B\_720s} & Disambiguated Full-Disk Vector Magnetic Field \\
{\sf hmi.sharp\_720s}\,$^\dag$ & Vector and Scalar Field HARPs $+$ Active-Region Indices \\
 & Region Tracked for Entire Disk Passage Every 720s \\
{\sf hmi.sharp\_cea\_720s}\,$^\dag$ & Remapped Field Patches and AR Indices \\
 & Cylindrical Equal Area Projection of $B_r, B_{\theta}, B_{\phi}$ \\
{\sf hmi.lookup\_ChebyCoef\_BNoise} & Velocity-Dependent Noise Masks for Vector Field \\
\hline
\multicolumn{2}{l}
{$^\dag$\, Series available in both definitive and near-real-time versions, e.g.  hmi.sharp\_720s\_nrt} \\
		\end{tabular}
		\end{center}
\end{table}

The HMI vector field data are of good and consistent quality and the random
noise characteristics (about 100 Mx cm$^{-2}$ in the total magnetic field
strength) in quiet regions are consistent with the design specifications of the
instrument and the uncertainties intrinsic to the VFISV Milne-Eddington 
inversion code. 
The dominant time varying systematic errors arise from the $\pm 3$~km\,s$^{-1}$
daily velocity shift of the spectral line due to the
geosynchronous orbit of SDO.
The effects can be mitigated; nevertheless, daily variations that 
depend on velocity, field strength, and disk position remain in 
the data. See Section \ref{sec:uncertainties}.

Comparison of the LoS magnetic field strength derived from HMI
LCP and RCP observations analyzed using the simpler MDI-like method
(see Section \ref{sec:LineOfSight} and \opencite{Couvidat-wavelength2012}) with
field strengths obtained using the {\tt fd10} inversion show differences at
values as low as 1000 G. These arise from difficulties in precisely modeling
the spectral line and the detailed instrument spectral characteristics. The
dynamic range limitations of the instrument do not become significant until
nearly 3000 G (lower in high radial-velocity regions). 
Comparison with higher $0.3^\arcsec$-resolution $Hinode$/SP data 
qualitatively match the $1^\arcsec$-resolution HMI observations in AR\,\#11158. However, the flux density and
intensity contrast reported by HMI are lower for a variety of reasons, as
may be expected, see Section \ref{sec:hinode}. 

\subsection{Paper Overview}

Section \ref{sec:Pipeline} gives more detail about the HMI pipeline processing
steps relevant to the quality and interpretation of the vector field
data through to the production of the Stokes observable. 
The active region identification and tracking analysis is summarized in
Section \ref{sec:HARP}.
Section \ref{sec:MEInversion} describes the vector field inversion, specifically
improvements to the VFISV code. 
Section \ref{sec:disambiguation} explains the disambiguation
processing. The SHARPs are described in Section \ref{sec:regions}. 
In Section \ref{sec:uncertainties} the sensitivity of the instrument, systematic errors,
and other limitations of the data are discussed. The final section 
provides a summary.
Three appendices provide details of the data
segments in the final vector field data series, describe differences between the 
pipeline processing and an earlier released version of the HMI vector field, and
give more detail about the space weather quantities in SHARPs.


%% file: VP2-Pipeline.txt
\section{The HMI Magnetic Field Pipeline Processing}
\label{sec:Pipeline}

\subsection{The HMI Instrument and Data Flow}
\label{sec:Instrument}

This section briefly describes the HMI instrument and data flow for
the basic observables. The HMI instrument design and planned processing 
scheme are described by \inlinecite{Schou-calib2012} and 
\inlinecite{Scherrer2012}.
The SDO mission, launched 11 February 2010, is described by \inlinecite
{Pesnell2012}. The Joint Science Operations Center (JSOC) 
serves both the HMI and \textit {Atmospheric Imaging Assembly} (AIA) instruments on SDO,
providing pipeline processing of the incoming data products as well
as archival and retrieval services for investigators who want to use the
data. Extensive on-line documentation for the JSOC data center can be accessed
at {\tt jsoc.stanford.edu}; see also Table \ref{tab:websites} in
Section \ref{sec:summary}.

The HMI telescope feeds a solar image through a series of 
bandpass filters onto two $4096^2$-pixel CCD cameras. Each camera records
a full-disk image of the Sun every 3.75 seconds in a 76 m\AA\ 
wavelength band selected by tuning the final stage of a Lyot filter and
two Michelson interferometers across the Fe~\small{I}~6173.34 \AA\ absorption line.
Six wavelengths are measured in different
polarizations in a sequence defined by a continuously repeating frame list. 
One camera measures right and left circular polarization at each wavelength,
completing a 12-filtergram set every 45s from which the Doppler velocity,
LoS magnetic field, and intensities can be determined. The second,
Vector camera measures six polarization states every 135s: nominally I$\pm$V, I$\pm$Q, 
and I$\pm$U, where I\,Q\,U\,V are the Stokes polarization parameters. From
these the vector magnetic field and other plasma parameters can be derived.

The filtergrams are cropped, effectively truncated to the
photon noise limit using a look-up table, and then losslessly compressed before
being
downlinked in real time to a ground station at
White Sands, NM from geosynchronous orbit with a nominal bit rate
of 55 Mb/s. The data are decoded and buffered and transmitted to the JSOC at
Stanford University with a delay of about a minute. As the first step of the
JSOC SDO data analysis pipeline, the telemetry are validated and checked for
completeness; they can be retransmitted within 60 days from data retained
at White Sands. Level~0 filtergrams are reconstructed and used together
with the housekeeping and flight dynamics data to produce calibrated Level~1
filtergrams.
The primary HMI observables are computed from these definitive Level~1 filtergrams.

The HMI observables are available since 1 May 2010. There are very few gaps
in coverage. However, because of SDO's inclined orbit, twice yearly, during two-week
intervals around March and September, the Earth eclipses the Sun for up to
72 minutes per day near local midnight. Calibrations briefly interrupt the
observing sequences each day at 6 UT and 18 UT and there are occasional disruptions due
to eclipses, station keeping, special calibrations, intense rain, ground
equipment problems, etc. As one measure, through the end of 2012 98.44\%
of the possible 1.876 million 45\,s Dopplergrams are available. Instrument
performance is constantly monitored and adjustments are made periodically to
account for changes in instrument performance, such as focus, filter drift,
and optical transmission.

\subsection{Stokes Vector Processing Description}
\label{sec:Stokes}\label{sec:level1}

\begin{figure}
\centerline{
\includegraphics[width=1.0\textwidth]{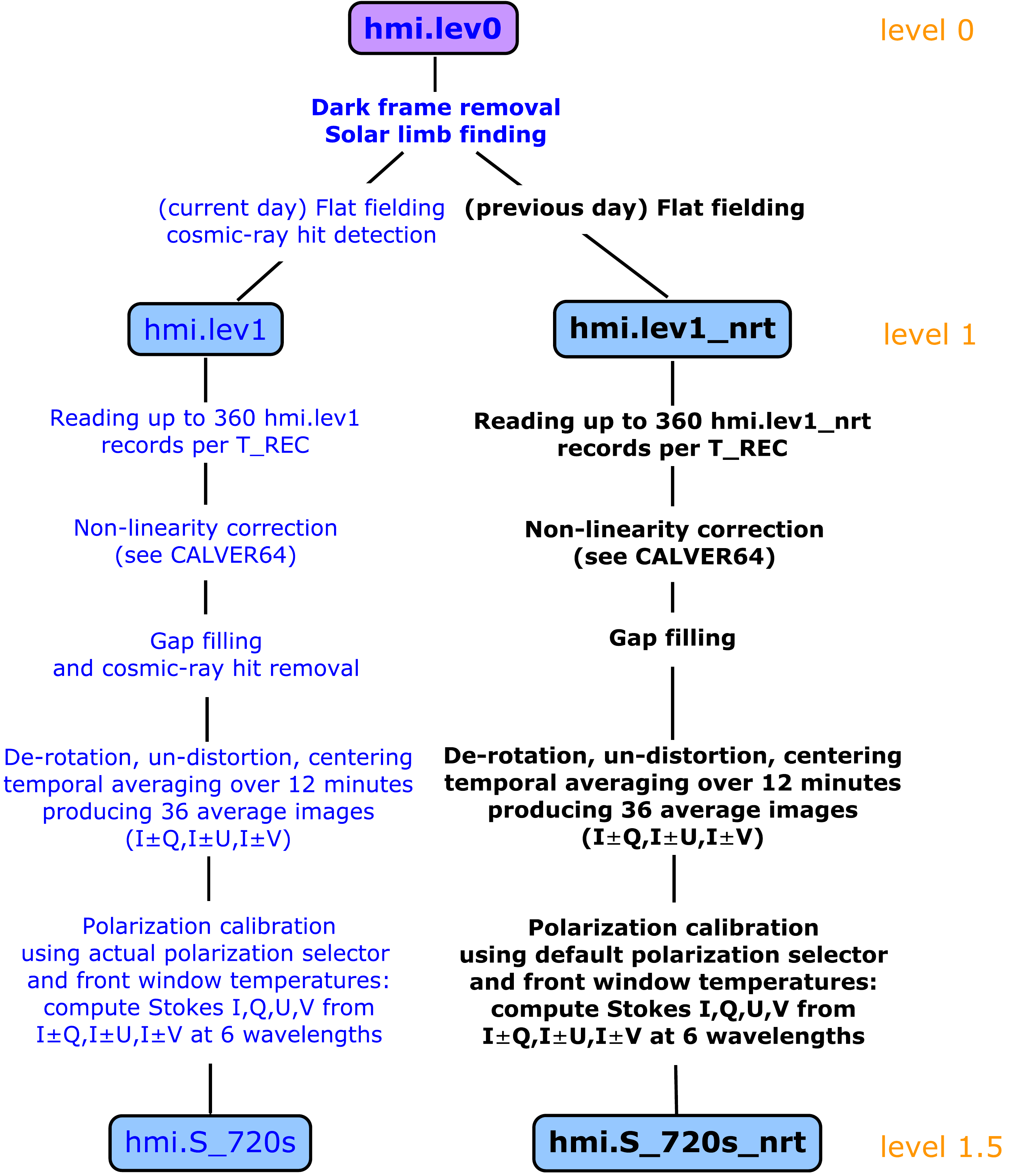}}
\caption{
The HMI vector field pipeline from Level~0 filtergrams to the Level~1.5
Stokes observable. Reconstructed Level~0 filtergrams are corrected for dark
current, the disk location is determined, and the images are flat fielded.
For each time record, the resulting Level~1 images are corrected for
temporally and spatially dependent instrumental non-linearities; gaps in
the time series are interpolated; bad pixels are corrected; and each
filtergram is undistorted, co-registered, and derotated to compensate
for solar rotation.
A tapered temporal average is performed every 720 seconds using observations
collected over 22.5 minutes (1350\,s) to reduce noise and minimize the effects of solar
oscillations. The averaged values for each wavelength and
polarization tuning are corrected for known spatially and temporally varying
polarization effects. The filtergrams are linearly combined to determine the
Stokes I\,Q\,U\,V values at each wavelength and archived in {\sf hmi.S\_720s}.
The near real time pipeline, right, uses preliminary calibration values and not all
the recovered data may be available.
}
\label{fig:StokesPipeline}
\end{figure}

The Stokes-vector observables code produces averaged I\,Q\,U\,V 
images at six wavelengths on a regular 12-minute cadence centered at the
time given in the data series keyword {\sc t\_rec}.
Figure~\ref{fig:StokesPipeline} summarizes the processing steps.
The vector field observing sequence captures six polarizations at
each wavelength according to a repeating 135\,s framelist. 
The code uses 360 Level~1 filtergrams taken by the Vector camera of HMI.
Level~1 filtergrams have had a dark frame
subtracted, been flat-fielded, and had the overscan rows and columns of the CCD removed. 
A list of 
permanently and temporarily bad pixels is stored in the {\sc bad\_pixel\_list} segment of
each {\sf hmi.lev1} record.  (See Appendix~\ref{app:SegmentDescriptions} for more
about JSOC data series nomenclature.) 
A limb fitting code finds the solar disk center
coordinates on the CCD and the observed solar radius. 
Because the formation height of the signal changes, the limb
position measured by HMI changes with wavelength away from the Fe\,\small{I} line
center. Consequently, the measured radius (corrected for the SDO-Sun
distance) varies as a function of the difference between the target wavelength
and the wavelength corresponding to the known SDO-Sun velocity, {\sc obs\_vr}.
We regularly verify our model of this variation from the measured radius.
Since the velocities at the east and west limbs differ due to solar rotation, an
offset also appears in the center position of the image.
The keywords for the disk center location ({\sc crpix1} and {\sc crpix2})
and solar radius at the SDO distance ({\sc rsun\_obs}) 
returned by the limb finder are corrected by up to about half a pixel
for the atmospheric height sampled at each wavelength; the 
reported image scale ({\sc cdelt1=cdelt2}) is consistent with these values.

Weekly flat fields are obtained by offsetting the HMI field of view (FOV) with the
piezo-electric transducers \cite{Wachter2009}. 
Periodically observables calculated for a {\sc t\_rec} close
to midnight use Level~1 filtergrams corrected with different
flat fields. Flat fields are stored in the {\sf hmi.flatfield} series. 
Every three months offpoint flat fields are obtained by moving the 
legs of the instrument. 

The HMI cameras are affected by a small nonlinearity in their response to
light exposure (see Figure 19 of \opencite{Wachter2012}). This correction, of
the order of 1\% for intensities below 12000 DN/s, is implemented separately
for Doppler and Vector camera data. A third-order polynomial, based on
intensities obtained from ground calibration sequences and averaged over
the four quadrants of the CCDs, is applied to Level~1 values. The version
of this calibration is indicated by the {\sc calver64} keyword.

After collecting the appropriate Level~1
records for the entire time window, a gap filling routine deals
with bad pixels and cosmic ray hits in each filtergram \cite{Schou2013}.
The observables code then calls a temporal averaging routine that performs
several operations:

\begin{itemize} 

\item De-rotation: 
each filtergram is re-registered to correct for the pixel shift caused by
solar differential rotation. This correction is done at sub-pixel accuracy
using a Wiener spatial-interpolation scheme. The time difference used to
calculate the pixel shift is the precise observation time of the filtergram,
{\sc t\_obs}.

\item Un-distortion: 
each filtergram is corrected for optical distortions that can be up to
a few pixels in magnitude. The instrumental distortion as a function of
field position is reconstructed from Zernike polynomials 
determined during pre-launch calibration using a random-dot
target mounted in the stimulus telescope \cite{Wachter2012}.
The solar disk center and radius of the undistorted images are recalculated.
This center and radius are slightly different from the Level~1
values calculated using distorted images.

\item Centering:
each filtergram is spatially interpolated to a common solar-disk center
and radius that are averages of the input Level~1 filtergrams used to produce the
I\,Q\,U\,V observables at {\sc t\_rec}. These first three steps are done in a
single operation.

\item Temporal Averaging: 
de-rotated, un-distorted, and re-centered filtergrams are combined with a weighted average for the target time {\sc t\_rec}.

\end{itemize}

Conceptually the temporal averaging is performed in two steps. First a temporal Wiener
interpolation of the observed filtergrams onto a regular temporal grid
with a cadence of 45\,s is performed; this results in a set of 25 frames for
each wavelength/polarization state constructed using the ten original 135s
framelists. The full time window over which the interpolation is performed
is 1350s, which is wider than the averaging window; a wider window is
required as the interpolation needs filtergrams before and after
the interpolated times. That is followed by the averaging of the
frames using an apodized window with a FWHM of 720s; the window is a boxcar
with cos$^2$ apodized edges that nominally has 23 non-zero weights, of which
the central nine have weight $1.0$. Temporal gap filling is also performed
if needed. 

The pipeline next calls the calibration routine
that converts the six polarizations taken by the observables sequence into a
Stokes [I\,Q\,U\,V] vector. To perform this conversion the polarimetric model
described in \inlinecite{Schou-calib2012} is used, including the corrections
dependent on the front window temperature and the polarization selector 
temperature.  At each point in the image a least squares fit is then 
performed to derive I\,Q\,U\,V from the six observed polarization states.

Two additional corrections based on post-launch analysis are applied to the 
model described in \inlinecite{Schou-polarization2012}. The
first compensates for what looks like telescope polarization, a 
spatially dependent
term proportional to I that appears in the demodulated Q and U
at the level of about a part in $10^4$.
The dependence of Q/I, U/I, and V/I on distance from
disk center was determined using the good-quality images from 3 May - 3 September
2010.
The effect on V is negligible, so no correction is performed on V. 
The coefficients of proportionality 
for Q and U
are given as fourth-order polynomials in the 
square of the 
distance from the center of the image 
and this allows for a correction to better than a few parts in $10^5$.
It may be noted that this is not
strictly a telescope polarization term, because it depends on the polarization
selector setting.

While the need for the first correction was anticipated, the second was not.
A perceptible (relative to the photon noise) granulation-like pattern appears
in Q and U (again, V is largely unaffected). This signal appears to
be caused by a point spread function (PSF) that differs with polarization
state. The consequence is that a contamination from I convolved with a
different PSF is added to the two linear polarization signals. This is corrected
by convolving I with a five by five kernel and subtracting the result. At
present a spatially independent kernel is used.
Details can be found in the pipeline code (See
\href{http://jsoc.stanford.edu/jsocwiki/PipelineCode}{{\sf PipelineCode}} in Table
\ref{tab:websites}).

The final result of the Stokes-vector observables code are the four
Stokes parameters at the six wavelengths in {\sf hmi.S\_720s}.

\subsection{Line-of-Sight Field Processing}
\label{sec:LineOfSight}

The line-of-sight observables code is called to compute the Doppler velocity
and LoS magnetic field strength, as well as the Fe~\small{I} line width, line depth
and continuum intensity, from the Stokes parameters at each time step.
The LoS observables are calculated with the MDI-like algorithm detailed in
\inlinecite{Couvidat-wavelength2012}. Briefly, discrete estimates of the first
and second Fourier coefficients of the solar neutral iron line are calculated
from the six wavelengths, separately for the I $\pm$ V polarizations. The Doppler
velocity is proportional to the phase of the Fourier coefficients; currently,
only the first Fourier coefficient is used to compute the velocity. Certain
approximations made in this calculation introduce errors in the results: the
HMI filter profiles are not delta functions, the discrete estimate of the
Fourier coefficients is not completely reliable due to the small number of
wavelength samples, and the spectral line profile is not Gaussian (an assumption
made to relate the phase to the velocity). Hence Doppler velocities returned by
the basic algorithm must be corrected.  This is accomplished in two steps. 

The first relies on tabulated functions saved in the {\sf hmi.lookup} data series. These
tables are based on calibrated HMI filter transmission profiles and on a more
realistic Fe~\small{I} line profile. They vary slightly across the HMI CCDs and with
time, primarily due to drifts in the Michelson interferometers. The values
also depend on the tuning of the instrument and must be re-calculated after
each HMI re-tuning.  However, uncertainties remain in the filter transmission
profiles and, perhaps more importantly, the Fe~\small{I} line profiles; consequently
the look-up tables do not completely correct the Doppler velocity. 

A second correction step has also been implemented. Every day the relationship
between the median Doppler velocity {\sc rawmedn} (measured over $99\%$
of the solar disk) minus the accurately known Sun-SDO radial velocity ({\sc obs\_vr})
is fitted as a function of {\sc rawmedn} with a 3rd-order polynomial. A linear
temporal interpolation of the polynomials closest in time to {\sc t\_rec}
is used to further adjust the Doppler velocities. The HMI LoS magnetic field
measurement is proportional to the difference between the corrected I$+$V and I$-$V
Doppler velocities. Only Doppler velocities and field strengths
are corrected in this way; no correction algorithm has been implemented for
the other LoS observables.  Table \ref{tab:DataSeries} lists the names of
data series for the HMI pipeline quantities.

\subsection{Quicklook / Near-Real Time Pipeline Processing}

Definitive data products are computed after all downlinked data have been retrieved
and all housekeeping, calibration, and trend data have been analyzed. Typically it takes a
couple days from the time data are observed until the definitive observables are
complete. For instrument health and safety monitoring and for quick-look 
space-weather purposes, much of the data are also processed in a near-real-time mode. 
During the last quarter of 2012 96\% of the NRT observables were available within 60 minutes
and an additional 1\% were processed in the second hour. Because of disk and
data base issues, 3\% of the NRT products became available more than 2 hours after the 
observations were recorded. Note that the higher level vector products derived
from the {\sf hmi.S\_720s\_nrt} data series require up to three hours of
additional processing time.

The algorithms for processing the NRT and definitive 720s data products are 
essentially the same. The differences arise due to the lack of robust cosmic 
ray detection, stale flat fielding maps, and the use of default or predicted 
instrument parameters and housekeeping information. Differences in the data are 
generally small or localized. Note that in the parallel 45s reduction pipeline, the 
processing of the definitive and NRT observables are less alike because fewer 
images are used to perform the temporal interpolation and gap filling. 
Higher level NRT products are generally not archived, with
the exception of the SHARP product described in Section \ref{sec:SHARP}.

%% file: VP2A-HARP.txt
\section{The Geometry of HMI Active Region Patches - HARPs}
\label{sec:HARP}

\begin{figure}
\centerline{
\includegraphics[width=1.0\textwidth]{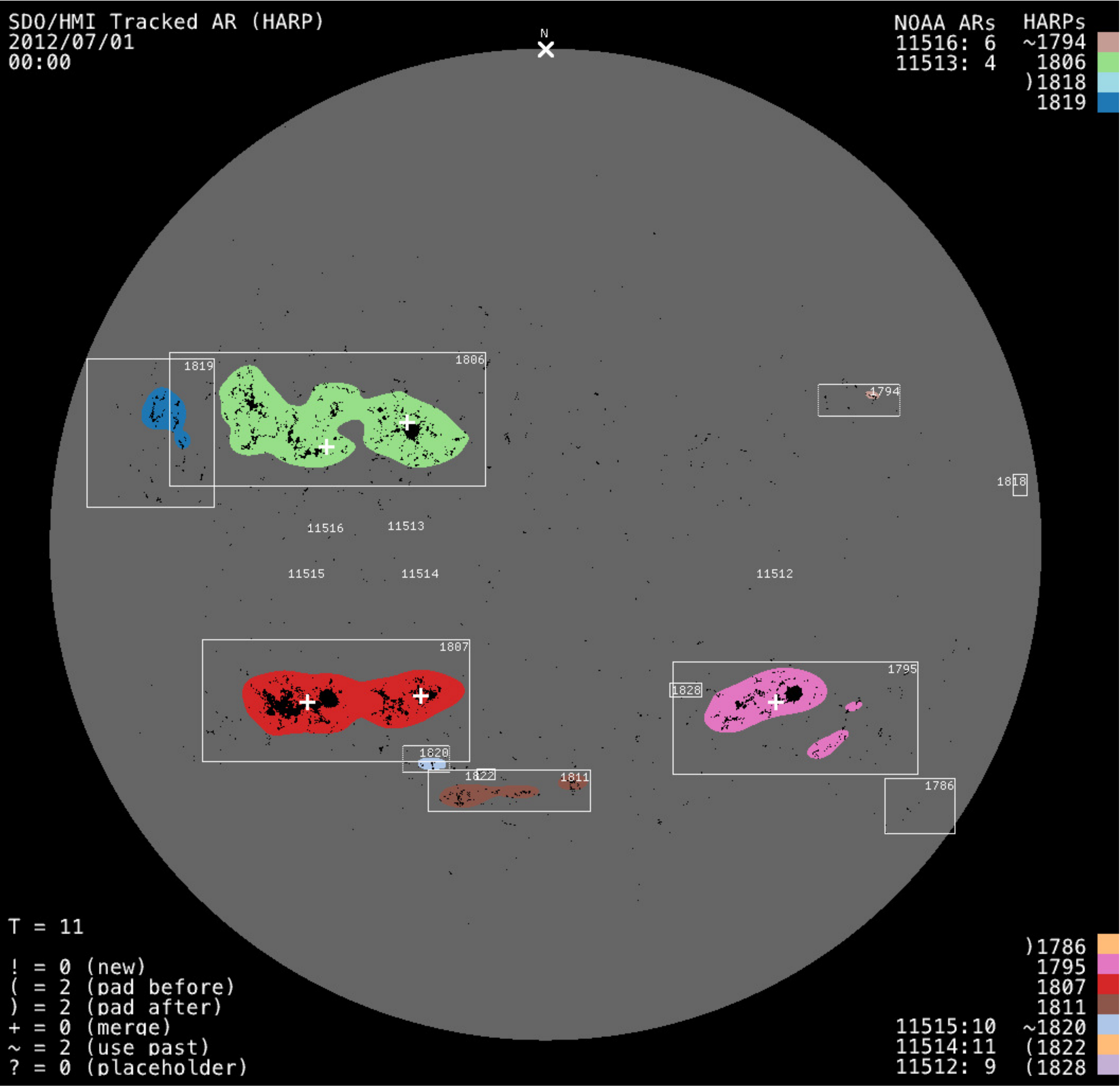}
}
\caption{The eleven HARPs present on the Sun at 00:00 TAI on 1 July 2012, 
four in the north and seven in the south, are shown with their life-time
maximum bounding boxes. Each box is labeled with the HARP number in the upper
right corner. The colored patch encloses the active region (with active 
pixels shown in black) associated with the HARP bitmap at the time of the image. 
White $+$ signs mark
the reported locations of NOAA Active Regions; the NOAA number appears at the
same longitude near the equator for identification purpsoses. 
HARPs may be associated with one or more
NOAA active regions (\textit{e.g.} HARP 1807 is associated with NOAA AR\,\#11515 and
\#11514), but most are not. HARP bounding boxes may overlap, as 1819 and 1806 do in
this image, or a small HARP may even be completely enclosed inside another,
\textit{e.g.} 1822 inside 1811. Colored patches denoting unique HARPs never overlap.  
Definitive HARPs are tracked both before emergence
(\textit{e.g.} 1822) and after decay (\textit{e.g.} 1786), as indicated by the parenthesis character
to the left of the HARP number in the legends in the two right corners.
The key for the HARP information appears in the lower left. Additional
information can be found at
{\tt jsoc.stanford.edu/data/hmi/HARPs\_movies/movie-note}.
}
\label{fig:HARP}
\end{figure}

Magnetic regions of various sizes emerge, evolve, and sometimes disappear
ra\-pid\-ly.  HARPs provide primarily spatial information about long-lived, 
coherent magnetic structures at the size scale of a solar active region.
The data series {\sf hmi.Mharp\_720s} catalogs HARPs that are automatically 
identified using HMI 720s LoS magnetograms and intensity images. 
HARPs are given an indentifying number, {\sc harpnum}, and followed
during their entire disk passage. 
Many HARPs can be associated with NOAA active regions (ARs). 
A typical set of definitive HARP patches inside their bounding
boxes is shown in Figure~\ref{fig:HARP} for 1 July 2012.
Since we could not afford to fully process all of the vector magnetic field data,
we concentrate some of the HMI vector pipeline analysis on the patches surrounding
active regions, particularly for the NRT observations (see Section \ref{sec:SHARP}).  
Definitive full-disk vector data collected after 19 December 2013 are being fully
processed.
Depending on available resources and performance, the complete 
full-disk vector field may ultimately be computed at a lower cadence 
or more intensively for selected earlier intervals.

There is not a 1:1 correspondence between HARPs and NOAA ARs.  Each HARP
is intended to encompass a coherent magnetic structure, so two or more NOAA
ARs may all belong to a single HARP region.  Coherent regions that are small
in extent or have no associated sunspot can be detected and tracked by our
code; such faint HARPs often have no NOAA correspondence. In the interval
from May 2010 through January 2012 there were 350 NOAA ARs and 1187
definitive HARPs.

HARPs are identified at each time step in the 720s HMI data series according
to the following steps. The module first identifies the magnetically
active pixels: approximately, those pixels with absolute LoS field greater
than 100\,G. Then, taking spherical geometry into account, active pixels
are grouped into instantaneous activity patches, which generally have
smooth contours, but may consist of several regions separated by quiet
Sun~\cite{Turmon2002,Turmon2010}. The instantaneous patches, which are the
colored blobs in Figure~\ref{fig:HARP}, are combined into a temporal track
by stringing together image-to-image associations of the patches. A simple
latitude-dependent motion model is part of the association metric, so data gaps
of less than two days do not cause association problems.  Note that there is
no requirement that the net flux be zero, so a HARP may be somewhat unipolar.

An individual HARP is therefore a list, over a time index, of all the
associated instantaneous patches, plus geometric metadata and selected
summary statistics.  The instantaneous patches are stored as coded bitmaps
coinciding with a rectangular bounding box within the CCD plane. The bitmap's
numerical code identifies those pixels that are part of the active region
patch and also indicates whether each pixel is magnetically active or quiet.
See Appendix \ref{app:bitmap} for details.  The potential exists to introduce
additional encodings, such as sunspot umbra/penumbra.

The geometric information in the HARP is used to determine how parts
of the solar disk are processed in the vector magnetic field pipeline. In
particular, the rectangular bounding box (shown in white
in Figure~\ref{fig:HARP}) is the smallest pixel-aligned box enclosing an inner
latitude/longitude box. The latter box has constant width in heliographic
latitude and
longitude, and rotates across the disk at a constant angular velocity (defined
for each HARP). This latitude/longitude box is just big enough to enclose all
instantaneous patches during the lifetime of the HARP. All these geometric
parameters are stored in HARP keywords. Some elementary summary characteristics
of each HARP at each time are also determined, \textit{e.g.} the total LoS flux, net
LoS flux, total area, and centroid, as well as any corresponding NOAA region
number (or numbers).  The keyword {\sc noaa\_ar} gives the number of the first
NOAA active region associated with a HARP.  If more than one NOAA region is
associated with a HARP, the total number is given by {\sc noaa\_num} and a comma
separated list of regions appears in {\sc noaa\_ars}. 
Maps of the observed solar quantities are not part of the HARP data series;
the observables are collected and additional parameters
are provided as part of the space-weather HARPs (SHARPs) 
described in Section \ref{sec:SHARP}.

There are in fact two types of HARP, definitive and near real time (NRT). The
main distinction is that the definitive HARP geometry is not finalized
until after the associated magnetic region has decayed or rotated off the
disk, so all geometric information referred to above is consistent for the
lifetime of the region.  In particular, the bounding box of a definitive
HARP encloses the same heliographic area during its entire lifetime. The
temporal life of a definitive HARP starts when it rotates onto the visible
disk or one day before the magnetic feature is first identified in the
photosphere. The HARP expires one day after the feature decays or when it
rotates completely off the disk. Steps are taken to ensure that activity
fluctuating around the threshold of detection is not separated into multiple
HARPs with temporal breaks. Additionally, active regions that grow over time,
eventually overlapping other regions, are merged into a single definitive HARP.

A near-real-time HARP is similar, but does not begin until the feature can be
identified (there is no one-day pad at the beginning). Also, the heliographic
size of the region may change during its life.  NRT HARPs may also merge, but
the merger is not seamless, because it results in termination of one or more
NRT HARPs and continuation of the larger, merged object.  Flags are provided to
identify mergers in the NRT product, because the sudden amalgamation of an NRT
HARP with another region is an artifact of the grouping and not of physical
origin. In the definitive product, regions do not merge because the entire
history of the HARP is known before the bounding box is determined. Note
that definitive and NRT HARP numbers are not the same.

\begin{sloppypar}
HARPs are available in the data series {\sf hmi.Mharp\_720s\_nrt} 
and {\sf hmi.Mharp\_720s}. 
More information about HARPs is available on-line
at \href{http://jsoc.stanford.edu/jsocwiki/HARPDataSeries}{{\sf HARPDataSeries}} in Table
\ref{tab:websites}) (see Table \ref{tab:websites})
and in \inlinecite{Turmon2013}.
\end{sloppypar}

%% file: VP3-Inversion.txt
\section{Milne-Eddington Inversion}\label{sec:MEInversion}

Spectral line inversion codes are tools used to interpret spectro-polarimetric
data. Given a set of observed Stokes spectral profiles, the purpose of an
inversion code is to infer the physical properties of the atmosphere where
they were generated.

In order to interpret HMI data, a relatively-simple spectral line synthesis
model, based on a Milne-Eddington (ME) atmosphere, is used. The ME model
assumes that all of the physical properties of the atmosphere are constant
with height, except for the source function, which varies linearly with
optical depth. The generation of polarized radiation is described by the
classical Zeeman effect. Under these assumptions, the polarized radiative
transfer equation has an analytical solution, known as the Unno-Rachkovsky
solution \cite{Unno1956,Rachkovsky1962,Rachkovsky1967}. In the ME context,
the forward model for spectral line synthesis traditionally has 11 free
parameters: three magnetic-related quantities (magnetic field strength, its
inclination with respect to the line of sight, and its azimuth with respect
to an arbitrarily chosen direction in the plane perpendicular to the LoS),
five thermodynamical ones (Doppler width, line core-to-continuum opacity
ratio, Lorentz-wing damping, and the source function and its gradient), two
kinematic variables (Doppler velocity and macroturbulent velocity), and a
geometrical parameter, the filling factor, that quantifies the fraction of
the pixel that is occupied by a magnetic structure.

VFISV (Very Fast Inversion of the Stokes Vector) is a ME spectral
line inversion code specifically designed to infer the vector magnetic field of
the solar photosphere from HMI Stokes measurements
\cite{vfisv2010,Centeno2013}.  VFISV uses a
Levenberg-Marquardt least-squares minimization \cite{Press1992} of a $\chi^2$
function. $\chi^2$ is a metric of the difference between the observed and
synthetic Stokes profiles (\textit{i.e.} the goodness of the fit). Given an initial
guess for the model atmosphere, the algorithm produces synthetic Stokes
profiles and compares them to the observations. The model parameters are
then modified in an iterative manner until the synthetic data are deemed to
be a good-enough fit to the observations. `Good enough' is defined by the
convergence criteria that determine when to exit the iteration loop.

Several limitations arise as a consequence of using the ME approximation. The
thermodynamical parameters are complicated, non-linear combinations of
actual physical properties (density, temperature, pressure, etc.), so one
cannot extract the corresponding real thermodynamical information using
only the ME assumption. Also, due to the lack of gradients in the model,
asymmetries present in the Stokes profiles cannot be properly fit by
the inversion code. However, the retrieved values of magnetic field and
LoS velocity are representative of the average conditions throughout the
spectral region of formation and averaged over temporal and spatial resolution
\cite{westendorp1998,LekaBarnes2011}.
HMI does not have an absolute wavelength reference. The zero point of the
derived velocity depends on filter profile calibrations determined relative to
the full-disk line center at the nominal wavelength of 6173.3433
\cite{Norton2006}; no correction is made for the effect of convective blue
shift on the observations. See Figure \ref{fig:vel} in Section \ref{sec:hmimdi} for
information about the daily variation of the velocity.
\inlinecite{Welsch2013} have developed a scheme to determine the velocity
zero point in active regions.

Despite being a general ME code with customizable settings, VFISV has been
optimized to work within the HMI data processing pipeline. The inversions
are typically applied to the {\sf hmi.S\_720s} series (full Stokes vector,
averaged over 720 seconds, for the full disk of the Sun, with $4096 \times 4096$
pixels), which requires being able to invert more than 20,000 pixels per
second. In order to meet the speed requirements, certain aspects of VFISV
have been hard-coded for performance purposes; for instance, VFISV uses the
HMI transmission filter profiles for each pixel in the computation of the
synthetic spectral line.  One important facet of this inversion code
is that it forces the filling factor, $\alpha$, to be unity.  This constraint, enforced
due to the sparse spectral sampling and moderate spatial resolution of the HMI
data, assumes that the plasma is uniformly magnetized in each pixel. In the
weak field regime, there is a strong degeneracy between the filling factor
and the magnetic field strength. However, the product of the two ($\alpha B$) 
is very well constrained by the observed Stokes profiles. Therefore,
assuming $\alpha = 1$ means that the quantity $B$ should be interpreted as
the average magnetic field of the pixel. Also, the stray light, \textit{i.e.} light
scattered from other parts of the Sun that contaminates any given pixel,
is currently not being considered. This generally leads to lower values of
the inferred magnetic field strength, especially in strong-field regions
\cite{straylight}.

Several changes and optimizations have been introduced in VFISV since its
release \cite{vfisv2010}. The HMI team has derived optimal settings for the
pipeline processing in order to improve the speed and general performance of
the code. These changes to the original VFISV code are described in detail
in \inlinecite{Centeno2013}. Some of the significant changes 
include: 
a) a regularization term added to the $\chi^2$ minimization merit 
function to bias the solution towards the more physical solution 
when encountering two nearly degenerate minima, 
b) an adjustment to the relative weighting applied to the Stokes parameters I\,Q\,U\,V, 
c) implementation of tailored iteration and exit criteria 
that results in an average of $\sim 30$ iterations per pixel,
d) use of a spectral line
synthesis module that computes the full radiative transfer solution 
only for the inner wavelength range of the HMI spectral line, and
e) a simpler initial guess scheme that does not use a neural net. 
In the HMI pipeline context, the inversions produced with these settings are referred to as {\tt fd10}.

Two keywords indicate the version of the {\tt fd10} code:
{\sc invcodev} and {\sc codever4}. A significant improvement was incorporated on
30 April 2013, when the code was enabled to use time-dependent information about the
HMI filter profiles. The differences are spatially dependent and systematic.
Over the full disk, a typical difference map between the two versions shows a mean of
0.1~G and an rms of 4~G, with a distinct east/west asymmetry and greater bias
in strong-field regions.
The filter profiles must be updated when the instrument tuning is adjusted,
as indicated by the {\sc invphmap} keyword. Inversions of data collected
between 1 August 2011 and 23 May 2013 use earlier, less accurate filter phase maps.
Notable changes to the pipeline modules and processing notes are
documented at \href{http://jsoc.stanford.edu/jsocwiki/PipelineCode}{{\sf
jsoc.stanford.edu/jsocwiki/PipelineCode}} as listed in Table
\ref{tab:websites}.

For each pixel in the field of view, VFISV returns its best estimate of
the atmospheric model together with the standard deviation associated with each
model parameter and the normalized covariances in the errors of selected pairs of parameters.
Appendix \ref{sec:e15w1332} describes differences between {\tt fd10} and a
previous version of the inversion associated with an earlier data release.


%% file: VP4-Disambiguation.txt
\section{The Disambiguation Algorithm}\label{sec:disambiguation}

Inversion of the Zeeman splitting to infer the magnetic field component
transverse to the line of sight results in an ambiguity of $180^\circ$ in its
direction \cite{Harvey1969}; some additional assumption(s) or approximation
must be made in order to completely determine the magnetic field vector. A
plethora of methods have been proposed to resolve this ambiguity (for an
overview, see \opencite{ambigworkshop}). For disambiguation of HMI data, a
variant on the ``Minimum Energy'' method \cite{Metcalf1994} called ME0 has been
implemented.

The original Minimum Energy algorithm minimizes the sum of the squares of the
magnitude of the field divergence and the total electric current density,
where the vertical derivatives in each term are approximated by the
derivatives of a linear force-free field. Minimizing
$|\mbox{\boldmath$\nabla$} {\mathbf{\cdot}} \mbox{\boldmath$B$}|$ gives a
physically meaningful solution and minimizing ${\rm{\bf J}}$ provides a smoothness
constraint. For computational efficiency, the algorithm applied to HMI data
uses a potential field to approximate the vertical derivative in
$|\mbox{\boldmath$\nabla$} {\mathbf{\cdot}} \mbox{\boldmath$B$}|$ and
minimizes only the magnitude of the normal component of the current density.
Note that now we minimize 
$\sum {\left| \mbox{\boldmath$\nabla$} {\mathbf{\cdot}} \mbox{\boldmath$B$} \right| + \left| \mbox{${\rm J}$}_z \right|} $
in place of 
$\sum {|\mbox{\boldmath$\nabla$} {\mathbf{\cdot}} \mbox{\boldmath$B$}|^2 
 + \left|{\mbox{\rm J}}\right|^2}$ 
as in the original Minimum Energy method.

This method has been shown to be among the best performing on a wide range of
tests on synthetic data for which the answer is known
\cite{ambigworkshop,ambig2}. However, \inlinecite{ambig2} showed that the
minimum energy state is not always the correct ambiguity resolution in the
presence of noise. Further, convergence of the optimization routine is slow in
pixels where the signal is dominated by noise, particularly noise in the
transverse component of the field. Thus, the final disambiguation for those
pixels that fall below a specified noise threshold is done with several
simpler and quicker methods in the pipeline. 

The steps needed for implementing the algorithm are:
\begin{itemize}
\item Determine spatially dependent noise mask;
\item Compute potential field derivative;
\item Determine confidence in result;
\item Minimize energy function;
\item Disambiguate noisy pixels.
\end{itemize}
These basic steps apply to disambiguation of both HARPs and full-disk images.
The code currently uses a planar approximation when computing the energy 
for small HARPs
($<20^\circ$ extent in both latitude and longitude), but uses spherical
geometry for larger HARPs or full disk. Further details of the algorithm as
used in the HMI pipeline are presented in \inlinecite{Barnesetal2012}. 

\subsection{The Noise Mask}

The HMI noise level varies with location on the disk and with relative
velocity between the spacecraft and Sun (see Section \ref{sec:magnoise} for a
discussion of the temporal and large scale spatial variations found in the
data). A constant is added to the derived noise value, and pixels with a
transverse field strength above this noise estimate are assigned to the
disambiguation mask. The mask is then eroded to eliminate isolated pixels above
the noise and this defines the high-confidence pixels. Intermediate confidence
is assigned to a buffer zone surrounding the clusters of pixels with a well
determined field. All pixels within that grown mask are included in minimizing
the energy.

\begin{figure}
\centerline{\includegraphics[width=0.99\textwidth]{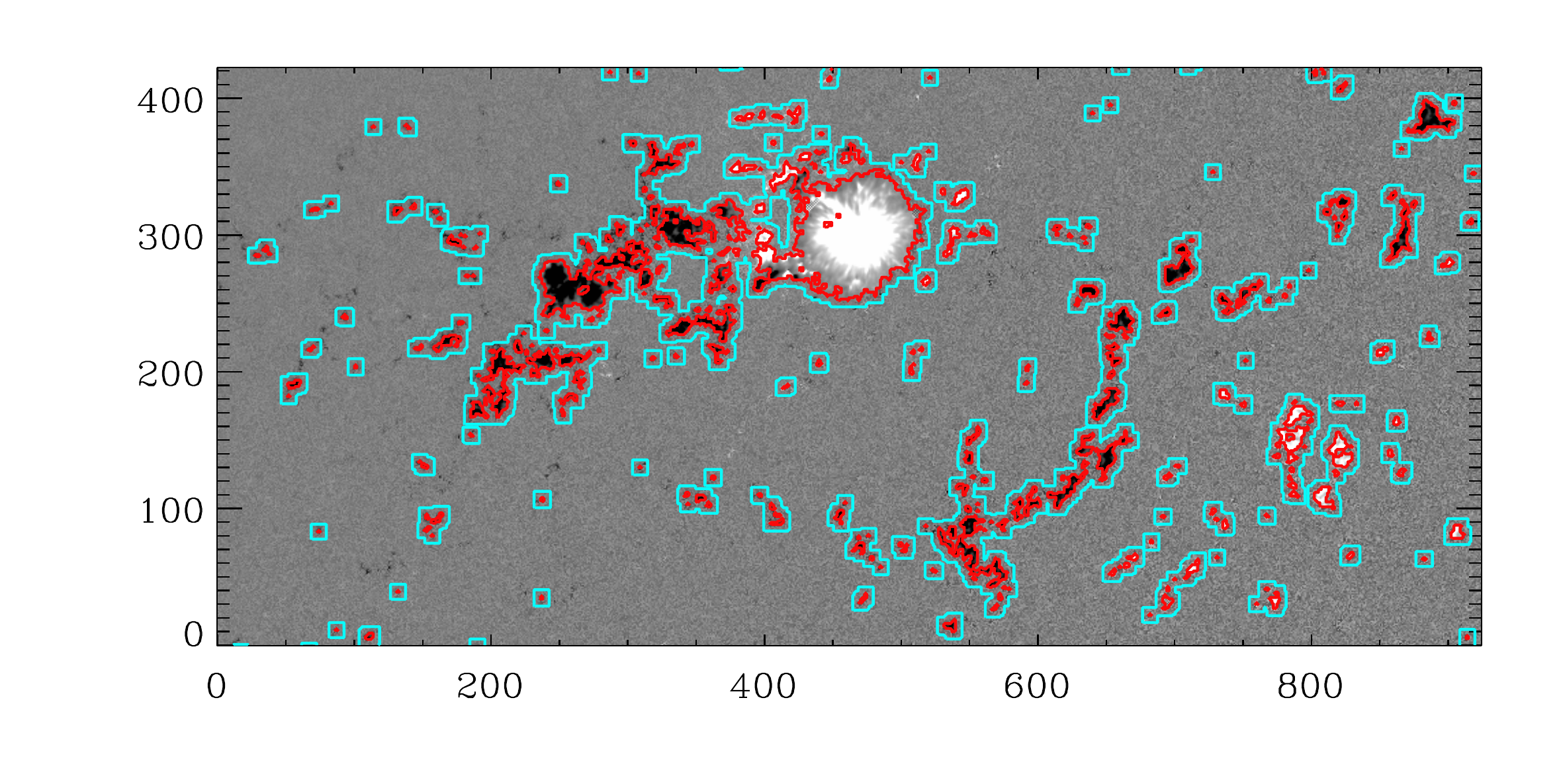}}
\caption{The radial component of the field for HARP~1795 (NOAA AR \#11512) on
1 July 2012 at 00:00\,TAI, saturated at $\pm$500\,G. Axes are in pixels. The
disambiguation result from simulated annealing is returned for pixels inside
the red contour. SHARPs observed before January 2014 are disambiguated 
in patches, and all pixels outside the red contour are annealed
and smoothed. In the case of full-disk disambiguation, pixels between the blue and 
red curves are annealed then smoothed to determine the disambiguation; 
pixels outside the blue contour have three disambiguation results returned: 
potential field acute angle, random, and most radial.}
\label{fig:disambig_conf}
\end{figure}

Figure~\ref{fig:disambig_conf} shows an example of the mask determined for
a sub-area of a full-disk disambiguation. The actual solution from the energy
minimization is returned only for pixels within the eroded mask, \textit{i.e.}, 
where the transverse field is well determined. As described in Section \ref{sec:weakfield},
in the surrounding five-pixel buffer area the annealed solution is smoothed, and 
in weak-field regions alternatives to the minimum-energy disambiguation are provided.

When only patches are disambiguated {\bf all} pixels outside the high-confidence
region are considered to be intermediate.
Currently all SHARPs observed before 15 January 2014 rely on such patch-wise disambiguation; 
consequently all pixels within the HARP bounding box have been annealed, and those outside 
strong-field high-confidence areas have also been smoothed.
This changed when the definitive pipeline switched over to computing the full disk 
disambiguation at each time step. 
The type of solution returned is recorded in the {\sc conf\_disambig} segment; 
see Appendix \ref{app:disambig} for more information.

\subsection{The Potential Field}

To determine the vertical derivative used in approximating
$\mbox{\boldmath$\nabla$} {\mathbf{\cdot}} \mbox{\boldmath$B$}$, the potential
field that matches the observed LoS component of the field is
computed in planar geometry \cite{Alissandrakis1981}.  By matching the
line-of-sight component, the potential field only needs to be computed once,
and by using planar geometry, Fast Fourier Transforms can be used.  When
planar geometry is used for the energy calculation, a single plane is used in
the potential field calculation.  For spherical geometry, the observed disk is
divided into tiles by taking a regular grid in an area-preserving Mollweide
projection and mapping it back to the spherical surface of the Sun.  Each tile
is treated as a separate plane for computing the potential field, although
neighboring tiles are overlapped, and a windowing function is applied to
reduce edge effects.

\subsection{Minimum Energy Methods}

The ambiguity is resolved by minimizing an energy function
\begin{eqnarray}\label{eqn:energy}
E &\propto& \sum (|\mbox{\boldmath$\nabla$} {\mathbf{\cdot}}
\mbox{\boldmath$B$}| 
+ |(\mbox{\boldmath$\nabla$} {\mathbf{\times}} \mbox{\boldmath$B$})
{\mathbf{\cdot}} \mbox{\boldmath$n$}|) 
\end{eqnarray}
where the sum runs over pixels in the noise mask, $\mbox{\boldmath$n$}$ is a
unit vector normal to the surface, and the second term on the right hand side
is proportional to the normal component of the current density.  The
calculation of the normal component of the curl, and the horizontal terms in the
divergence is straightforward, requiring only observed quantities in the
computation and a choice of the ambiguity resolution; a first order forward
difference scheme is used to calculate all the horizontal derivatives.
However, these finite differences mean that the computation is not local.  To
find the permutation of azimuthal angles that corresponds to the minimum of
$E$, a simulated annealing algorithm \cite{Metropolis1953,Kirkpatrick1983}, as
described in \inlinecite{Barnesetal2012}, is used.  This method is an
extremely robust approach when faced with a large, discrete problem (there are
two and only two possibilities at each pixel) with many local minima
\cite{Metcalf1994}.

\subsection{Treatment of Areas Dominated by Noise}\label{sec:weakfield}

The utility of minimizing the energy given by Equation~(\ref{eqn:energy})
depends on being able to reliably estimate the divergence and the vertical
current density.  Where noise dominates the retrieved field, alternate
approaches are used.

A neighboring-pixel acute-angle algorithm based on the University of Hawai'i
Iterative method (UHIM; \opencite{Canfieldetal1993}; \opencite{ambigworkshop})
that locally maximizes the sum of the dot product of the field at one pixel
with the field at its neighbors is used.  In areas close to pixels with well
determined fields, the method performs well, as it quickly produces a smooth
transition to noisy pixels. While the result is always a smooth solution, the
algorithm can get trapped in a smooth but incorrect result, which tends to be
propagated outwards from well measured field \cite{ambig2}.  Thus this method
is currently used for SHARPs, and in small intermediate areas around well 
determined fields for the full-disk.

Further away from well measured field in full-disk mode, where the annealing method is not
applied, the pipeline module disambiguates the azimuth in three ways,
each with its own strengths and weaknesses. Method 1 selects the azimuth which
is most closely aligned with the potential field whose derivative is used in
approximating $\mbox{\boldmath$\nabla$} {\mathbf{\cdot}} \mbox{\boldmath$B$}$.
Method 2 assigns a random disambiguation for the azimuth. Method 3 selects the
azimuth that results in the field vector being closest to radial.  Methods 1
and 3 include information from the inversion, but can produce large-scale
patterns in azimuth.  Method 2 does not take advantage of any information
available from the polarization, but does not exhibit any large scale
patterns, and reflects the true uncertainty when the inversion returns purely
noise.  The results of all three methods are recorded in {\sc disambig}; 
see Appendix \ref{app:disambig} for specifics.

%% file: VP5-HARP.txt
\section{Space-Weather HARPs -- SHARPs}
\label{sec:SHARP}
\label{sec:regions}

The Space-weather HARP (SHARP) data series collects most of the relevant
observables in HMI active region patches and computes from them a variety of
summary active-region parameters that evolve in time. The SHARP data series,
{\sf hmi.sharp\_720s}, is quite complete, comprising 31 maps of the active
patch at each
time step with all of the disambiguated field and thermodynamic parameters
determined in the inversion, along with uncertainties and error cross-correlation
coefficients, as well as the continuum intensity, Doppler velocity, and
line-of-sight magnetic field \cite{Bobra2013}.

There is an extensive literature describing potentially useful active-region
indices (\textit{e.g.} \opencite{LekaBarnes2003I}, \opencite{LekaBarnes2003II}, \opencite{BarnesLeka2006},
\opencite{Leka1}, \opencite{Leka2}, \opencite{Falconer2008}, \opencite{Schrijver2007},
\opencite{Mason2010}, \opencite{Georgoulis2007}). The SHARP series
computes many of these quantities that are averaged or summed over the entire
region. A list of computed indices can be found in Appendix
\ref{app:SHARP} and includes total unsigned flux, mean inclination angle
(relative to vertical), mean values of the horizontal gradients of the total, 
horizontal, and vertical field, means of the vertical current density, current helicity,
twist parameter, and excess magnetic energy density, the total excess
energy, the mean shear angle, and others. 
See \inlinecite{Bobra2013} for details.

The alternate SHARP series {\sf hmi.sharp\_cea\_720s} includes a smaller number 
of maps (eleven) remapped to cylindrical equal area (CEA) heliographic coordinates 
centered on the HARP center point. 
The coordinate remapping and vector transformation are decribed in Section \ref{sec:RemapTrans}. 
The CEA series includes maps of the three vector magnetic field components and the expected
standard deviation of each component, along with the LoS magnetic field, 
Dopplergram, continuum intensity, disambiguation confidence, and HARP bitmap
(see Apppendix \ref{app:SHARP} for details).

\subsection{Coordinate Remapping and Vector Transformation}
\label{sec:RemapTrans}

Two coordinate systems are involved in the representation of the vector
magnetograms. The first refers to the observation's physical location on the
Sun or in the sky, 
for which HMI uses World Coordinate System (WCS) standards for 
solar images \cite{thompson2006}.
The second refers to the direction of the three-dimensional
field vector components at a particular location.  In this paper, we call the
conversion from one 
WCS
format to another ``remapping'' in the first geometrical
case and ``vector transformation'' in the second.

\subsubsection{Remapping}

Currently SHARP data are available in CCD image
coordinates or in a cylindrical equal area projection
centered on the patch with the WCS projection type specified in {\sc ctype1} and
{\sc ctype2}.  When data are requested from the JSOC,
users will eventually have the option to customize the data using various
remapping and vector transformation schemes.\footnote[1]{Relevant modules have
been developed and tested. However, their integration with the JSOC export web
tools has yet to be implemented.} For a more detailed account of different
map projections, one can refer to the work of \inlinecite{calabretta2002}
and \inlinecite{thompson2006}.

Care must be taken with image coordinates. The standard SHARP
image segment cut-outs,
like most HMI 
image
data series, are stored in 
helioprojective
CCD 
image
coordinates,
registered and corrected for distortion, but
without any remapping. Each pixel
represents the same projected area on the plane of the sky, but corresponds
to a different area on the solar surface. We note that the $+y$ direction
of the HMI CCD array 
in the data series
does not generally correspond to solar north. The
counterclockwise (CCW) angle needed to rotate the image to the north ``up''
format is recorded in the keyword {\sc crota2}. For 
most standard
HMI 
images
{\sc crota2}
is close to $180^\circ$,
\textit{i.e.} raw HMI CCD images appear upside down.

The vector magnetic field in SHARPs is also provided in a cylindrical equal
area projection data series, {\sf hmi.sharp\_cea\_720s}.
A thorough discussion is available as a technical note \cite{Sun2013}. 
Each pixel in the remapped CEA image represents
the same surface area on the photosphere. For the CEA remapping, the vector
magnetogram is first converted into three images, each frame representing one
component of the 3D field vector (see next section). A target coordinate grid
is defined for a CEA coordinate system in which the origin of the map
projection is the center of the HARP. 
The CEA grid is centered on the region of interest in order to minimize geometric
distortion in the result. The remapping is done individually for
each vector component, as well as on the scalar observables. 
The uncertainties, {\sc bitmap}, and confidence maps are derived a little 
differently: first, the center of each pixel
in the remapped coordinate system is located in the original image; then
the nearest neighboring pixel in that original image is identified and the
value for that nearest original pixel is reported.

\subsubsection{Vector Transformation}

In the ``native format'' the field vector at each pixel is represented by
three values derived from the inversion (Section \ref{sec:MEInversion}): the total field strength
($B$), the inclination ($\gamma$), and the azimuth ($\psi$). 
Inclination is defined with respect to the HMI line of sight, so
the longitudinal ($B_l$) and transverse field ($B_t$) components can be
easily separated as $B_l=B\cos\gamma$ and $B_t=B\sin\gamma$.

As mentioned in the previous section, we decompose the vector magnetogram
into three component images before remapping using the following basis vector:
(\textbf{\textit{e}}$_{\xi}$, \textbf{\textit{e}}$_{\eta}$, \textbf{\textit{e}}$_{\zeta}$), where 
\textbf{\textit{e}}$_{\xi}$ refers
to the $+x$ direction of the native format image, \textbf{\textit{e}}$_{\eta}$ to $+y$, and
\textbf{\textit{e}}$_{\zeta}$ to $+z$ out of the image. Note that \textbf{\textit{e}}$_{\zeta}$ coincides with the
LoS direction. Remapping is done subsequently on the three component images.

The vector transformation is done as the last step, following
Equation~(1) of \inlinecite{gary1990} after determining the heliographic
latitude and longitude at each pixel.  For the CEA map-projected data we
transform the vector into standard heliographic spherical coordinates with a basis vector
(\textbf{\textit{e}}$_{r}$,
\textbf{\textit{e}}$_{\theta}$, \textbf{\textit{e}}$_{\phi}$), where \textbf{\textit{e}}$_{r}$ is the radial direction normal to the
solar surface, 
\textbf{\textit{e}}$_{\theta}$ points southward along the meridian at the point of interest
and \textbf{\textit{e}}$_{\phi}$ points westward in the direction of solar rotation
along a constant latitude line. 
The $r$ component corresponds to the
vertical component, $\theta$ and $\phi$ the horizontal components.

\subsection{A Comparison of Definitive and Quick-look SHARP Data}
\label{sec:HARP-QL}

Definitive SHARPs provide the most complete and best-calibrated data. Because
the extent of the definitive bounding box is determined only after a region
completes its disk passage in order to ensure uniform geometric coverage,
there is a $\sim 35$-day delay in data availability. Fortunately quick-look
data are available in near real time (NRT), and the SHARP indices are computed
as data become available, usually in less than three hours. Occasional data
gaps longer than an hour are simply bypassed. Otherwise, the processing of
the definitive and NRT data differs in three significant ways.

\begin{figure}
\includegraphics[width=0.98\textwidth]{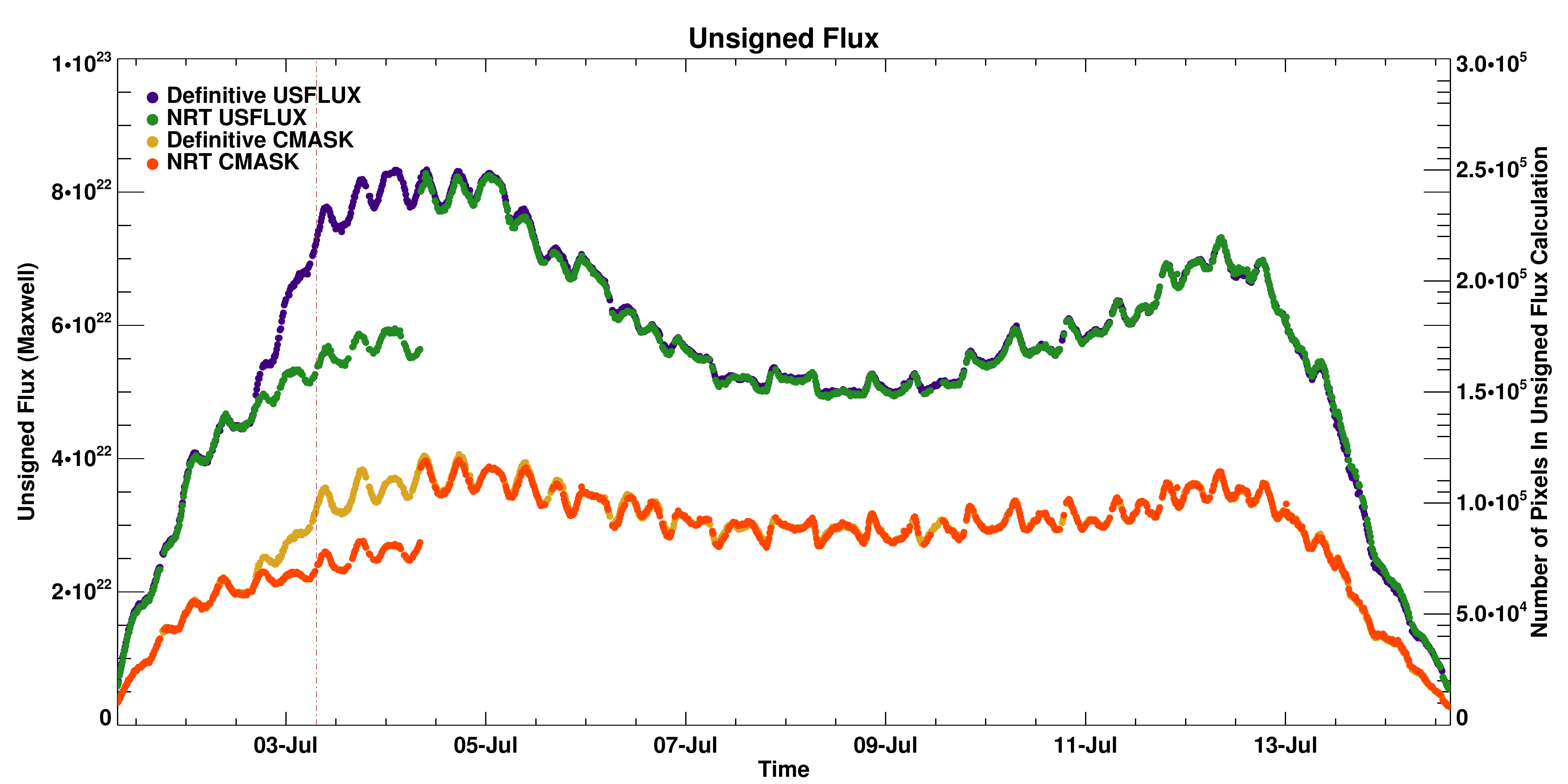}
\includegraphics[width=0.98\textwidth]{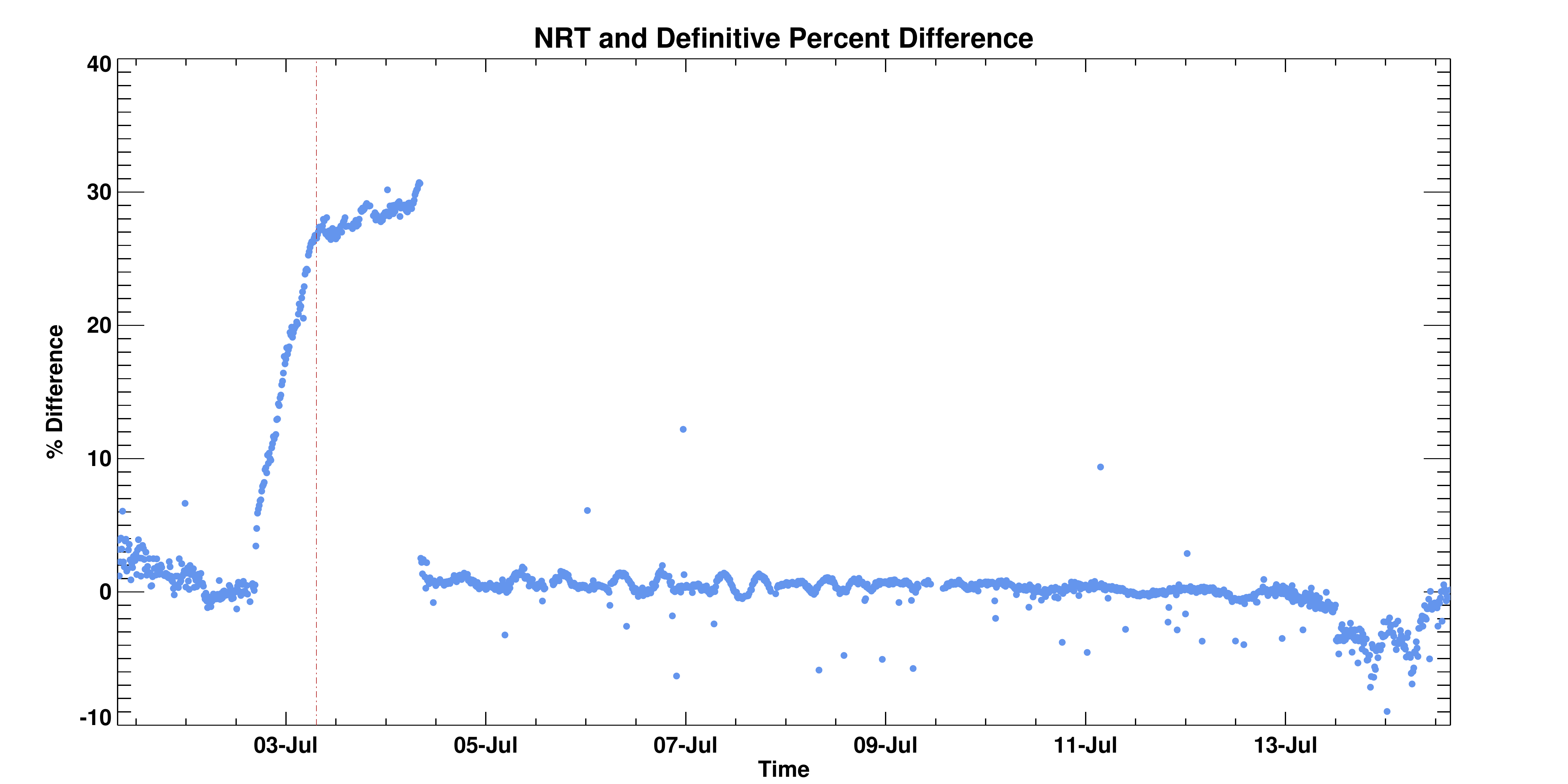}
\caption{
The top panel shows the total unsigned flux, {\sc usflux}, during the
disk passage of SHARP\,2920 in purple and the same region, NRT SHARP\,1407,
in green from 1 -- 14 July 2013. The region crossed disk center $\sim$ 7:48~UT 
on 8~July. The region is ultimately associated with NOAA ARs\,\#11785,
\#11787, and \#11788. The vertical line at 07:18~UT on 3~July indicates the
GOES X-ray flux peak associated with an M\,1.5-class flare in NOAA AR\,\#11787.
Differences between the definitive and NRT processing are described in the
text. The heliographic area inside the bounding box of the defininitive
SHARP does not change, whereas the green line jumps on 4~July when two NRT
SHARP regions merge. Also shown are the number of high-confidence pixels
contributing to the {\sc usflux} for the definitive (yellow) and NRT (orange)
SHARP. Variations in {\sc usflux} track the changes in the area of the region
due to evolution and to systematic effects, see text. The total flux increases
as the region moves away from the limbs. The broad peaks in total unsigned
flux $\sim 60^\circ$ from central meridian are attributed to increased noise
in the field away from disk center. The bottom panel shows the percentage
difference between the {\sc usflux} computed for the definitive and NRT SHARPs.
The differences between the two are small after accounting for the merger.
Outliers in the values correspond to times when the {\sc quality} of the
images was not good. The poor quality images were excluded in the upper panel.
}
\label{fig:DefNRTCompare}
\end{figure}

\begin{itemize}
\item
As described in Section \ref{sec:level1} a preliminary flat field is used,
gap filling is less sophisticated, cosmic ray correction is not performed,
and calibrations are based on default or preliminary instrument parameters.
All of the NRT 720s observables (vector and scalar) are calculated from the 
{\sf hmi.S\_720s\_nrt} Stokes series.

\item
The NRT HARP bounding box is based on the data available at that point in time,
thus the heliographic size may change and regions may even be merged by the
time the definitive data are created. Obviously no NRT HARP data are available before 
the region emerges and no association with a NOAA AR can be made until NOAA
assigns a number. The NRT and definitive HARP numbers may be different.

\begin{sloppypar}
\item
The NRT {\tt fd10} VFISV inversion is computed only for identified NRT
HARPs. The inverted region extends beyond the NRT HARP bounding box by 50
pixels in each direction. Consequently, for NRT the size of the padding
region outside the HARP used to compute the initial field parameters is
reduced from 500 to 50 (or 20 in some intervals, check keyword {\sc ambnpad}).
NRT disambiguation parameters are adjusted to speed up the calculation by
increasing the effective temperature scale factor, {\sc ambtfctr},
and decreasing the number of reconfigurations, {\sc ambneq}
(see \inlinecite{Barnesetal2012} for details.)
The results for particularly small and (until September 2013) particularly
large NRT SHARPs are not computed in order to limit the computing load.
All NRT SHARPs are disambiguated using planar geometry rather than spherical.
The same noise threshold map is used in both pipelines, subject to some delays
in updating. The SHARP parameters are computed using only pixels 
both within the HARP (the colored blobs in Figure~\ref{fig:HARP})
and above the noise threshold (see Section \ref{sec:NoiseValue}); the number of
pixels is given in the keyword {\sc cmask}.
\end{sloppypar}

\end{itemize}

Figure~\ref{fig:DefNRTCompare} illustrates some of the differences between
SHARP\,2920 and NRT SHARP 1407 during the interval 1--14 July 2013.
The region was associated with NOAA ARs\,\#11785, \#11787, and \#11788.
The total unsigned flux, {\sc usflux}, is computed by integrating the vertical
flux, B$_r$, in the high-confidence areas of the active region. For computing
the SHARP indices, high-confidence regions are those above the noise-mask,
where the confidence in the disambiguation is highest ({\sc conf\_disambig}
= 90, see Appendix~\ref{app:SHARP}). As the region first rotates onto the
disk on 1 July and subsequently grows, the {\sc usflux} increases, as shown
by the overlying purple and green points in the upper panel. A second region
emerges nearby on 2 July that is initially classified as an independent NRT
SHARP that does not contribute to the {\sc usflux} of 1407.  That region
grows quickly on 2-3 July.  An M\,1.5-class flare was observed at 07:18 UT
on 3 July. On 4 July the two NRT SHARPs merge. The definitive SHARP evolves
smoothly because its heliographic boundaries are determined only after the
region has completed its disk transit and so includes both regions. The
SHARP eventually rotates off the disk on 14 July.

The number of high-confidence pixels is shown by the yellow (definitive) and
orange (NRT) points in the upper panel of Figure~\ref{fig:DefNRTCompare}. The
number reflects both the size of the evolving active region as well as
systematic variations. The clear 12-hour periodicity is associated with the
spacecraft orbital velocity, V$_r$, relative to the Sun and is discussed
in Section \ref{sec:Periodicity}. The broad, roughly symmetric peaks centered
$\sim 60^\circ$ from central meridian are due to the spatially dependent
sensitivity of the instrument (see Section \ref{sec:NoiseValue}). The noise
level in low and moderate field strength regions changes as a function of
center-to-limb angle and V$_r$ and consequently the
number of pixels contributing to the {\sc usflux} index changes. Histograms of two
representative active regions show a shift in the peak of the distribution
of low-to-moderate field pixels toward higher values as the regions move
away from central meridian; the width of the distribution also broadens.
An increase of a few tens of Gauss in the field strength distribtuion at $60^\circ$
increases the number of pixels in the 250--750\,G range by a few tens of
percent. The field strength in higher field regions is more stable, but,
because the area covered by strong field is much smaller, the variation in
{\sc usflux} is significantly affected.

The lower panel shows the percentage difference in the definitive and NRT
{\sc usflux} index as a function of time. The two match within a few percent
except during the 2--4 July interval when there were two NRT SHARP regions. In
the upper panel we have excluded all images where the {\sc quality} keyword
exceeded 0x010000. Bits in the {\sc quality} keyword identify potential problems with the
measurements (in this case typically missing filtergrams due to calibration
sequences). For comparison, the lower panel includes all of the points and the isolated points lying well
off the line warn of the consequences of using less-than-ideal measurements.

%% file: VP6-Uncertainties.txt
\section{Uncertainties, Limitations, Systematics, Sensitivities}
\label{sec:uncertainties}

The HMI vector field data are of high quality and more than meet their
original performance specifications. Nevertheless, the measurements are far from
perfect and proper care must be taken when using the data. 
This section discusses various uncertainties, limitations, and known systematic
issues with the data produced in the HMI vector field pipeline. Sources of
error include imperfections and limitations of the instrument and observing
scheme, as well as imperfections and limitations of the analysis methods.
Variances of the inverted variables and covariances between their errors are obtained during the
VFISV inversion (see Section \ref{sec:MEInversion}). These are recorded 
as estimated errors (square roots of the variances) for these quantities 
and their correlation coefficients (related to the covariances). 
Having characterized as best we can the known uncertainties and errors, the ultimate
user must take this information into account when performing research.

\subsection{Temporal and Spatial Variations of the Inverted Magnetic Field}
\label{sec:magnoise}

\begin{figure}
\centerline{
\includegraphics[width=0.92\textwidth]{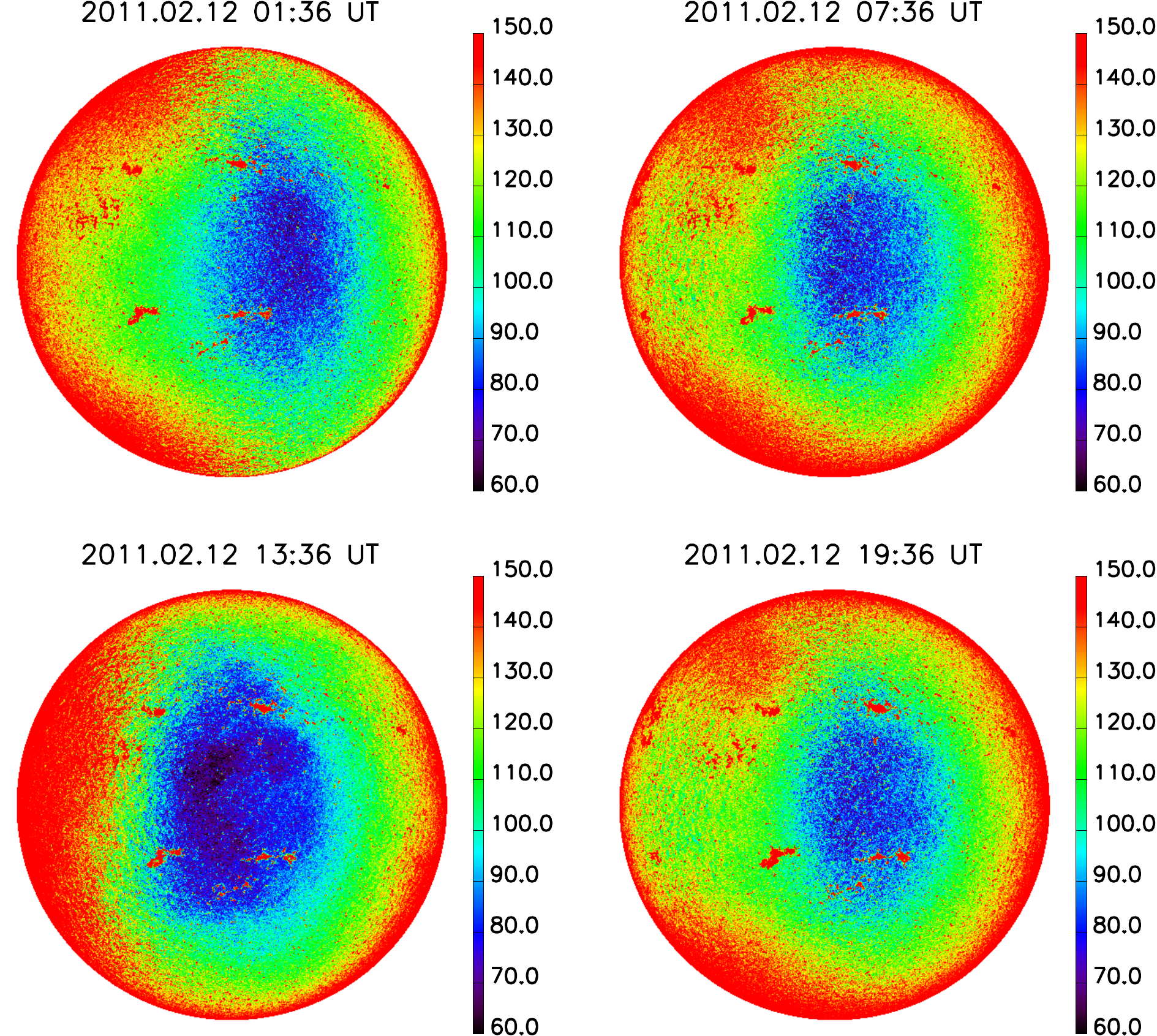}
}
\caption{ 
From left to right then top to bottom: the total magnetic field strength 
at 01:36 TAI (local dusk, maximum relative radial velocity, V$_r$ = 2980
m\,s$^{-1}$), 
07:36 TAI (noon, near zero V$_r$ = 360 m\,s$^{-1}$), 
12:36 TAI (dawn, minimum V$_r$ = -2372 m\,s$^{-1}$), 
and 18:36 TAI (midnight, near zero V$_r$ = 263 m\,s$^{-1}$) on 12 February 2011. 
The field strength should be uniform in quiet Sun if measured perfectly.
The large-scale quiet-Sun noise pattern varies with time of day, \textit{i.e.} relative
velocity. The color table shows field strength in Mx cm$^{-2}$.
}
\label{fig:btotal4}
\end{figure}

The dominant driver of time-varying errors in the magnetic field is the Doppler
shift of the spectral line. By far the largest contributor is the daily period
due to the geosynchronous orbit of SDO. On longer time scales there are other
orbital and environmental variations, along with changes in the instrument itself, 
changes in the way the instrument is tuned or operated, and variations in the 
calibration. Most of the short-term variation from solar oscillations is
eliminated by the 12-minute averaging. 
This section concentrates primarily on the daily variability.

\subsubsection{Time Varying Noise Mask}
\label{sec:NoiseValue}

\begin{figure}
\centerline{
\includegraphics[width=0.92\textwidth]{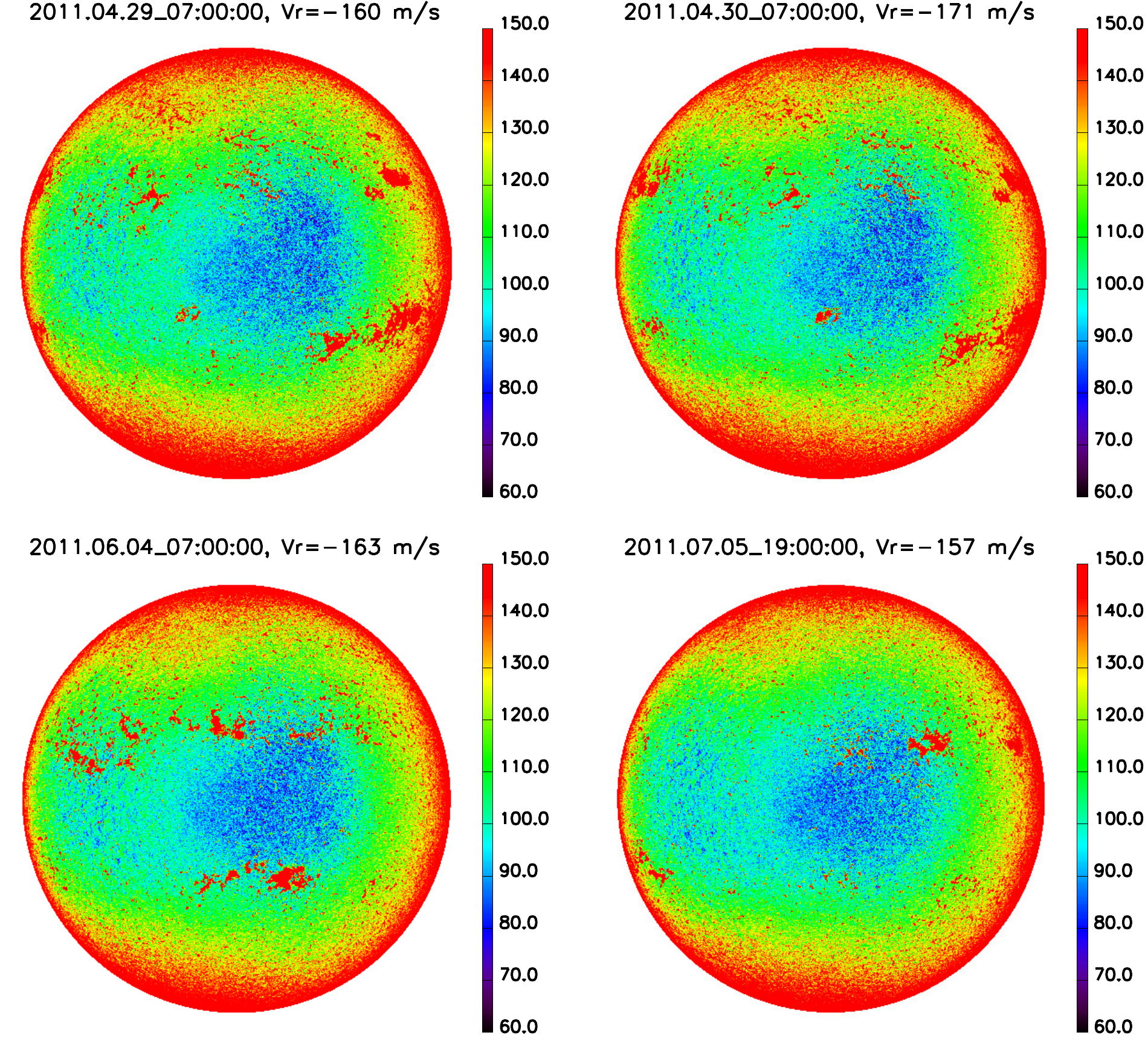}
}
\caption{Field strengths observed at 07:00 TAI on 29 April 2011 (top left),
30 April 2011 (top right), 4 June 2011 (bottom left) and at 19:00 TAI on 5
July 2011 (bottom right). SDO's radial velocity relative to the Sun, V$_r$,
is close to -165 m s$^{-1}$ at those four times. Excluding strong-field
regions, the patterns are similar. The color table shows field strength in
Mx cm$^{-2}$.
}
\label{fig:samev}
\end{figure}

The noise level of the inverted magnetic field shows large-scale spatial
variations over the entire disk due to effects of irregularities in the 
HMI instrument and viewing angle.
This pattern changes with the orbital velocity of the geosynchronous satellite
relative to the Sun, V$_r$ (keyword {\sc obs\_vr}). V$_r$ ranges $\pm 3$ km~s$^{-1}$ 
from local dusk ($\sim 1$ UT) to dawn ($\sim 13$ UT). The corresponding Doppler shift moves the
spectral line back and forth by about one tuning step every 12 hours. This
combined with the fixed velocity pattern of solar rotation leads to the temporal and
spatial variations of the inverted magnetic field over the Sun's disk every
24 hours.

\begin{figure}
\centerline{
\includegraphics[angle=0, scale=0.6]{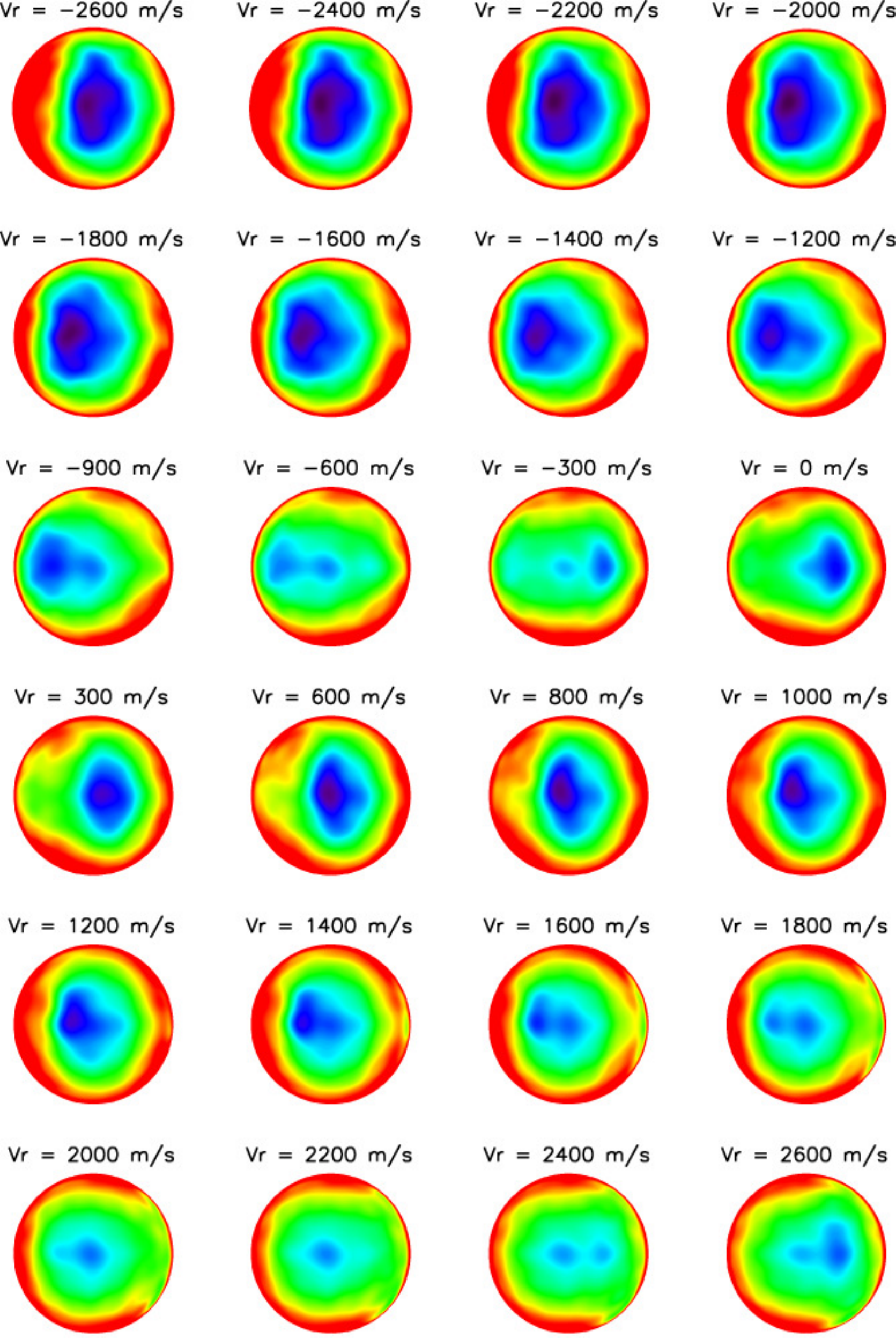}
}
\caption{Derived noise masks for a range of orbital velocities, V$_r$. The data used
for these reference masks were collected between 01 February and 
11 March 2011. The colors represent values between 60\,-\,150\,Mx\,cm$^{-2}$,
as in Figure \ref{fig:samev}.
}
\label{fig:mask24}
\end{figure}

\begin{figure}
\centerline{
\includegraphics[width=0.92\textwidth]{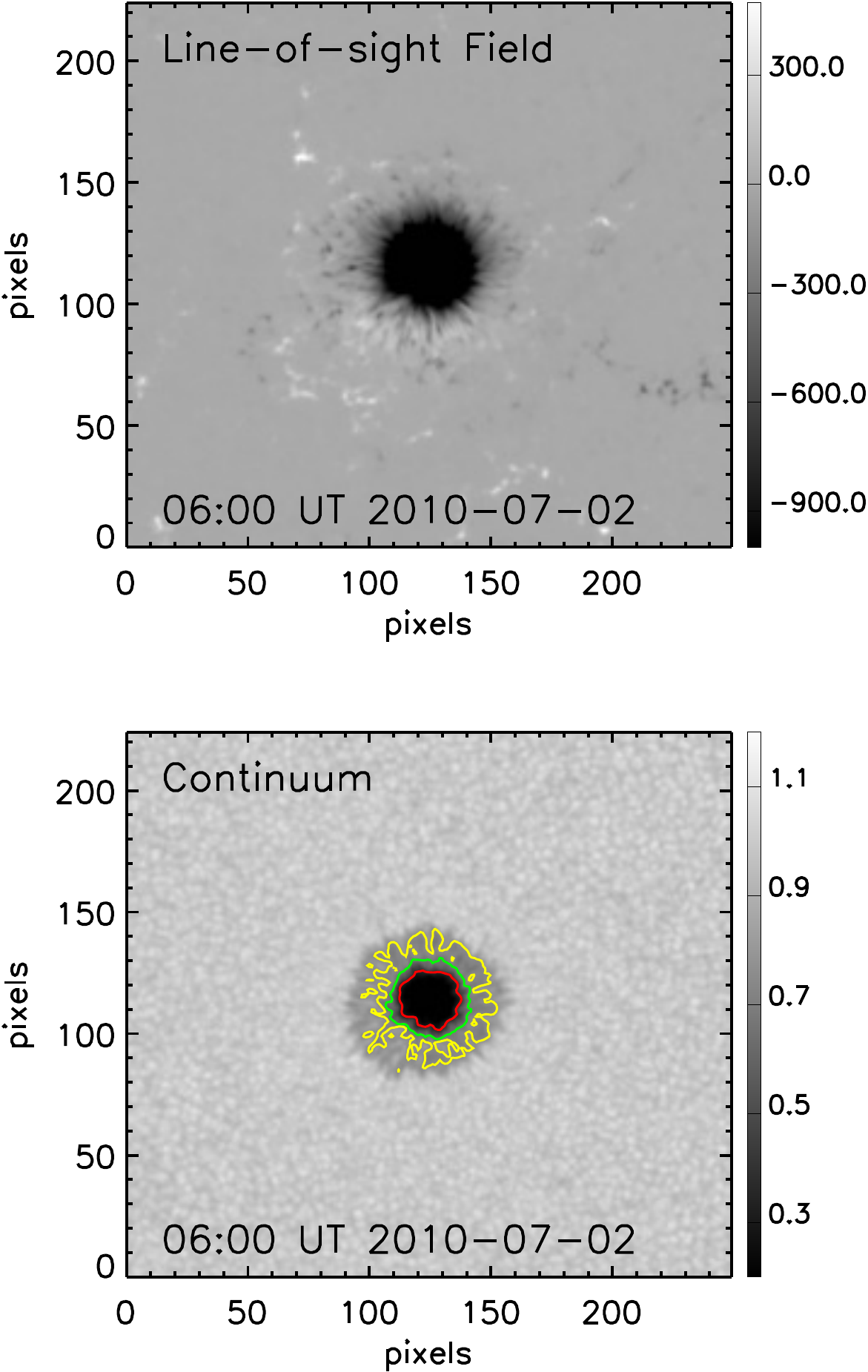}
}
\caption{HMI LoS magnetic field, B$_{\rm{Los}}$ (top), and continuum intensity (bottom) 
of AR\,\#11084 at 06:00 TAI on 2 July 2010 when the region was at S19\,E01. 
The pixel size is 0.504 arc sec. The units of field are Mx cm$^{-2}$ and
intensity is shown in relative units of the mean nearby quiet-Sun 
continuum intensity.
Intensity contours are drawn at 0.35, 0.65, and 0.75 of the
quiet-Sun continuum intensity in red, green,
and yellow, respectively. 
}
\label{fig:ar11084img}
\end{figure}

\begin{figure}
\centerline{
\includegraphics[width=0.95\textwidth]{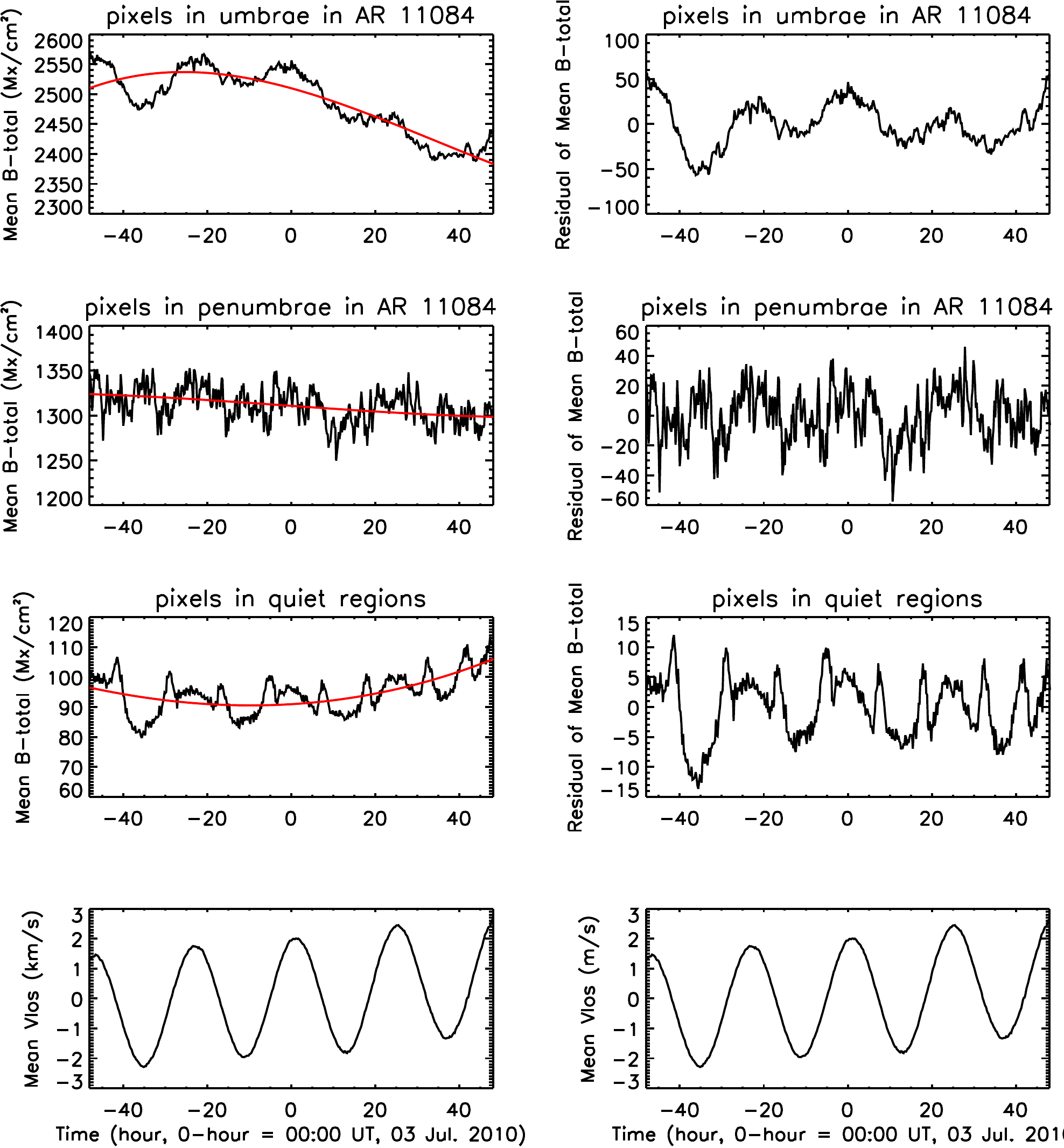}
}
\caption{The black curves in the left column in the top three panels are
4-day temporal profiles for AR\,\#11084 of the mean magnetic field strength
determined for the umbra, penumbra, and quiet Sun, respectively. The red 
curves are 3$^{rd}$-order polynomial fits.  Residuals are plotted in the top three 
panels on the right.  The residual is the difference between the mean field 
strength and the polynomial fit. For reference, the mean LoS 
velocity observed in the quiet-Sun region is plotted in the two bottom panels.
For this analysis the umbra includes pixels where, compared with the quiet
Sun, the continuum intensity, I$_c < 0.35$ when corrected for limb darkening. 
In the penumbra $0.65 < $ I$_c < 0.75$.
}
\label{fig:variation}
\end{figure}

\begin{figure}
\centerline{
\includegraphics[width=0.40\textwidth,height=12cm]{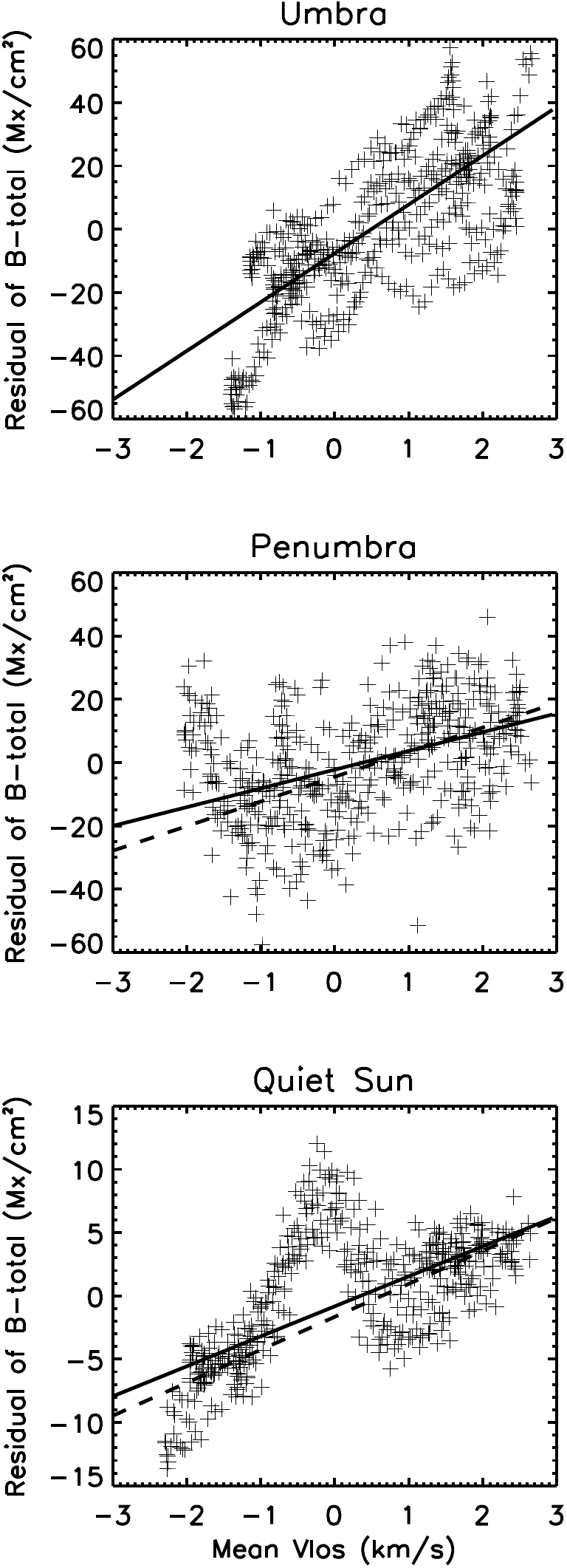}
\includegraphics[width=0.60\textwidth,height=12cm]{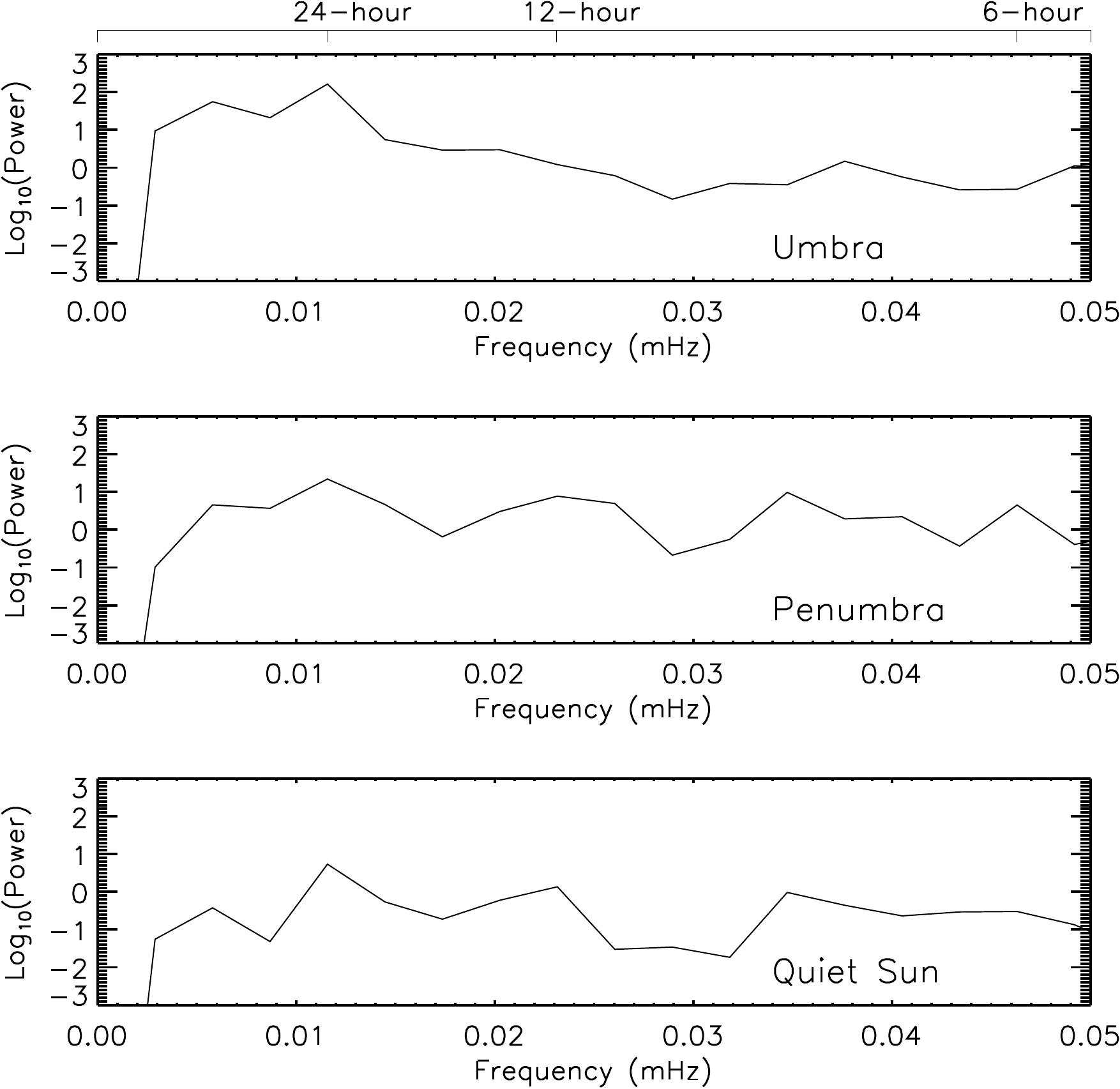}
}
\caption{Scatter plots on the left show the relationship between the residual of 
the mean field strength and LoS velocity in the AR\,\#11084 umbra 
(top), penumbra (middle), and nearby quiet Sun (bottom).  Note the scale change 
for the quiet Sun. Linear fits are described in the text.
The right panels show the power spectra of the mean field strength residuals for
the sunspot umbra (top), penumbra (middle), and quiet Sun (bottom).
}
\label{fig:powers}
\label{fig:scatters}
\end{figure}

\begin{figure}
\centerline{
\includegraphics[width=0.50\textwidth]{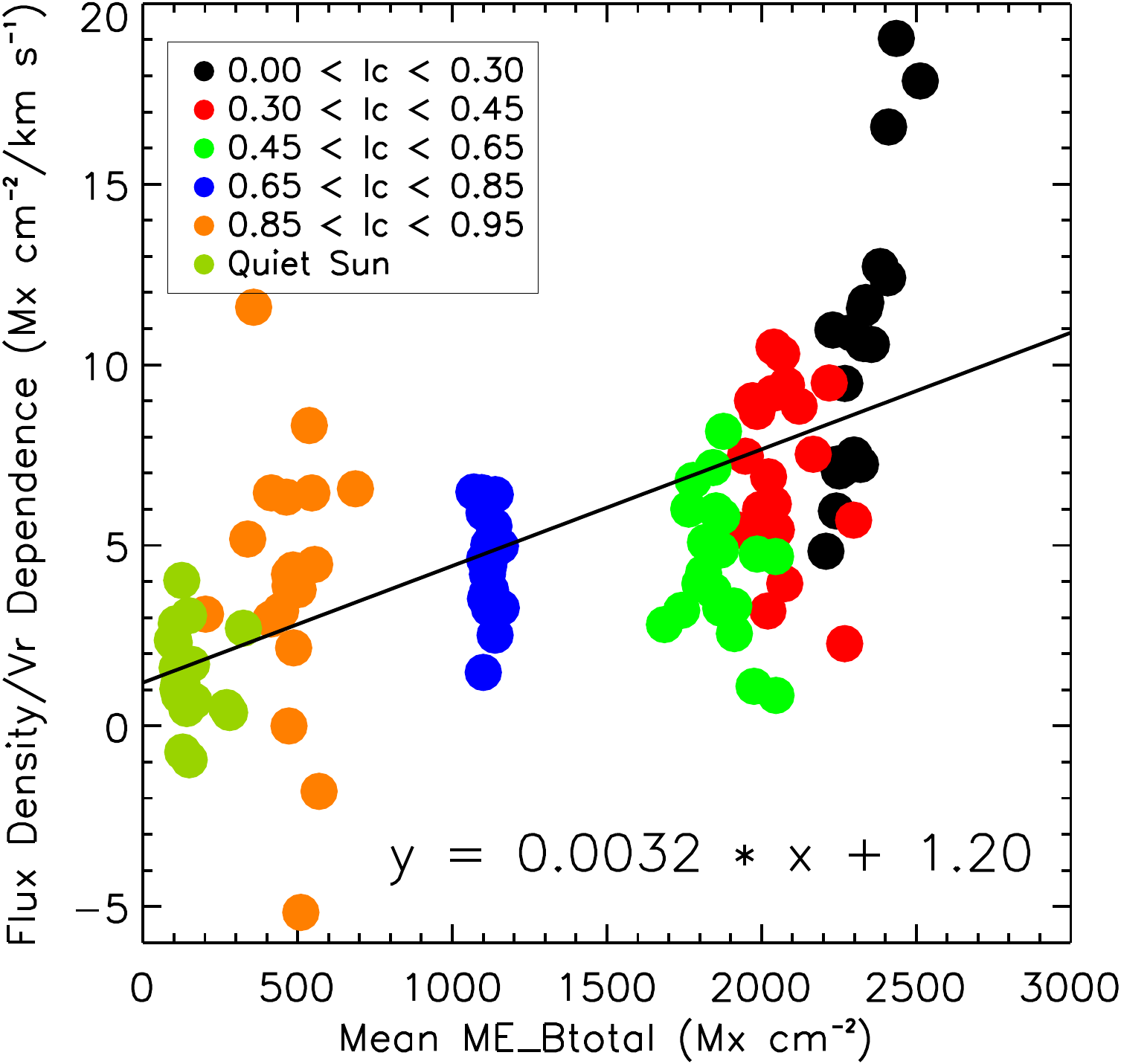}
\includegraphics[width=0.50\textwidth]{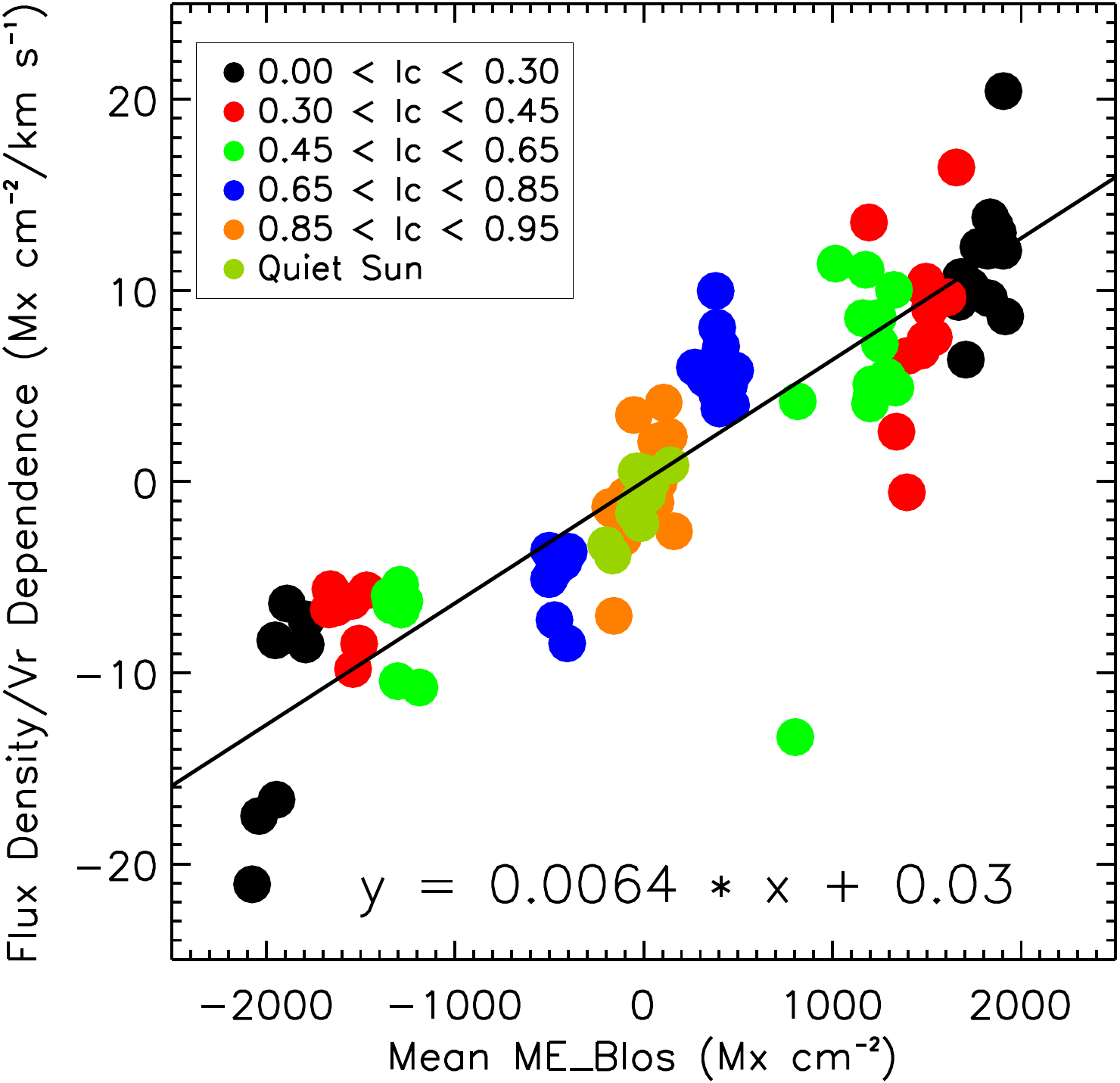}
}
\caption{The change in measured magnetic field strength with radial velocity
is plotted as a function of field strength for a selection of 20 simple
sunspots. Each spot is divided into continuum intensity bins to segregate
by field strength. The bins are tracked for several days and the dependence
of measured field strength on V$_r$ determined. That slope is plotted for
each sunspot in each bin as a function of the measured field strength. The
left panel shows the slope versus total magnetic field magnitude. There is a
significant zero offset and a relationship of 0.0032 Mx-cm$^{-2}$/km-s$^{-1}$
per Mx-cm$^{-2}$. The LoS component of the vector magnetic field, B$_{\rm{Los}}$,
can be determined and that relationship is shown in the right panel. Note
the scale change. There is a smaller zero offset and a relationship of 0.0064
Mx-cm$^{-2}$/km-s$^{-1}$ per Mx-cm$^{-2}$.
}
\label{fig:slopes}
\end{figure}

Figure~\ref{fig:btotal4} shows the field strength determined at 01:36 TAI
(top left), 07:36 TAI (top right), 13:36 TAI (bottom left), and 19:36 TAI 
(bottom right) on 12 February 2011, times when SDO was close to local dusk,
midnight, dawn and noon. At these times, the relative orbital velocity
reaches a maximum, minimum, or is close to zero. The large-scale patterns
differ with V$_r$. However, when the SDO-Sun velocities are approximately
the same, the patterns are very similar, as shown in the right column of
Figure~\ref{fig:btotal4} and in Figure~\ref{fig:samev}.
The spatial pattern in the magnetic field strength noise
is stable in time but varies with V$_r$.

As described in Section \ref{sec:weakfield} the disambiguation requires
an estimate of the noise, as may researchers analyzing the data. 
To describe this temporally and spatially varying
magnetic strength noise we construct a noise mask. Each mask 
characterizes the noise level in the magnetic field strength for a particular
relative velocity, V$_r$. Reference noise masks are first determined
at 50~m\,s$^{-1}$ steps using sets of full-disk magnetic field strength 
data ($\sim$50 magnetograms) observed in 100~m\,s$^{-1}$ intervals.
The median of the field strength for each on-disk pixel is calculated; excluded 
from the fit are strong field pixels
where $B > 300$~Mx\,cm$^{-2}$, roughly 3$\sigma$ of the nominal field
strength noise. To that map of median values a 15$^{th}$-order 2D polynomial
is fit. The basis functions are Chebyshev polynomials of the first kind. The
reference coefficients are saved in the data series {\sf hmi.lookup\_ChebyCoef\_BNoise}. 
When a noise mask for a specific
velocity is required, the coefficients are linearly interpolated for that
V$_r$ and used to reconstruct a full disk noise mask. Figure~\ref{fig:mask24}
shows the reference noise masks for a range of orbital velocities.

The noise pattern changes somewhat with instrument parameters, for example
when the front window temperature is adjusted. Such instrument changes
have occurred at a few known times: 13 December 2010, 13 July 2011, 18
January 2012, and 15 January 2014. 
See references in
\href{http://jsoc.stanford.edu/jsocwiki/PipelineCode}{PipelineCode}, Table \ref{tab:websites}.
Reference noise masks coefficients are generated specifically for each time interval. 
 
\subsubsection{Periodicity in the Inverted Magnetic Field Strength}
\label{sec:Periodicity}

The daily variation manifests itself differently in strong and weak fields.
To quantitatively describe this V$_r$-dependent variation in the inverted
magnetic field we consider active region AR\,\#11084. AR\,\#11084 is a mature
early-cycle southern hemisphere active region with a single, stable, 
circular, negative-polarity sunspot (see Figure~\ref{fig:ar11084img}). The
umbra and penumbra are clearly seen. The
active region was tracked for 4 days, from 01\,--\,04 July 2010. Shown in
black curves in the left column of Figure~\ref{fig:variation} are the mean
magnetic field strength as a function of time in the umbra of AR\,\#11084
(top panel), the penumbra (second), and a quiet-Sun region (third); for reference
the mean LoS velocity in a quiet-Sun region appears at the bottom 
of both columns. 
The quiet-Sun region is a 40$\times$30 pixel rectangle in the bottom-right corner of
the images in Figure~\ref{fig:ar11084img} where no evident magnetic features
appear. The fixed-size quiet region moves with AR\,\#11084. The red curves
in the three upper panels on the left are a 3$^{rd}$-order polynomial fit to the
evolving mean field strength. The residuals are plotted in the right column. 
The variations in the residuals show a strong correlation to the relative velocity.

Scatter plots of the mean field strength residuals and the relative velocity
better illustrate this dependence. The left panels of Figure~\ref{fig:scatters}
show that a $\sim\pm 2$~km~s$^{-1}$ relative radial velocity variation
causes a $\pm$1\% change in the umbral field strength through the day, or
if a linear fit is made, the field strength varies $\sim15$~G/km-s$^{-1}$.
In the penumbra the daily variation has slightly smaller magnitude, but is
a larger fraction of the mean value, $\sim\pm$2\%.  A linear fit to velocity
gives about 6~G/km-s$^{-1}$, though there seems to be a concentration of high
positive-value pixels below -1.5~km~s$^{-1}$ that may be an artifact of the
polynomial fitting. Excluding those points, the linear fit for the penumbra
is $\sim8$~G/km-s$^{-1}$ (dashed line).  An analysis of the LoS
and transverse fields separately confirms that this velocity-dependent
variation is only seen in the line-of-sight component of the strong field
(in the penumbra and umbra). Strong field shifts either left or right circular
polarization away from one of the wavelength tuning positions, whereas it only
broadens the linear polarization that contribute to Q and U. This may provide
a partial explanation for the velocity-dependent variation in the strong field.

Variability of the mean field strength in the quiet Sun is as much as 5\% of
the typical weak field magnitude and is dominated by daily variations and a
reproducible twice-daily excursion centered near 0~km~s$^{-1}$. That excursion
appears only in the transverse field, \textit{i.e.} Stokes Q and U, and is attributed
to sensitivity that depends on the line position relative to the HMI tuning
wavelengths.  The linear trend is just 2.3~G/km~s$^{-1}$.  Excluding the range
-1.0 to +0.5~km~s$^{-1}$, a linear fit gives 2.6~G/km-s$^{-1}$ (dashed line).
Note that the velocity range does not extend to as highly negative velocity
values in the strong field regions because of the suppression of the convective
blue shift.  Power spectrum analysis shows a peak at 24 hours in all three
region classes (see right panels of Figure~\ref{fig:powers}).  The 24-hour
peak is strongest in the umbra and less so in weaker field regions, which
also have a significant 12-hour periodicity due to the 0~km\,s$^{-1}$ excursion.

Does the sensitivity of the magnetic field strength measurement to radial
velocity itself vary with field strength? The dependence of the magnitude
of the V$_r$-dependent variation of magnetic field to the measured magnetic
field strength is shown in Figure~\ref{fig:slopes}. The linear dependence
on Doppler velocity of the residual magnetic field is determined for a
sample of 20 simple, stable sunspots. For each sunspot the data are sorted
using the continuum intensity ({\sf hmi.Ic\_720s}) as a proxy for magnetic field strength. 
As above, the average magnetic field measured in each intensity bin forms a several-day time
series from which a 3$^{rd}$-order polynomial fit is removed. The magnetic
field residual is fit to the Doppler velocity, as in Figure~\ref{fig:powers},
to determine the sensitivity of the magnetic measurement to velocity. The
y-axis in Figure~\ref{fig:slopes} is the slope of this linear fit in units
of Mx-cm$^{-2}$/km-s$^{-1}$. The x-axis indicates the mean magnetic field
strength of the region in each intensity bin. The left panel shows the slopes as
a function of total field strength. Colors denote bins of continuum intensity
with the indicated range. ``Quiet Sun'' reflects data in a fixed 40$\times$30
pixel rectangle at the lower right corner of the HARP bounding box where
there is no significant flux concentration. The solid black line is a linear
fit to the points. The variation increases linearly from 1.2 G/km-s$^{-1}$
at 0 field strength to 4.4 G/km-s$^{-1}$ at 1000 G. From the vector
we can also determine the LoS component of the magnetic
vector, B$_{\rm{LoS}}$. The right panel shows how the velocity dependence of B$_{LoS}$
changes with B$_{\rm{Los}}$. A relationship between B$_{\rm{Los}}$
and the magnitude of the V$_r$-dependent variation is clearly
shown. There is virtually no variation with velocity at low field values,
but the variation increases linearly to 6.4 G/km-s$^{-1}$ at 1000 G.

\subsection{Known Issues}

\subsubsection{Bad Pixels}

Occasionally, bad inversion results are obvious in the inverted magnetic field
data. They are particularly noticeable in sunspots, where pixels with abnormal
values differ significantly from those adjacent to them or at nearby time
steps. This usually happens when the inversion code fails to converge.

\begin{figure}
\centerline{
\includegraphics[angle=0, scale=0.58]{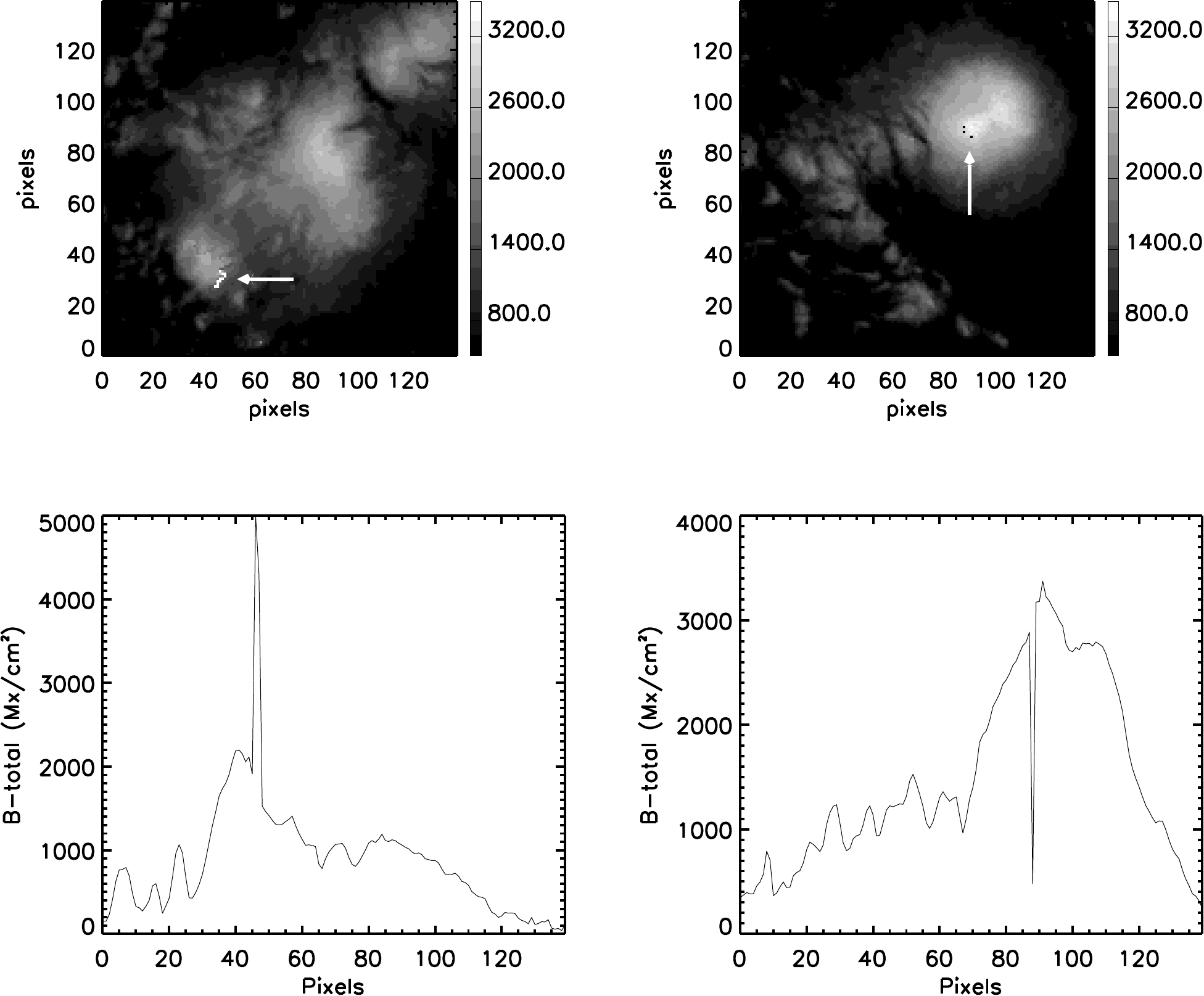}
}
\caption{Two types of bad pixels. Top left: magnetic field strength of
AR\,\#11476 at 12:00\,TAI 8~May 2012 at N10\,E34. The erroneous values are
denoted by an arrow. The field strengths are much higher than in adjacent
pixels. Bottom left: field strength along a horizontal row including a bad
pixel. Note, 5000\,G is the maximum possible value allowed by the VFISV
code. Top right: field strength of AR\,\#11515 at 02:24\,TAI 1~July 2012
at S17\,E29. Bad pixels are denoted by an arrow. The field strengths at bad
pixels are much lower than in adjacent pixels. Bottom right: field strength
along a horizontal row including a bad pixel.}
\label{fig:badpixelimgs}
\end{figure}

\begin{figure}
\centerline{
\includegraphics[angle=0, scale=0.58]{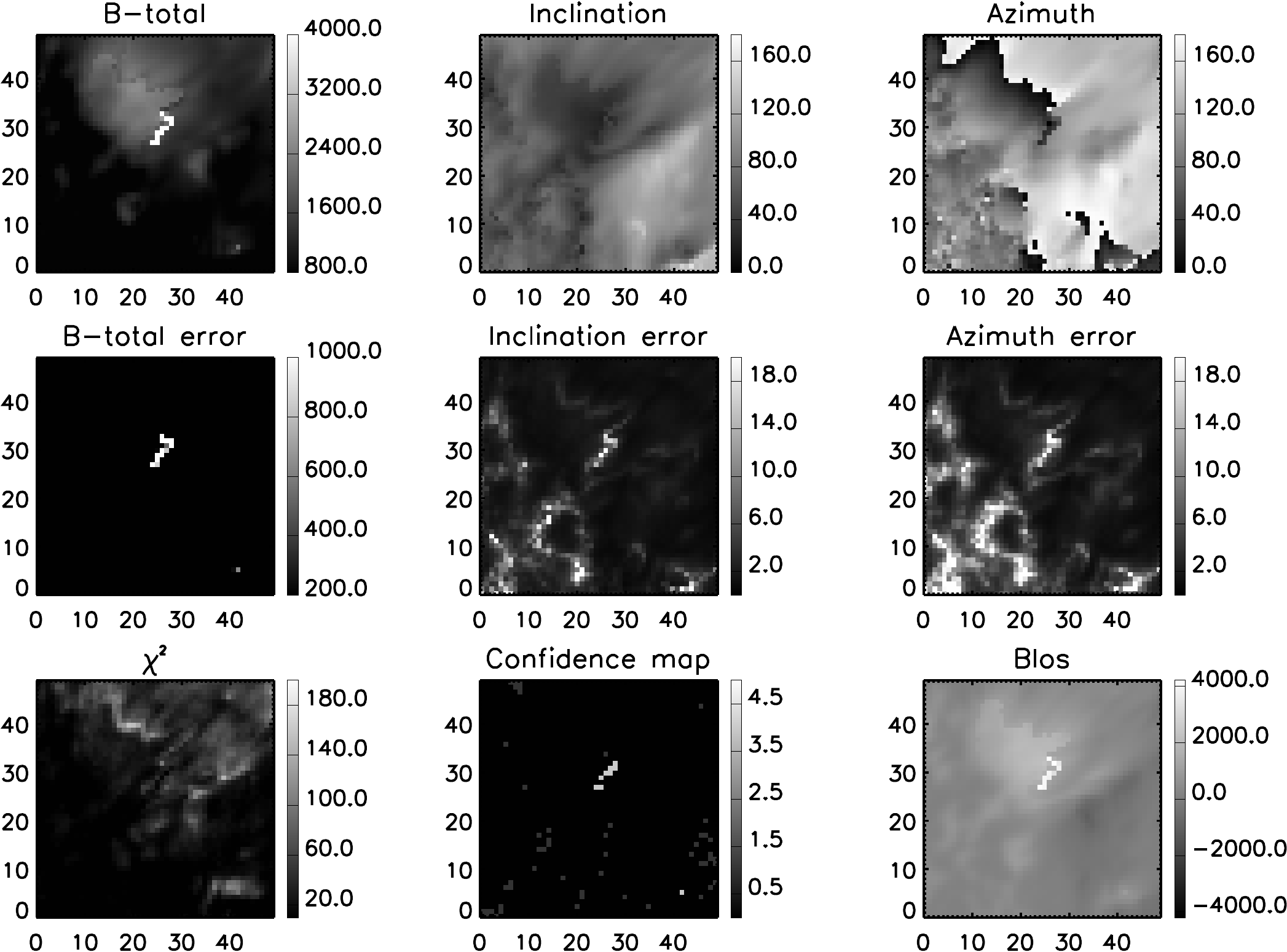}
}
\caption{CCD cut-outs of inverted data for a small area in HARP\,1638, AR\,\#11476 showing
bad pixels. Panels from left to right and top to bottom are field strength,
inclination, azimuth of magnetic field, their errors, $\chi^{2}$ for the 
least-square-fit, the confidence map, and LoS magnetic field. Erroneous values can
be clearly seen in field strength, error of field strength, and
LoS field. They are also visible in errors of inclination and azimuth. 
Units along the x and y axis are pixels.}
\label{fig:typeone}
\end{figure}

Some examples are shown in Figure~\ref{fig:badpixelimgs}. In the top left
panel, the field strength in HARP\,1638, AR\,\#11476 at 12:00~TAI 8 May 2012
at N10\,E34 is shown. Pixels with magnetic fields significantly stronger
than the surroundings are denoted by an arrow. A horizontal cut through
this image is plotted in the lower left panel, where it is easy to see an
abrupt jump from 2000 Mx cm$^{-2}$ to 5000 Mx cm$^{-2}$ (the VFISV maximum). 
The panels in the right column display another instance of a failed convergence. 
In this case, HARP\,1807, AR\,\#11515 at 02:24 TAI 1 July 2012 at S17\,E29 presents a few
pixels with rather low field strengths. Again, a horizontal cut through
this image in the lower right panel shows a drop from 3000 Mx\,cm$^{-2}$
to 400 Mx\,cm$^{-2}$ for an adjacent pixel.

A number of reasons can lead the inversion code to fail to converge. Inside
very dark sunspots, where the magnetic field is very strong, the Zeeman
splitting of the spectral line can be large enough to push one (or both) of
the Stokes V (Q and U) lobes outside of the measured spectral range observed
by HMI. This problem is exacerbated by the Doppler shift that arises as a
combination of the orbital velocity of SDO, the solar rotation, and the
intrinsic plasma motions at the photosphere.
In areas with strong gradients of the magnetic field or the LoS velocity (such
as the penumbra of a sunspot, an emerging region, or a flaring region) 
the Stokes profiles tend to show large asymmetries. The inversion code often 
fails to converge in these
cases because the ME approximation is strongly violated.

Besides having abnormal field strengths, these pixels typically have
out-of-the-ordinary values in other quantities computed with VFISV. This
provides a way to identify them. Figure~\ref{fig:typeone} shows a cluster of
anomalous inversion values in AR\,\#11476. They are clearly seen in the error
of the field strength, and are also visible in the errors of inclination
and azimuth and the LoS field. Some also appear in the confidence maps,
although they do not always show up in $\chi^2$.  A description of the data
confidence map is in Appendix \ref{sec:ConfMAP}.

\subsubsection{Reliability of the Disambiguation}

There are two main failure modes for the simulated annealing part of the
disambiguation: 1) the minimization algorithm fails to find the global minimum
of the energy, or 2) the global minimum does not correspond to the correct
disambiguation. In the first case, the result can be improved by changing the
cooling schedule used for the annealing, but it is computationally impractical
to use a cooling schedule in the pipeline that will find the global minimum.
Typically, however, the pixels that differ from the global minimum
configuration occur in areas where the noise in the transverse field is high.
The second case can occur for several reasons. As in the first case, when the
transverse field is dominated by noise or is not spatially resolved, then
finite differences do not accurately represent the derivatives of the field.
Alternatively, if the field is sufficiently far from potential, then the
approximation to the divergence of the magnetic field will not be accurate. 

\begin{figure}
\centerline{\includegraphics[width=0.6\textwidth]{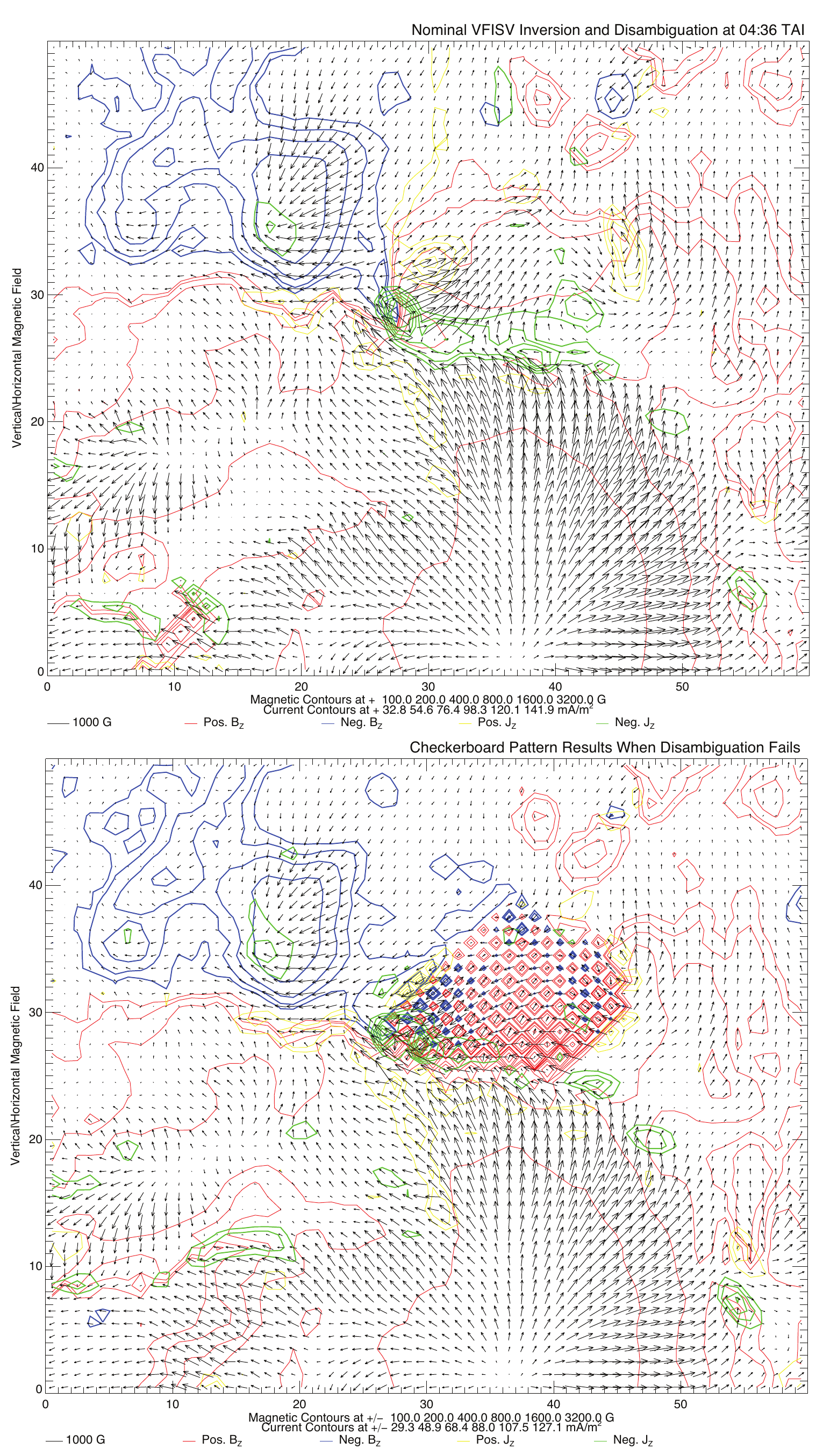}}
\caption{Closeup of magnetic field and currents in AR\,\#11158 for two consecutive 
time steps at 2011.02.16\_04:36 and 04:48 TAI. The red and blue contours show 
the vertical magnetic field and the arrows indicate the strength and direction of 
the horizontal field component.  Yellow and green contours show where the computed 
vertical current is strong.  Axes are in pixels.  The upper panel at 04:36\,TAI 
shows a fairly normal field configuration.  The central region of the lower panel 
at 04:48\,TAI with diamond-shaped contours shows a case where the disambiguation 
module gives conflicting results in adjacent pixels. Certain areas of an active 
region may be more susceptible to the instability, which will come and go from 
one time step to the next.  See text for details.}
\label{fig:checkerboard}
\end{figure}

One common manifestation of a failure of the annealing to reach the global
minimum is the ``checkerboard'' pattern illustrated in
Figure~\ref{fig:checkerboard}. In this case, the azimuthal angle alternates
between neighboring pixels in being in the range $0^\circ \le \psi <
180^\circ$ and being in the range $180^\circ \le \psi < 360^\circ$. This
particular pattern is an artifact of the four-point stencil used for the
finite differences, but similar artifacts occur when other stencils are used.
It results because the energy is very similar in the checkerboard
configuration and in the configuration in which all the azimuths are in one or
the other of the above ranges.  Strong-field regions where the disambiguation
is less stable are often regions where the inversion also gives variable
results from frame to frame, suggesting that the magnetic field is complex in
such regions.

\subsubsection{Comparing HMI Velocity and Magnetic Field Using VFISV and the MDI-like Algorithms}
\label{sec:hmimdi}

\begin{figure}
\centerline{
\includegraphics[angle=0, scale=0.63]{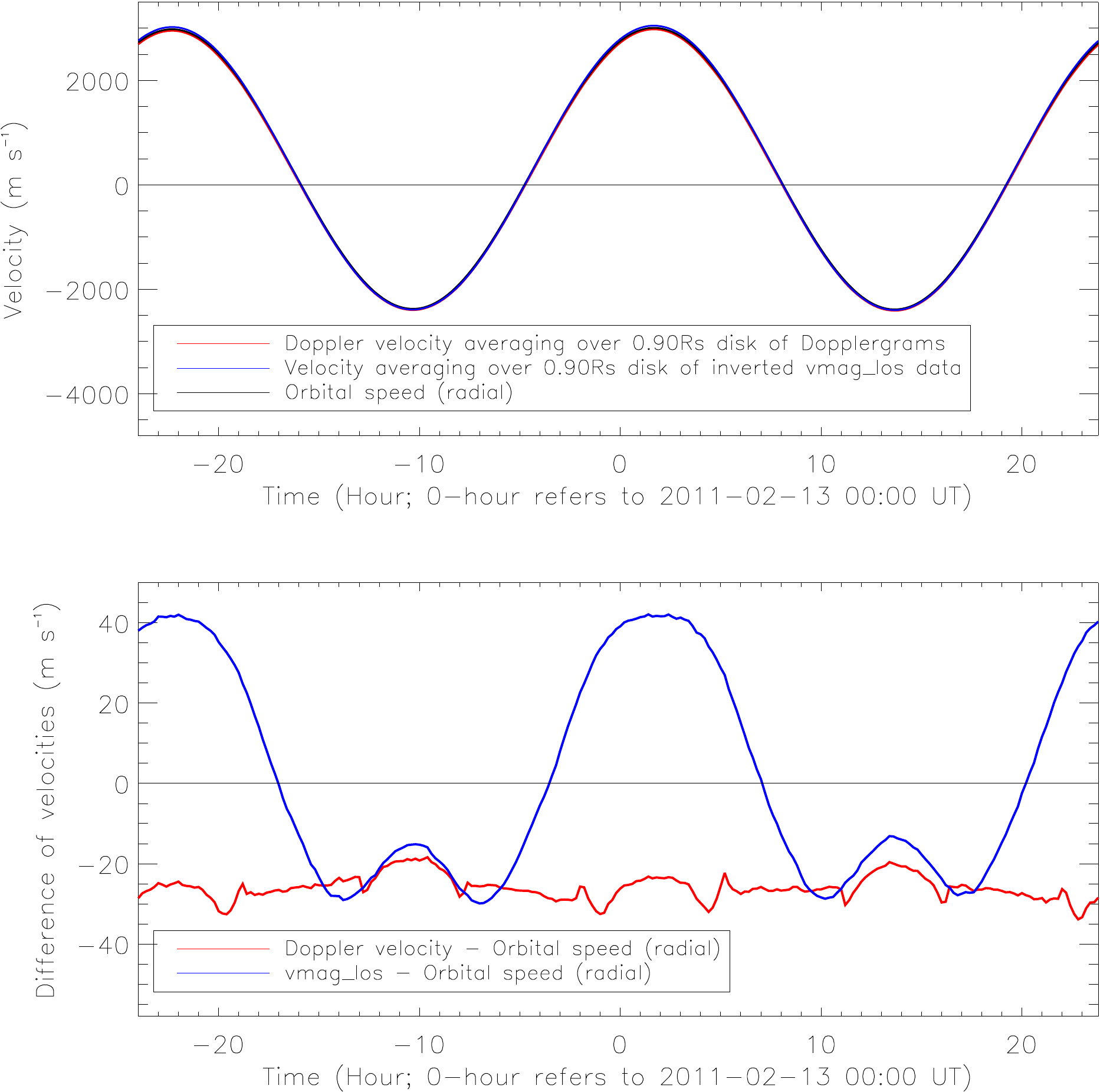}
}
\caption{\small Top: 48-hour temporal profiles of the known orbital radial
speed relative to the Sun, V$_r$ (black); the average over the inner 0.9 
of the solar disk of Dopplergrams, V$_{\rm Dop}$ (red); and a similar disk average 
of the LoS velocity from inverted VFISV data, V$_{\rm Inv}$ (blue). 
The maximum redshift of $\sim 3000$\,m\,s$^{-1}$ occurs at dusk, a little after 1\,UT.
Bottom: Offset difference V$_{\rm Dop}$~-~V$_r$ (red) and the offset difference
V$_{\rm Inv}$~-~V$_r$ (blue) as a function of time.
The average Doppler velocity, V$_{\rm Dop}$, is blue shifted by about
30~m\,s$^{-1}$.
V$_{\rm Inv}$ shows a clear daily variation correlated with V$_r$.}
\label{fig:vel} 
\end{figure}

A comparison is made between the precisely known satellite speed relative to
the Sun, V$_r$, the calibrated HMI Doppler velocity determined using the standard 
MDI-like algorithm, V$_{\rm Dop}$, and the LoS velocity obtained with the VFISV inversion 
V$_{\rm Inv}$.  The three curves in the top panel of Figure~\ref{fig:vel}
show 48-hour temporal profiles of V$_r$ (black), the velocity averaged
over the inner 0.90 $R_s$ from HMI Dopplergrams computed with the same
algorithm and corrections used for helioseismology (V$_{\rm Dop}$, red), and the LoS
velocity averaged over the inner 0.90 $R_s$ determined with VFISV (V$_{\rm Inv}$,
blue). $R_s$ refers solar disk radius. Those three measurements are nearly
identical. 
The bottom panel of Figure~\ref{fig:vel} displays
differences of averaged V$_{\rm Dop}$ - V$_r$ (red) and of averaged V$_{\rm Inv}$ - V$_r$
(blue).
The calibrated Doppler velocity, V$_{\rm Dop}$, is blue shifted $\sim
-30$~m\,s$^{-1}$
relative to V$_r$; 
the calibration depends on a smoothed daily fit to the
45\,s V$_{\rm Dop}$ observed with the other HMI camera.
The offset in the {\tt fd10} velocity, V$_{\rm Inv}$, has a
daily periodicity with amplitude $\sim 40$~m\,s$^{-1}$ associated with the wavelength
shift caused by V$_r$. The error in V$_{\rm Inv}$ changes character at dawn ($\sim 14$\,UT), when
the spacecraft is moving most rapidly toward the Sun. \inlinecite{Welsch2013}
discuss the sources of offsets associated with instrumental effects and the
convective blue shift.

\begin{figure}
\centerline{
\includegraphics[angle=0, scale=0.75]{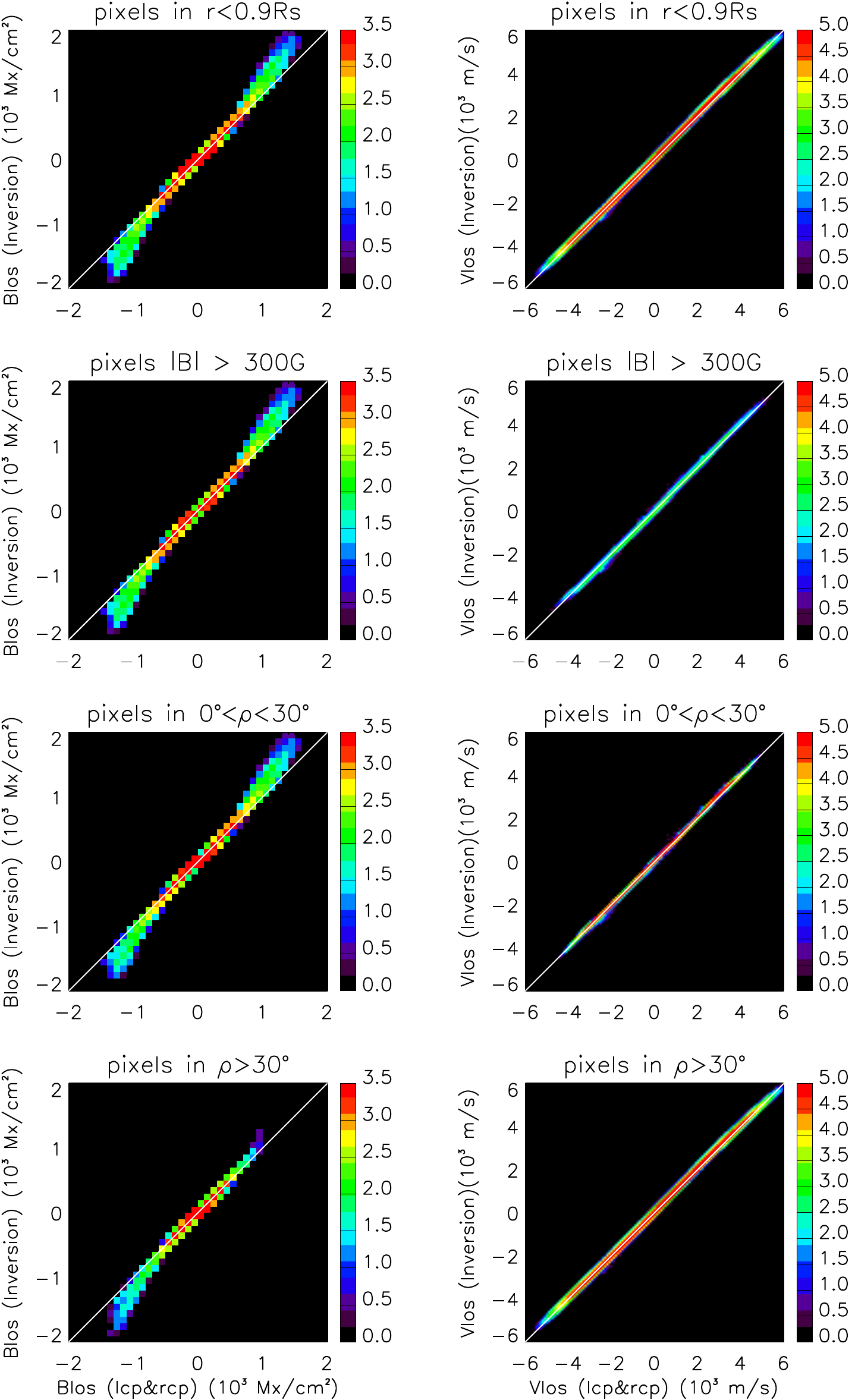}
}
\caption{\small Comparison of VFISV inversion and the data derived using
the MDI-like algorithm. The data used are the 720s {\tt fd10} inverted data
(ME), Dopplergrams and LoS magnetograms taken 12 February 2011
at 00:48, 06:48, 12:48, and 18:48 TAI. Comparison is made between
LoS magnetic fields from the MDI-like algorithm (x-axis in left
column) and from the ME inversion (y-axis in left column). The right column
compares LoS velocities from calibrated Dopplergrams (x-axis in right column)
to velocity from the ME inversion (y-axis in right column). The color table
shows the logarithmic density of the number of scatter plot points. From top to
bottom panels show comparisons for pixels within 0.90 R$_s$, strong-field pixels 
with B$_{Tot} > 300$ Mx cm$^{-2}$, disk center pixels (\(\rho<30^\circ\)), and limb pixels
(\( \rho>30^o \)), where $\rho$ is the center-to-limb angle.
}
\label{fig:comp}
\end{figure}

Figure~\ref{fig:comp} shows a more detailed comparison between both the
velocity and magnetic fields. The inverted velocity, V$_{\rm Inv}$, and inverted
LoS magnetic field, B$_{\rm{Los}}$, are determined using the {\tt fd10} inversion.
The Doppler velocity, V$_{\rm Dop}$, and LoS magnetic field, M$_{\rm{Los}}$, are computed
from LCP and RCP using the MDI-like algorithm (see Section \ref{sec:LineOfSight}).  
The data are all observed with the same HMI camera. 
Four pairs of observations, taken 12 February 2011 at 00:48, 06:48,
12:48, and 18:48 TAI are used. The
magnetic comparison is shown in the left column of Figure~\ref{fig:comp}
and the velocity comparison is shown in the right column. The MDI-like
LoS magnetogram and Dopplergram values are shown on the x-axis versus the
{\tt fd10} magnetic and velocity values on the y-axis. Results for
different parts of the disk and for strong-field regions are shown.
Note the color
scales in the left and right columns are different in order to better show
the details of the distribution of scatter-plot points. For reasons
discussed in Section \ref{sec:LineOfSight} and by \inlinecite{Couvidat-wavelength2012}
and \inlinecite{Liu2007}, 
the MDI-like algorithm is not as accurate in strong field
regions and underestimates the field strength. The LoS velocities from the
two algorithms agree very well.

\subsection{Limits and Validation}

The inference of solar photospheric magnetic field vector maps is at best an
estimate that represents the average field over height, time, and space,
as deduced using a record of photons averaged in the spectral domain
as well. In practice vector magnetogram maps are the result of inverting
data that do not well constrain an optimization method based on physical
assumptions known to be incorrect. 
There are additional instrumental artifacts that may or may not be mitigated
fully. In this section, some rudimentary evaluation of the evolution
of AR\,\#11158 features are discussed in light of the limitations of the
data, and a brief qualitative comparison is presented with simultaneous
observations from the {\it Hinode}/\textit{SpectroPolarimeter}
\cite{hinode,hinodesp,Lites2013}.
Comparisons beyond the qualitative presentations made herein are beyond the
scope of this paper.

\subsubsection{Validation and Limitations of HMI Processing and Data}

Characteristics of both the HMI instrument and the processing can
lead to systematic and possibly subtle effects of which users need to at
least be aware.

First, there is no treatment of scattered light, either as part of the data
reduction or the inversion procedure. This can lead to systematically
lower inferred field strengths and incorrect inferred angles of the magnetic
vector \cite{ivm3}. It will also provide simply incorrect relative photometry 
between different solar structures. In the future a treatment of scattered light
may be performed; this is a feature available in the original VFISV scheme. 

Second, the sparse and limited spectral sampling precludes a clean continuum
point from being sampled consistently in the same part of the spectrum.
For segments of the daily orbit, either the first or the last filter position
should sample near the spectral continuum, but rarely far enough from the
spectral line to be uncontaminated, and never for more than a few hours
at a time. Again, this can impact interpretation of relative photometry
especially for analysis of temporally varying features.

Third, as summarized in Section \ref{sec:Stokes} there is overlap
in the contribution of filtergrams to each 12-minute average, such
that a single magnetogram is not a completely independent measurement from
its temporal neighbors. The effects of this may be subtle, however may
be detectable in the velocity tracking codes, during times of strong temporal
evolution (\textit{e.g.} flares or emergence), and may contradict some assumptions made by 
statistical analysis.

\subsubsection{HMI Vector Magnetograms in Context}\label{sec:hinode}

Every instrument and observing scheme makes compromises to balance the limitations
imposed by a finite number of photons and the performance of real optics. 
HMI's vector magnetogram
capability, as with data from other vector-spectro\-polari\-met\-ric instruments,
enables a broader spectrum of physical interpretation than instruments such
as MDI that provide solely LoS measurements.
Comparing with other vector magnetogram data, HMI observations have superior
spatial and temporal coverage as compared to {\it Hinode}/SP, whereas 
the spatial and spectral resolution, sampling, and polarization sensitivity 
of \textit{Hinode}/SP are significantly better.  
HMI data have superior spatial and temporal coverage as compared to the 
\textit{Imaging Vector Magnetograph} (IVM)
archive, but worse spectral sampling, in addition to the
advantages and disadvantages of space-based vs. ground-based observing.
HMI's full-disk imaging capability makes its global spatial coverage superior
to any data except the SOLIS/SPM and Huairou/HSOS.  The HMI spatial resolution
of $1^\arcsec$ is comparable to the IVM and the optical
potential of SOLIS/SPM, but significantly worse than {\it Hinode}/SP and 
certain high-resolution ground-based instruments. HMI's combination of 
cadence, coverage, and continuity are unmatched.

\begin{figure}[t]
\centerline{
\includegraphics[width=0.95\textwidth]{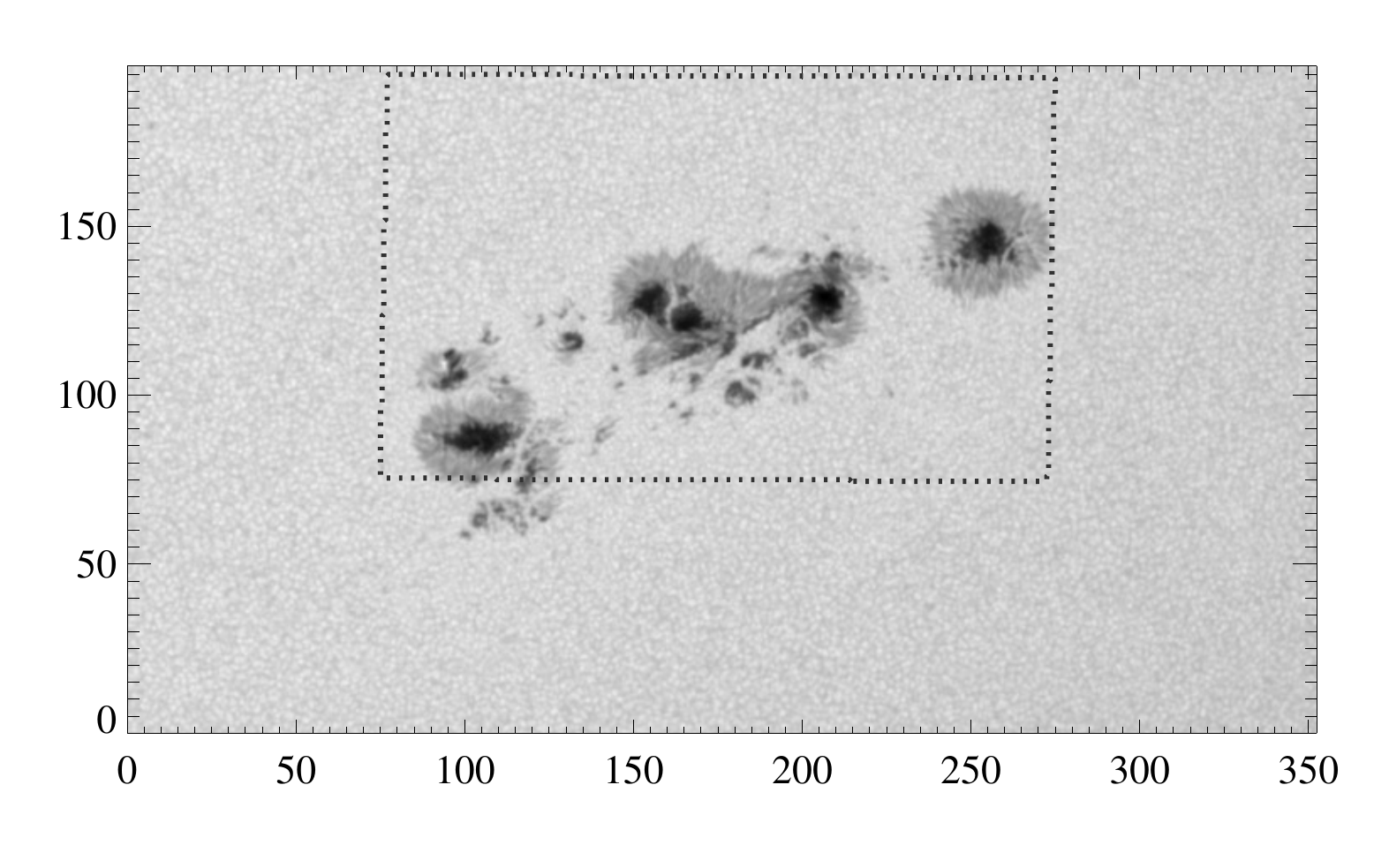}
}
\vspace{-0.2cm}
\caption{\small 
Continuum image from {\sf hmi.sharp\_720s} centered on NOAA AR\,\#11158,
from 15 February 2011 at 10:36, with an outline of the field of view covered
by the {\it Hinode}/SP fast-scan obtained 10:11--10:55 TAI, which had a center
time approximately the same as the HMI data. Axes labels are in arc seconds.}
\label{fig:hsphmiolay}
\end{figure}

\subsubsection{HMI Vector Magnetograms {\it vs.} {\it Hinode}/SP Vector Magnetograms}

\sloppy
A simple statistical comparison is performed between co-temporal
HMI and {\it Hinode}/SP data. The HMI data from {\sf hmi.sharp\_720s}
were co-aligned with the {\it Hinode}/SP fast-scan in {\tt
20110215\_101126.fits}\footnote{\href{http://sot.lmsal.com/data/sot/level2d}{sot.lmsal.com/data/sot/level2d}}, which had a
center-scan time approximately two minutes different from the time-stamp in the
HMI data. The {\it Hinode}/SP map was inverted using the MERLIN inversion
scheme as the data are available from the Level-2 production; no additional
processing was performed.  Figure~\ref{fig:hsphmiolay} shows the HMI cutout
area, with the field-of-view covered by the {\it Hinode}/SP scan indicated.
\fussy

\begin{figure}[t]
\centerline{
\includegraphics[width=0.50\textwidth]{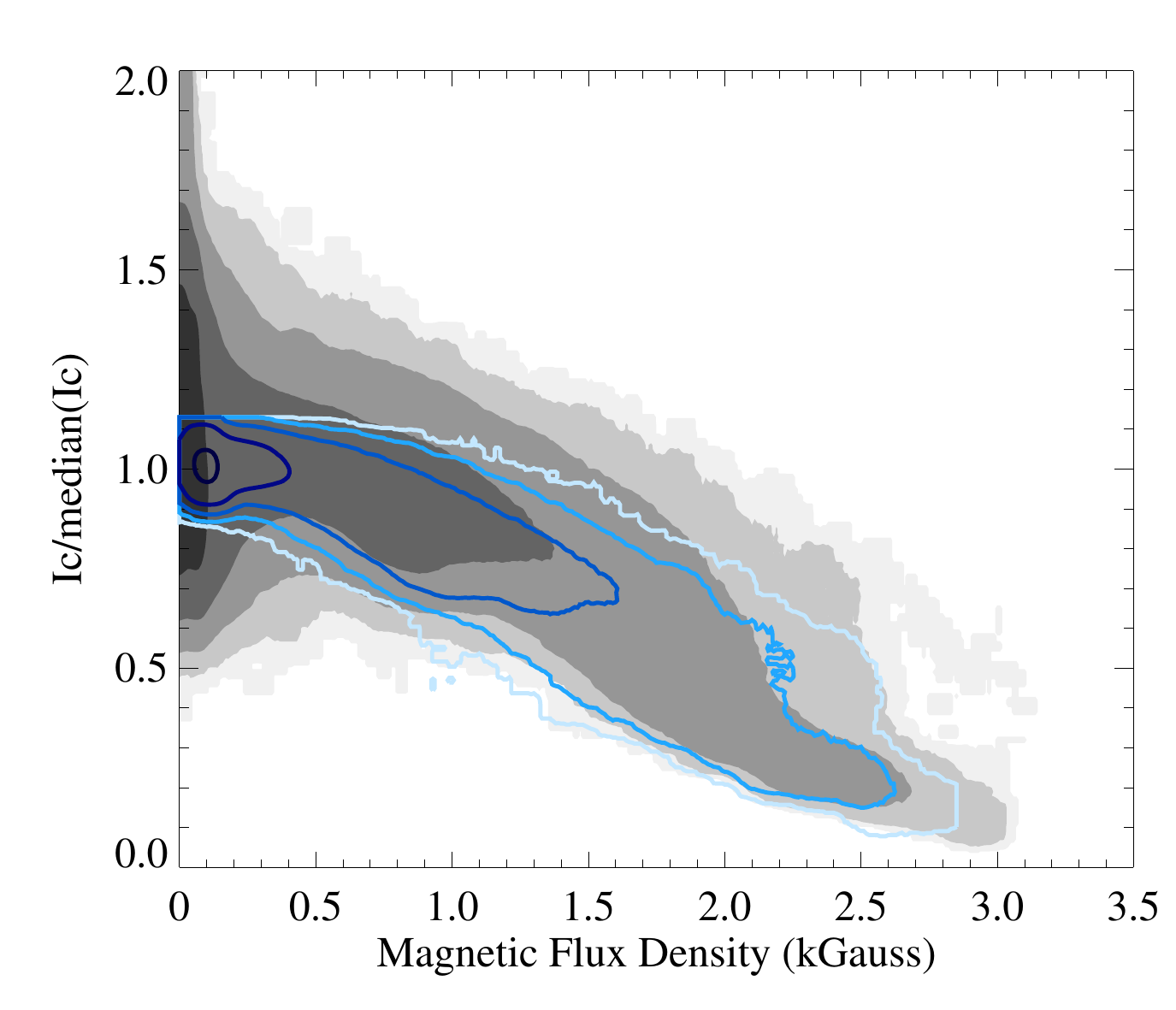}
\includegraphics[width=0.50\textwidth]{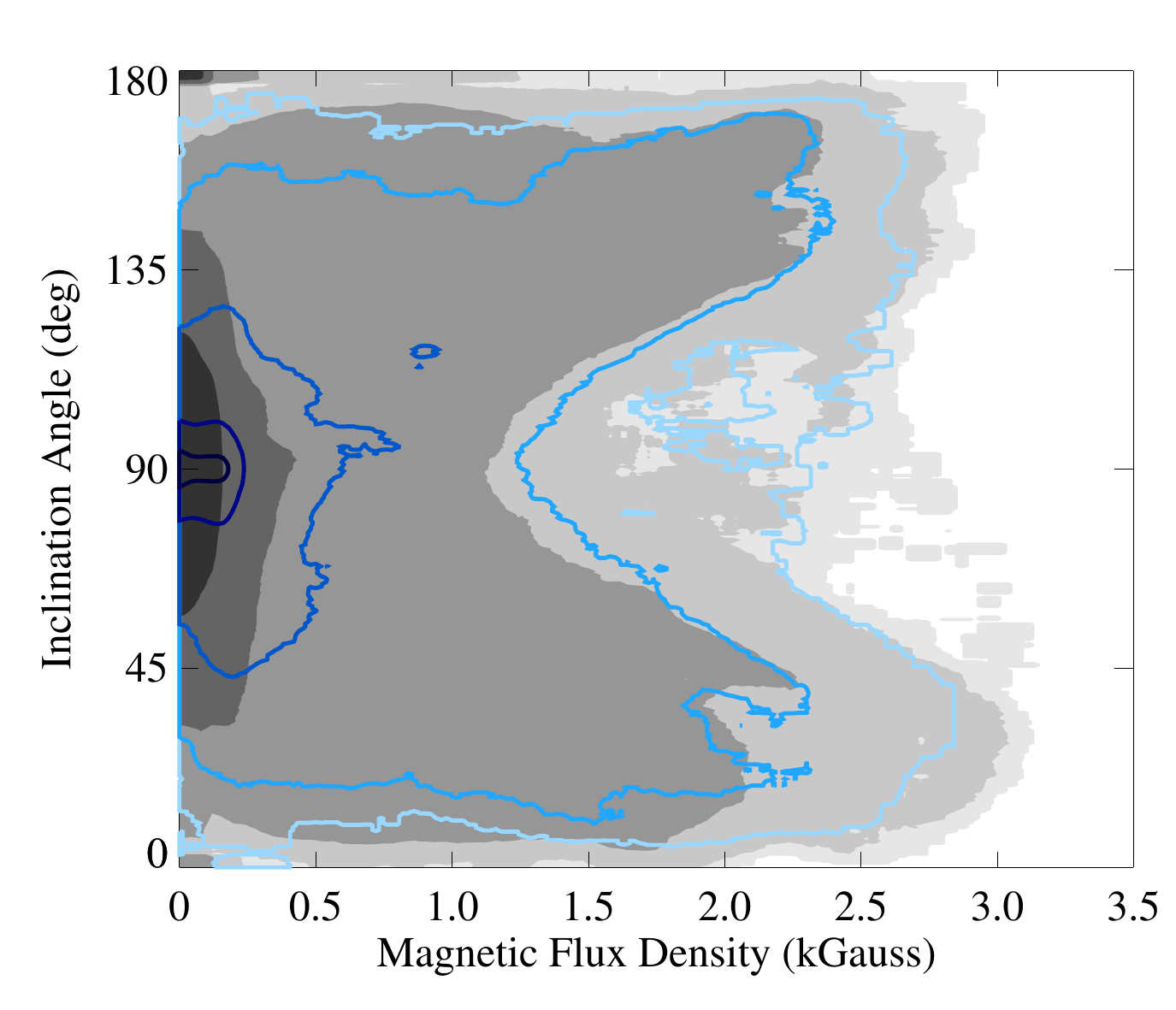}
	}
\vspace{-0.2cm}
\caption{\small Left: density histogram of the scatter plots of the flux density 
(field times the inferred magnetic filling factor for {\it Hinode}/SP) 
{\it vs.} continuum intensity (normalized by the median). 
Right: flux density {\it vs.} the magnetic vector inclination relative to the LoS. 
Gray scale: {\it Hinode}/SP data, blue contours: HMI data.
Contour levels are logarithmic in density, normalized for the number of pixels considered,
at 0.01,\,0.1,\,1,\,10,\,100 (except the highest contour is 50 for the intensity plot).
Only the area co-observed by the two instruments is considered.
The fraction of pixels inside each contour decreases from over 99.9\% within
contour level 0.01 to between roughly 75\%--99\% at level 1.0 and to 30\% -- 50\% at the highest contour.
If the distributions were identical, the contours and color-boundaries would coincide. 
The figure shows that the flux density measured by HMI is generally lower than {\it Hinode}/SP.
}
\label{fig:hsphmiscats}
\end{figure}

A comparison between continuum intensity (normalized by the median continuum
intensity) and the inferred field strength for the two datasets is shown in
Figure~\ref{fig:hsphmiscats}.  The HMI data are reported in their native
$0.5^\arcsec$ sampling reflecting the $\approx 1.0^\arcsec$
spatial resolution of the HMI telescope.  The {\it Hinode}/SP fast-scan data
are sampled to match the $\approx 0.3^\arcsec$ resolution of that telescope.
The flux density is presently reported for HMI data, as the magnetic fill
fraction is not a free parameter in the inversion and is fixed at unity.
The field and magnetic fill fraction are fit and returned separately for
{\it Hinode}/SP data inverted with MERLIN; for these comparisons, those two
quantities are multiplied.  No further manipulation of the data has been
performed for this comparison.

The distribution plots demonstrate that the data are qualitatively very
similar, but there are quantitative and systematic differences.  The HMI flux
density generally falls short of that reported by {\it Hinode}/SP, as does the
contrast and its variation.  Both of these effects are expected with lower
spatial resolution \cite{LekaBarnes2011} and the presence of scattered light
\cite{straylight}.
In these plots the flux-density effects are clearest in the sunspot umbrae,
whereas the contraction of the contrast distribution is easiest to see in the
quiet-Sun or plage areas.  The lack of a scattered light correction may also
contribute to both the field strength and continuum-intensity differences.
The inclination angles display the same morphology in the distribution plots;
however, the HMI data generally fail to achieve the extreme inclination
angles at all flux densities and tend to lie closer to the plane of the
sky. This again may be a consequence of scattered light \cite{ivm3} or
noise \cite{borrerokobel11},
or it may be attributable to the better wavelength sampling of {\it Hinode}/SP,
which effectively leads to an enhanced signal-to-noise ratio.

%% file: VP7-Summary.txt
\section{Summary}\label{sec:summary}

HMI vector magnetic field observations have been collected continuously
since 1 May 2010.  Examples have been shown to demonstrate the capabilities
of the HMI instrument and explain how the observables are generated routinely 
in the vector field pipeline analysis.  HARP\,377, AR\,\#11158 was highlighted in part 
because it produced the first X-class flare of Cycle 24 and in part because 
an earlier version of the HMI vector field data were released for this region.

A sequence of full-disk, polarized, narrow passband filtergrams are collected
every 135 seconds.  A Level 1 data analysis pipeline corrects the filtergrams
for a variety of instrumental and observational effects, co-aligns them, and
averages data from ten sequences together on a 720s cadence \cite{Schou2013}.
The basic magnetic observables, Stokes [I\,Q\,U\,V], are determined at six
wavelengths.  Other scalar quantities provided at the same cadence include
Doppler velocity, line-of-sight magnetic field, continuum intensity, line
depth, and line width. The data series are listed in Table
\ref{tab:DataSeries}.

HMI active region patches are identified automatically and tracked during
their entire disk passage \cite{Turmon2013} based on analysis of the
HMI LoS magnetic field and continuum intensity. Each HARP is
assigned a number and may in many cases be associated with one or more
NOAA active regions. There are also many HARPs that contain no sunspot
and therefore have no NOAA AR number. 
The geometric information in the HARP data series
is used to select regions for further processing.

VFISV, a very fast Milne-Eddington inversion code, is applied to determine
the magnetic field and other plasma parameters. The {\tt fd10} version of the
code described in this paper and in more detail in \inlinecite{Centeno2013}
has been run on all HMI full-disk vector field data since 1 May 2010.

The inverted {\tt fd10} data require disambiguation to resolve the $180^\circ$
azimuthal uncertainty in the transverse field direction. We have applied
the method described by \inlinecite{Barnesetal2012} to a growing selection
of active region patches. The disambiguation minimizes a quantity proportional 
to the sum of the absolute values of the field divergence and the electric current density. 
The pipeline disambiguation code is being run
routinely on active region patches observed before 15 January 2014. 
Full-disk disambiguated vector magnetograms are being used for more recent
observations.

The SHARP (Space-weather HARP) data series gathers together for each HMI
active region patch at each time step the disambiguated vector field, other
inversion outputs, and most of the HMI scalar observables. The
SHARP data series also provides a number of active-region quantities computed
at each time step for each patch, such as total flux, mean current helicity,
and weighted measures of the field gradient \cite{Bobra2013}. Maps of select
inversion outputs are remapped into heliographic coordinates using a cylindrical
equal area projection centered on the HARP.

The pipeline is applied to definitive HMI data that become available several
days after the observations are collected. The definitive HARP analysis is
not completed until after the region has rotated off the visible solar disk,
introducing delays as long as several weeks. For most scientific purposes
these high-quality data are the most useful. However, for space-weather forecasting and
evaluative purposes, quick-look near-real-time data for many observables are
generally available within minutes or at most a few hours of observation. These
data include the full-disk scalar quantities and everything required to make
SHARPs.

In Section \ref{sec:uncertainties} and elsewhere many of the known limitations,
uncertainties, and issues associated with the vector field data are described.
Prudence must be exercised when using the data quantitatively.  HMI vector
field measurements compare favorably with other observations, but of course
have been optimized for the capabilities and constraints of the instrument,
satellite, and processing.

Table \ref{tab:websites} provides links to additional
documentation relevant to the HMI magnetic field data.

\begin{table}
\begin{center}
\caption{Useful Web Sites With Information About HMI Magnetic Field Data}
\label{tab:websites}
\begin{tabular}{r p{0.5\textwidth}}
\hline
Web Address & HMI Data Product\\
\hline
\urlurl{hmi.stanford.edu/magnetic} & HMI Magnetic Field Images and Data Products \\
\urlurl{jsoc.stanford.edu} & Introduction to the Data in the JSOC \\
\urlurl{jsoc.stanford.edu/jsocwiki} & Introduction to JSOC Data Series and Access \\
\urlurl{jsoc.stanford.edu/jsocwiki/MagneticField} & HMI Magnetic Product Information \\
\urlurl{jsoc.stanford.edu/jsocwiki/HARPDataSeries} & Description of HARP Data Series \\
\urlurl{jsoc.stanford.edu/jsocwiki/sharp\_coord} & SHARP Coordinates and CEA Remapping \\
\urlurl{jsoc.stanford.edu/jsocwiki/PipelineCode} & Guide to HMI Pipeline Code and Processing Notes \\
\urlurl{jsoc.stanford.edu/ajax/lookdata.html} & Data Exploration and Request Tool for JSOC \\
\urlurl{jsoc.stanford.edu/ajax/exportdata.html} & Data Export Tool for JSOC Data \\
\urlurl{www.lmsal.com/sdouserguide.html} & Comprehensive Guide to SDO Data\\
\urlurl{sdo.gsfc.nasa.gov} & SDO Mission Web Site\\
\hline
\end{tabular}
\end{center}
\end{table}

%% file: VA1-Segments.txt
\section{Explanation of Certain Vector Field Data Segments}
\label{app:SegmentDescriptions}

\sloppy
A JSOC {\it data series} consists of {\it keywords}, {\it segments}, and {\it links} for a set of {\it records}. A
{\it record} holds the information for specific {\it primekeys}. The {\it primekey}
is typically a time, or in the case of HARPs and SHARPs, the two {\it
primekeys} are HARP number and time.  
A {\it record} can have one or more {\it segments} that generally contain image data.
A number of {\it keywords} contain metadata about the data series or about particular {\it records}. 
For example, in most data series the keyword {\sc t\_rec} gives the time of
the {\it record} and the keyword {\sc quality} holds information about possible issues in the 
collection or processing of the data 
(see \urlurl{jsoc.stanford.edu/doc/data/hmi/Quality\_Bits/QUALITY.txt} for details).
In the vector field data series the segment {\sc field} is
the array of total magnetic field strength. Various {\it links} provide pointers to
segments in other data series; for example the link {\sc hmi.Mharp\_720s} in the {\tt
hmi.sharp\_720s} series is a link to the full disk line-of-sight magnetogram at
time {\sc t\_rec}. The elements of the series are stored independently until
they are assembled on export.  Detailed explanations of how data are stored
and accessed can be found at the JSOC wiki page, \urlurl{jsoc.stanford.edu/jsocwiki/DataSeries}.
Brief definitions of keywords, segments, and links for each particular series are available 
using the {\tt lookdata} utility at \urlurl{jsoc.stanford.edu/ajax/lookdata.html} 
in the ``Series Content'' tab.
\fussy

Most data segments in the vector field data series are self-explanatory,
\textit{i.e.}
they contain maps of an observable or inverted physical quantity, an uncertainty, 
or the correlation of the errors in two inverted parameters. For example the data 
segment {\sc inclination} contains a map of the magnetic field inclination angle 
relative to the line of sight; the data segment {\sc azimuth\_err} is the formal 
error of the magnetic field azimuth value, and {\sc field\_inclination\_err} is 
the cross correlation of the field strength and inclination errros as determined by the 
{\tt fd10} VFISV inversion. 

Certain other data segments (see Table \ref{tab:SpecialSegments}) that are more obscure in their meaning are
explained in the following sections. For the most part they are the same in
all of the vector field data series, but in some cases (as noted) the name is
changed or the values are updated. Keywords generated by upstream modules in
the analysis pipeline are generally not described here.

\begin{table}
\begin{center}
\caption{Vector Data Segments Requiring Additional Explanation}
\label{tab:SpecialSegments}
\begin{tabular}{r l}
\hline
Segment Name & Description\\
\hline
{\sc confid\_map} & Quantitative estimate of data quality \\
{\sc info\_map} & Encoded confidence indicator for data quality in B and SHARP \\
{\sc qual\_map} & Same as {\sc info\_map} for ME series\\
{\sc conv\_flag} & Status of the VFISV inversion code result\\
{\sc disambig} & Code for azimuthal disambiguation \\
{\sc conf\_disambig} & Confidence indicator for disambiguation \\
{\sc bitmap} & HARP region map of active pixels inside and outside patch \\
\hline
\end{tabular}
\end{center}
\end{table}

\subsection{Data segment {\sc confid\_map}}
\label{sec:ConfMAP}

Data segment {\sc confid\_map} quantifies the data quality of each pixel in a
very simple way. Values range from 0 to 6 with lower numbers indicating
higher confidence in the reported quantities resulting from the VFISV
inversion and disambiguation. 

\medskip
\small
\begin {tabular}{r l}
\hline
\multicolumn{2}{c}{Meaning of {\sc confid\_map} Values}\\
\hline
Value & Meaning \\
\hline
0 & Pixel with no known problems \\
1 & One of Stokes [\,Q\,U\,V\,] or LoS field was lower than noise threshold \\
2 & Two of Stokes [\,Q\,U\,V\,] or LoS field were lower than noise threshold \\
3 & Stokes [\,Q\,U\,V\,] and LoS field all lower than noise threshold \\
4 & VFISV inversion failed to converge properly, see {\sc conv\_flag} \\
5 & Bad pixel, defined as aberrant compared to adjacent pixels \\
6 & Missing data in at least one of input Stokes [\,I\,Q\,U\,V\,] \\
\hline
\end{tabular}
\normalsize

\subsection{Data segments {\sc qual\_map} and {\sc info\_map}}
\label{sec:QUALMAP}

The segment {\sc qual\_map} or {\sc info\_map} describes qualities of the
vector magnetic field data with defined codes as described in Table
\ref{tab:QUALMAP}.  {\sc qual\_map} is used for {\tt fd10} products and
{\sc info\_map} is used for disambiguated vector products. The contents are
equivalent. These segments are 32-bit arrays that contain bitwise encoded
information about the VFISV {\tt fd10} inversion and disambiguation for each
pixel. 

Bit 27 indicates a questionable pixel, which is defined as one having inversion 
values that are abnormal compared with adjacent pixels. 
For example, the field strength value in a bad pixel is often much lower 
than the adjacent pixels. Currently we use an algorithm to identify those 
bad pixels.  We evaluate \( |M_{\rm LoS}| - |B_{\rm LoS}| \), where M$_{\rm LoS}$ refers 
to the LoS field derived using the MDI-like algorithm (see
Section \ref{sec:LineOfSight}), and B$_{\rm LoS}$ is LoS field from the VFISV inversion.
If the difference is greater than a threshold value, \textit{i.e.} the LoS inverted
field is much lower than the MDI-like field, it is likely to be a bad
pixel. The threshold value is currently set to 500 Mx\,cm$^{-2}$, which successfully
identifies bad pixels in sunspot umbrae, but sometimes fails to find bad
pixels outside of sunspots.

\begin{table}
\begin{center}
\caption{Description of {\sc qual\_map} and {\sc info\_map} Bits }
\label{tab:QUALMAP}
\begin{tabular}{rp{0.85\textwidth}}
\hline
Bit & Description\\
\hline
0-7 & Reserved for disambiguation information (TBD - to be defined)\\
8-23 & TBD\\
24-26 & Binary code indicating status of VFISV computation and inputs.\\
      & See {\sc conv\_flag} Table in next section for bit assignments. \\
27 & Bad pixel; 0=NO, 1=YES. (VFISV module)\\
28 & LoS field lower than noise threshold; 0=NO, 1=YES. (VFISV module)\\
29 & Stokes Q or U lower than noise threshold; 0=NO, 1=YES. (Stokes Module)\\
30 & Stokes V lower than noise threshold; 0=NO, 1=YES. (Stokes Module)\\
31 & Any missing Stokes [\,I\,Q\,U\,V\,] data; 0=NO, 1=YES. (Stokes Module)\\
\hline
\end{tabular}
\end{center}
\end{table}

\subsection{Data Segment {\sc conv\_flag}}
\label{app:convflag}

The status of the {\tt fd10} VFISV inversion module processing is recorded
in the character array {\sc conv\_flag}, which is of the same size as the
magnetic field map. The value is a 3-bit code that indicates if there was a
problem in the inversion of the pixel. This information is repeated in bits
24-26 of the {\sc qual\_map} segment.

\medskip
\small
\begin {tabular}{r l}
\hline
\multicolumn{2}{c}{VFISV {\sc conv\_flag} Error Codes}\\
\hline
Code & VFISV Error Code  \\
\hline
{\tt 000} & VFISV inversion module completed nominally \\
{\tt 001} & Continuum intensity below required threshold; pixel not inverted\\
{\tt 010} & Maximum number of iterations reached without converging\\
{\tt 100} & Value of $\chi^2$ too large; exited loop\\
{\tt 101} & Singular Value Decomposition (SVD) failed to compute errors\\
\hline
\end{tabular}
\normalsize

\subsection{Data Segment {\sc disambig}}
\label{app:disambig}

\begin{sloppypar}
The segment {\sc disambig} is a character array the size of the magnetic field
map that appears in disambiguated data series in native CCD coordinates,
\textit{e.g.} {\sf hmi.sharp\_720s}, {\sf hmi.Bharp\_720s}, and {\sf hmi.B\_720s}. 
If $180^\circ$ must be added to the result of the {\tt fd10} inversion reported in {\sc azimuth}, 
a bit is set in the {\sc disambig} segment.
Only in the four SHARP data series are the {\sc azimuth} or field component data actually modified.

Pixels are considered to be high confidence, intermediate confidence, or weak confidence 
as reported in the {\sc conf\_disambig} segment described in the next section.
In high-confidence pixels the minimum-energy method is always used. 
In intermediate pixels the minimum energy result is smoothed using the acute angle method.
In weak pixels the results of three disambiguation methods are provided.

For the patch-wise SHARP analysis applied to HMI data collected before 15
January 2014, 
the pixel disambiguation is generally considered high-confidence if the transverse 
field strength exceeds the velocity and position-dependent HMI noise mask by 
a threshold, {\sc doffset}  = 20\,G. 
The threshold is currently set to 50\,G for full disk disambiguation.
Adjustments are made by eroding the region to eliminate isolated pixels.
See Section \ref{sec:NoiseValue} for more description of the noise mask. 

Pixels below the threshold but near strong-field regions are considered to be intermediate.
In patch-wise disambiguation analysis all pixels that are not strong are considered to be intermediate.
In the full-disk disambiguation the intermediate pixels are those within five pixels of the 
high-confidence region.
The minimum-energy disambiguation is smoothed using the nearest-neighbor acute-angle 
method in intermediate regions.

In weak field regions the full-disk disambiguation module reports the results of three
alternative methods by setting bits as described in the accompanying {\sc disambig} 
table.  The minimum energy method is not applied in weak regions. Rather we provide
the disambiguation consistent with a potential field, a random assignment,
or a radial-field acute-angle method. The SHARP azimuth reported for full-disk 
disambiguation in weak regions is the radial-field acute-angle determination. 

In some {\sf hmi.sharp\_720s} series processed before August 2013 additional
information is provided for intermediate strength pixels. Bit 0 is the
standard smoothed minimum energy result, bit 1 is the result of a random assignment, and bit
2 is the radial acute-angle determination. Use of this information is
deprecated.

The {\sc conf\_disambig} segment indicates the pixel type.
\end{sloppypar}

\medskip
\small
\begin {tabular}{l c l}
\hline
\multicolumn{3}{ c }{{\sc disambig} Bit Codes} \\
\hline
Confidence & Bit & If Set, Add $180^\circ$ to Value in {\sc azimuth}  \\
\hline
High & All & Add $180^\circ$ to azimuth per minimum energy method \\
\hline
Intermediate & 0 & Add $180^\circ$ per smoothed minimum energy method \\
\hline
Weak & 0 & Add $180^\circ$ to azimuth per potential field method \\
Weak & 1 & Add $180^\circ$ to azimuth per random assignment method \\
Weak & 2 & Add $180^\circ$ to azimuth per radial-acute angle method \\
\hline
\multicolumn{3}{p{0.92\textwidth}}{The minimum energy disambiguation is reported for all high-confidence pixels above the noise threshold.}\\
\multicolumn{3}{p{0.92\textwidth}}{All pixels are considered to be either strong or intermediate in patch-wise disambiguation.}\\
\hline
\end{tabular}
\normalsize

\subsection{Data Segment {\sc conf\_disambig}}
\label{app:confdisambig}

Data segment {\sc conf\_disambig} is an indicator of the confidence in the
disambiguation result reported in {\sc disambig}. The 
value depends on field strength and proximity to strong-field regions.
The {\sc conf\_disambig} segment is a character array the size of the magnetic 
field map. If the pixel was not disambiguated, the value is 0, indicating no 
confidence in the {\sc disambiguation} segment.  
The value is set to 90 if the annealed results of
the minimum energy method are used directly, \textit{i.e.} in strong-field pixels. In
intermediate confidence regions, where smoothing is applied to the annealed result, the
value is 60. When disambiguation is performed on HARP patches, all on-disk pixels within 
the rectangular HARP bounding box have a value of 90 or 60. 
When full-disk disambiguation is performed, locations
more than five pixels from a high-confidence region are considered to be
lower-confidence and a value of 50 is assigned. Weak pixels are currently identified only
in full-disk disambiguation.
In principle other values may be assigned if another method is used.

\subsection{Data Segment {\sc bitmap}}
\label{app:bitmap}

The HARP module defines a rectangular array on the CCD called a bounding box. 
The bounding box encloses the identified active region. The active region
consists of one or a few contiguous regions with smooth borders within the bounding box. 
The bounding box may enclose an area substantially larger than the actual active region 
at some time steps. 
Within the bounding box the {\sc bitmap} also distinguishes between active and quiet pixels. 
Active pixels are those above a certain defined threshold in field strength. 
The {\sc bitmap} array has the same dimensions as the bounding box. 

\medskip
\small
\begin{tabular}{r l}
\hline
\multicolumn{2}{ c }{Meaning of HARP {\sc bitmap} Values}\\
\hline
Value & Meaning \\
\hline
0 & Off-limb pixel anywhere inside HARP bouding box \\
1 & Quiet pixel outside the HMI active region patch \\
2 & Active pixel outside the HARP \\
33 & Quiet pixel inside the HARP \\
34 & Active pixel inside the HARP \\
\hline
\end{tabular}
\normalsize

%% file: VA2-1332e15.txt

\section{The Earlier Version of HMI Vector Field Data: {\tt e15w1332}}
\label{sec:e15w1332}

\begin{sloppypar}
The first HMI vector field data released for scientific analysis in late
2011 provided information about NOAA AR\,\#11158, the first X-class flare producing
region of Solar Cycle 24. Subsequently, data for several other regions were released, all
processed with the same earlier version of the pipeline code, designated {\tt e15w1332}. 
Those data, described at \urlurl{jsoc.stanford.edu/jsocwiki/ReleaseNotes} and
in \href{http://jsoc.stanford.edu/jsocwiki/ReleaseNotes2}{ReleaseNotes2}, are archived at JSOC in the data series {\sf hmi.B\_720s\_e15w1332}.
and {\sf hmi.B\_720s\_e15w1332\_CEA}.  The current data being generated are labeled {\tt fd10}.
Section~\ref{sec:MEInversion} of this paper gives a brief description of the
{\tt fd10} pipeline processing, but an in-depth description can be found in
\inlinecite{Centeno2013} and \inlinecite{Barnesetal2012}. 
This Appendix briefly summarizes the differences between the processing that generated 
the {\tt e15w1332} data and the current pipeline that generates the {\tt fd10} and SHARP data.  
\end{sloppypar}

\begin{itemize}

\item The input HMI observables, the {\sf hmi.S\_720s} Stokes parameters, are unchanged
between the {\tt e15w1332} and {\tt fd10} data. 

\item The VFISV code that generated the {\tt e15w1332} data ran relatively slowly 
and did not converge as reliably.  The code was subsequently optimized for speed 
as described below.  

\item First, the forward calculation of the spectral line was changed 
to be a non-explicit calculation for wavelengths beyond $\pm$0.65\,\AA\ from the core of
the line.  This is done under the assumption that there are no significant spectral features
beyond the central core region of the line. This change sped up the inversion by  
$\sim$64\%.  

\item Second, instead of inverting the source function ($S_0$) and its gradient ($S_1$) separately, 
$S_0$ was fit and $(S_0+S_1)$ were fit together in the production of {\tt fd10} data, improving the 
efficiency of the inversion algorithm since these parameters are strongly coupled.    
Similarly, eta0 ($\eta_0$) and doppler width ($\Delta\lambda_D$) were found to 
be degenerate, in the sense that different combinations of the two parameters resulted
in very similar goodness of fit.  
The earlier inversion code fit $\eta_0$ and $\Delta\lambda_D$ separately, 
but {\tt fd10} fits $\left( \Delta \lambda_D \times \sqrt{\eta_0 } \right) $ and $ \Delta \lambda_D $. 
These variable changes account for a $\sim$10\% reduction in computing time for a full disk map.  

\item Third, the convergence criteria were altered to reduce
the average number of iterations from 200 for {\tt e15w1332} data 
to $\sim$30 for {\tt fd10}.
Ordinarily when the $\chi^2$ values of two consecutive successful
iterations (\textit{i.e.} $\chi^2$ is decreasing) differ by less than a tolerance value,
the algorithm exits the iteration loop.  
Although for most pixels the inversion results do not change after 30
iterations, a small percentage ($<1\%$) benefit from
increasing the number of iterations to 200. The convergence criteria
for {\tt e15w1332} was set very conservatively, $\epsilon < 10^{-15}$, so
that at each point the maximum 200-iteration limit was reached; this is the
origin of the {\tt e15} in the series name.  A new diagnostic procedure
was implemented for the {\tt fd10} data that identified the more difficult
pixels (the $<1\%$ of pixels) and enforced restarts on the problematic
pixels while reducing the average number of iterations overall.  

\item A set of weights for the Stokes profiles that provided the smoothest possible 
solution inside active regions (and performs less well in weak field regions)
was chosen for both the {\tt e15w1332} and {\tt fd10} data.  
The values are [1:3:3:2] for [\,I\,Q\,U\,\,V] for the {\tt e15w1332} data. 
A normalization of the weights by relative photon noise of the Stokes 
profiles was implemented in the {\tt fd10} data so that the 
equivalent weights are [1:2.89:2.89:2.02] for [\,I\,Q\,U\,\,V], as described in
\inlinecite{Centeno2013}.

\item A regularization term that penalizes high values of $\eta_0$ was 
added to the merit function in the
{\tt fd10} version of VFISV. This minimizes a double minima problem 
in the parameter space and helps prevent the code from converging to 
unphysical solutions. This was not present in the production of the  
{\tt e15w1332} data.

\item Limits on the range of each atmospheric parameter are set to 
prevent the algorithm from probing extreme unphysical values. These limits
were set to different values for the production of {\tt e15w1332} data.  

\item  The disambiguation for {\tt e15w1332} data is
essentially as described in Section \ref{sec:disambiguation}, but used
different annealing parameters.  The cooling speed was 0.998 for 
{\tt e15w1332} and 0.98 for {\tt fd10}.  The number of configurations attempted
at each cooling speed was 50 for {\tt e15w1332} and 100 for the SHARPs. 
The {\tt e15w1332} data relied on a simpler (not time-varying) treatment of 
the background noise. 

The value of the threshold field strength is primarily determined by the noise
level in the field. Due to the challenges in determining a realistic noise
value from the inversion 
(see Section \ref{sec:NoiseValue} for a discussion of the
temporal and large scale spatial variations found in the data),
the following simple expression was used for the threshold value
\begin{equation}
B_{\rm thresh} = B_1 + (B_0 - B_1)\sqrt{1-\mu^2},
\end{equation}
where $\mu=\cos \theta = r/r_s$ measures the distance from disk center; thus, the
parameter $B_0$ determines the threshold at disk center, while the parameter
$B_1$ determines the threshold at the limb. For the {\tt e15w1332} data,
we set $B_0=200$\,G and $B_1=400$\,G. Section \ref{sec:NoiseValue} 
explains the derivation of the threshold set for {\tt fd10}. 

\sloppy
\item No SHARP data were generated using the {\tt e15w1332} processing.  The 
geometry for NOAA AR\,\#11158 generated with the {\tt e15w1332} processing was 
based on HARP analysis with the bounding box padding adjusted manually to 
include a slightly greater area than was automatically generated.  
\fussy

\end{itemize}

%% file: VA3-SHARPDetails.txt
\section{SHARP Space-Weather Quantities}\label{app:SHARP}

With a regular time series of vector magnetic field data, the time
evolution of each active region can be characterized.
Table \ref{tab:SWIndices} lists the indices computed every 12 minutes that
are currently available as SHARP keywords. These quantities are generally mean 
values, sums, or integrations computed for the tracked HMI Active Region Patches using the
high-confidence values in the remapped CEA series. The high-confidence pixels
are those with {\sc conf\_disambig} = 90 (see Appendix \ref{app:confdisambig}) and
the number of pixels is given in keyword {\sc cmask}.
The formal uncertainty associated with each
parameter is given its own keyword. WCS-standard FITS header keywords {\sc
cdelt1}, {\sc rsun\_obs}, and {\sc rsun\_ref} were used to convert to the
units specified in the units column.  NRT data can be used for forecasting 
and definitive data for studies of how to make predictions. 
Details are provided in \inlinecite{Bobra2013}.

{\begin{table}
\caption{SHARP Active-Region Parameter Formulae}
\begin{flushleft}
\renewcommand{\arraystretch}{1.5}
\renewcommand{\tabcolsep}{0.1cm}
\begin{tabular}{ l p{4.96cm} c p{3.7cm}}
\hline
Keyword & Description & Units & Basic Formula\tabnote{Summation and normalization 
over the CMASK active pixels for mean quantities are implied; fundamental constants 
and constant multiplicative factors are generally not shown in the formulae.} \\
\hline
{\sc usflux} & Total Unsigned Vertical Flux & {\tiny Mx}  
 & $\sum|B_{z}|dA$ \\
{\vspace{4pt}\sc meangam} & Mean Deviation of Field from Radial & {\tiny Degrees} 
 & ${\rm tan}^{-1}\Big(\Big\vert \frac{B_h}{B_z}\Big\vert\Big)$ \\
\vspace{4pt} 
{\sc meangbt} & Mean Horizontal Gradient of Total Field & {\tiny G\,Mm$^{-1}$} 
 & $ \sqrt{\Big(\frac{\partial B}{\partial x}\Big)^2 + \Big(\frac{\partial B}{\partial y}\Big)^2}$ \\
\vspace{4pt}
{\sc meangbz} & Mean Horizontal Gradient of Vertical Field & {\tiny G\,Mm$^{-1}$} 
 & $ \sqrt{\Big(\frac{\partial B_z}{\partial x}\Big)^2 + \Big(\frac{\partial B_z}{\partial y}\Big)^2}$ \\
\vspace{4pt}
{\sc meangbh} & Mean Horizontal Gradient of Horizontal Field & {\tiny G\,Mm$^{-1}$} 
 & $ \sqrt{\Big(\frac{\partial B_h}{\partial x}\Big)^2 + \Big(\frac{\partial B_h}{\partial y}\Big)^2}$  \\
\vspace{4pt}
{\sc meanjzd} & Mean Vertical Current Density, $J_z$ & {\tiny mA\,m$^{-2}$} 
 & $ J_z = \Big(\frac{\partial B_y}{\partial x} - \frac{\partial B_x}{\partial y}\Big) $\\
\vspace{4pt} 
{\sc totusjz} & Total Unsigned Vertical Current & {\tiny A}  & $ \sum|J_{z}|dA$ \\
\vspace{4pt} 
{\sc meanalp} & Characteristic twist parameter, $\alpha$ & {\tiny Mm$^{-1}$}
 & $ \frac{\sum J{_z} \cdot B{_z}} {\sum B{_z}^2} $ \\
\vspace{4pt} 
{\sc meanjzh} & Mean Vertical Current Helicity & {\tiny G$^{2}$\,m$^{-1}$}
 & $h_c = B_{z} J_{z} $\\
\vspace{4pt} 
{\sc totusjh} & Total Unsigned Vertical Current Helicity & {\tiny G$^{2}$\,m$^{-1}$}
 & $ \sum|h_{c}|$ \\
\vspace{4pt}
{\sc absnjzh} & Absolute Value of the Net Vertical Current Helicity & {\tiny G$^{2}$\,m$^{-1}$}
 & $\Big|\sum h_{c}\Big|$ \\
\vspace{4pt}
{\sc savncpp} & Sum of the Absolute Value of the Net Vertical Currents Per Polarity & {\tiny A} 
 & $ \Big\vert \displaystyle\sum\limits^{B{_z}^{+}} J{_z}dA \Big\vert + \Big\vert \displaystyle\sum\limits^{B{_z}^{-}} J{_z}dA \Big\vert $ \\
\vspace{4pt}
{\sc meanpot} & Mean Photospheric Excess Magnetic Energy Density Proxy & {\tiny erg\,cm$^{-3}$} 
 & $ \rho = \frac{\Big({\bf B}^{\rm Pot} - {\bf B}^{\rm Obs}\Big)^2}{8 \pi} $\\
\vspace{4pt}
{\sc totpot} & Total Photospheric Magnetic Free Energy Density Proxy & {\tiny erg\,cm$^{-1}$}
 & $ \sum\rho\:dA $ \\
\vspace{4pt}
{\sc meanshr} & Mean Shear Angle & {\tiny Degrees}
 & cos$^{-1}\Big(\frac{\bf{B^{\rm Pot} \cdot B^{\rm Obs}}}{B^{\rm Pot} B^{\rm Obs}}\Big)$ \\
\hline
\end{tabular}
\end{flushleft}
\label{tab:SpaceweatherFormulae}
\label{tab:SWIndices}
\end{table}}

\subsection{SHARP Data Segments}

SHARPs are provided in standard CCD image coordinates and also remapped
and projected onto cylindrical equal area (CEA) coordinates.  Table
\ref{tab:SHARPSegments} lists all of the data segments provided in {\sf
hmi.Sharp\_720s} and {\sf hmi.Sharp\_720s\_nrt}. The 31 quantities include
line-of-sight observables as well as results of the VFISV inversion and
disambigutation. These
data are CCD pixel cut-outs that include the entire HARP bounding box in
the original solar disk pixels.  The inversion code provides estimates of
uncertainties as well, including $\chi^2$, the computed standard deviations
($\sigma$), and the correlation coefficients of the errors in the derived
parameters.

\begin{table}
\caption{The 31 Maps Included as SHARP Series Segments}\label{tab:SHARPSegments}
\begin{tabular}{|l|c|p{0.39\textwidth}|}
\hline
Segment Name & Units & Data (CCD Coordinates) \\
\hline
{\sc magnetogram}	& Gauss	& Line-of-Sight Magnetic Field (MDI Method) \\
{\sc bitmap}	& Code	& Bounding Box Mask for the HARP Patch \\
{\sc Dopplergram}	& m\,s$^{-1}$	& Line-of-Sight Velocity (MDI Method) \\
{\sc continuum}	& DN/s	& Continuum Intensity (MDI Method) \\
{\sc inclination}	& Degrees	& Inclination Angle of the Field to the LoS\\
{\sc azimuth}	& Degrees	& Azimuth Angle CCW from Up on the CCD \\
{\sc field}	& Mx\,cm$^{-2}$	& Total Magnetic Field Flux Density \\
{\sc vlos\_mag}	& cm\,s$^{-1}$	& LoS Velocity of Plasma with Field \\
{\sc dop\_width}	& m\AA	& Doppler Width \\
{\sc eta\_0}	& & Source Function, $\eta_0$ \\
{\sc damping}	& Doppler Width Units	& Damping Function \\
{\sc src\_continuum}	& DN/s	& Continuum Intensity \\
{\sc src\_grad}	& DN/s	& Gradient \\
{\sc alpha\_mag}	& & Field Filling Factor, currently 1.0 \\
{\sc chisq}	&	& $\chi^2$ of the VFISV Solution \\
{\sc conv\_flag}	& Code	& Flag and Index of VFISV Convergence \\
{\sc info\_map}	& Code	& Updated Quality Map \\
{\sc confid\_map}	& Code	& Updated Confidence Index \\
{\sc inclination\_err}	& Degrees	& Computed Standard Deviation ($\sigma$) of Inclination Angle \\
{\sc azimuth\_err}	& Degrees	& $\sigma$ of Azimuth \\
{\sc field\_err}	& Mx cm$^{-2}$	& $\sigma$ Field \\
{\sc vlos\_err}	& cm\,s$^{-1}$	& $\sigma$ V$_{\rm LoS}$ \\
{\sc alpha\_err}	& & $\sigma$ Alpha \\
{\sc field\_inclination\_err}	&	& Correlation of Field and Inclination Errors \\
{\sc field\_az\_err}	&	& Correlation of Field and Azimuth Errors \\
{\sc inclin\_azimuth\_err}	&	& Correlation of Inclination and Azimuth Errors \\
{\sc field\_alpha\_err}	&	& Correlation of Field and Alpha Errors \\
{\sc inclination\_alpha\_err}	&	& Correlation of Inclination and Alpha Errors \\
{\sc azimuth\_alpha\_err}	&		& Correlation of Azimuth and Alpha Errors \\
{\sc disambig}	& Code	& Flag for 180$^\circ$ Change in Disambiguated Azimuth \\
{\sc conf\_disambig}	& Code	& Confidence of Disambiguation Result \\
\hline
\end{tabular}
\end{table}

It is often useful to have the data remapped to heliographic coordinates and
the vector field projected onto three orthogonal components with uncertainties
for each component.  The CEA versions of the SHARP data series provide data
remapped onto a cylindrical equal area coordinate system centered on the
tracked center of the HARP. For a definitive SHARP the bounding box at every
time step encloses the minimum and maximum Carrington latitude and longitude
attained by the HARP during its entire lifetime. The heliographic center of
the HARP at central meridian passage is uniformly tracked at a nominal 
differential rotation rate appropriate for that latitude as given 
in {\sc omega\_dt} $= 13.6144 - 2.2$\,${\rm sin}^2({\rm lat})$ degrees per day. 
The NRT series are computed before
the full life cycle of the region is known, so the size and heliographic center 
of the NRT HARP bounding box will generally change as the region evolves, even if
regions do not merge.

Table \ref{tab:CEASegments} lists the 11 segment maps in the {\sf hmi.Sharp\_cea\_720s} and the {\sf hmi.Sharp\_cea\_720s\_nrt} series. The
$r$, $\theta$, and $\phi$ components of the field are given along with the
scalar field, velocity and continuum intensity. The line-of-sight field in
{\sc magnetogram} is remapped to the proper location, but the value is not
corrected for projection.  The uncertainty reported for each vector field
component is the one computed at the CCD pixel in the un-remapped SHARP
nearest to the center of the CEA pixel location.  Similarly, the {\sc bitmap}
and {\sc conf\_disambig} values refer to the nearest neighbor.  Because of the
remapping from a sphere to a plane and the conversion to a CEA coordinate
system, there is a slight difference between the
direction of the magnetic field vector coordinates and the coordinate system
unit vectors because the CEA coordinates do not strictly follow standard spherical
heliographic directions $r$, $\theta$, and $\phi$ except at the center of
the window \cite{Sun2013}.

\begin{table}
\caption{The 11 SHARP Cylindrical Equal Area Data Segment Maps}\label{tab:CEASegments}
\begin{tabular}{ l c p{0.6\textwidth} }
\hline
Name & Units & Data (CEA Coordinates) \\
\hline
{\sc magnetogram}	& Gauss	& Line-of-sight Magnetogram (MDI Method) \\
{\sc bitmap}	& Code	& Mask for the HARP patch \\
{\sc Dopplergram}	& m\,s$^{-1}$	& Line-of-sight Velocity (MDI Method) \\
{\sc continuum}	& DN\,s$^{-1}$	& Continuum Intensity (MDI Method)  \\
{\sc Bp}	& Gauss	& $\phi$ Component of the Magnetic Field, positive westward \\
{\sc Bt}	& Gauss	& $\theta$ Component of the Field, positive southward \\
{\sc Br}	& Gauss	& Radial Component of the Field, positive upward \\
{\sc Bp\_err}	& Gauss	& Computed $\sigma$ of B$_\phi$ \\
{\sc Bt\_err}	& Gauss	& Computed $\sigma$ of B$_\theta$ \\
{\sc Br\_err}	& Gauss	& Computed $\sigma$ of B$_r$ \\
{\sc conf\_disambig}	& Code & Confidence in Dis\-am\-bi\-gu\-ated Result. \hfill \\
\hline
\end{tabular}
\end{table}

\clearpage